\documentclass[apj,tighten,iop]{emulateapj}  % double-column

\usepackage{
	amsmath,
	amssymb,
	amsthm,
	booktabs,
	comment,
	epsfig,
	footnote,
	graphicx,
	grffile,
	import,
	morefloats,
	microtype,
	natbib,
	subfigure,
	times,
	ulem,
	url,
	verbatim,
	wasysym,
	xspace,
}

\usepackage[usenames,dvipsnames]{xcolor}
\definecolor{red}{rgb}{0.75, 0, 0}
\definecolor{green}{rgb}{0, 0.5, 0}
\definecolor{blue}{rgb}{0, 0, 0.75}
\usepackage[
    colorlinks=true,
    urlcolor=blue,
    filecolor=blue,
    linkcolor=red,
    citecolor=blue,
    pagebackref,
    bookmarks,
    bookmarksopen=true,
]{hyperref}

\newcommand{\degree}{\mbox{$^{\circ}$}}
\newcommand{\am}{\mbox{\arcmin}}
\newcommand{\as}{\mbox{\arcsec}}

\newcommand{\kms}{\mbox{\ km s$^{-1}$}} % km/s

\newcommand{\um}{\mbox{\ $\mu {\rm m}$}}
\newcommand{\kelvin}{\mbox{\ ${\rm K}$}}

\def\lsim {$\rlap{\raise.4ex\hbox{$<$}}\lower.55ex\hbox{$\sim$}\,$}
\def\gsim {$\rlap{\raise.4ex\hbox{$>$}}\lower.55ex\hbox{$\sim$}\,$}

\newcommand{\akari}{{\it Akari}}
\newcommand{\bgpsfull}{Bolocam Galactic Plane Survey}

\newcommand{\bgps}{BGPS}
\newcommand{\bolocat}{{\tt Bolocat}}
\newcommand{\cutex}{{\tt CuTeX}}

\newcommand{\gbtidl}{{\tt GBTIDL}}
\newcommand{\glimpse}{GLIMPSE}
\newcommand{\herschel}{\textit{Herschel}}
\newcommand{\higal}{Hi-GAL}
\newcommand{\hii}{H\textsc{ii}}
\newcommand{\hisa}{H\textsc{i}SA}
\newcommand{\hi}{H\textsc{i}}
\newcommand{\irac}{IRAC}

\newcommand{\mipsgal}{MIPSGAL}

\newcommand{\msx}{\textit{MSX}}

\newcommand{\spitzer}{\textit{Spitzer}}

\newcommand{\uchii}{UCH\textsc{ii}}

\newcommand{\lsun}{\mbox{L$_\odot$}}% Lsun
\newcommand{\msol}{\mbox{M$_\odot$}} % solar mass
\newcommand{\msun}{\mbox{M$_\odot$}} % solar mass

\newcommand{\tk}{\mbox{$T_{\rm K}$}}

\newcommand{\tpk}{\mbox{$T_{\rm pk}$}}

\newcommand{\vlsr}{\mbox{$v_{\rm LSR}$}}
\newcommand{\mmsd}{\mbox{$\Sigma^{\rm Total}_{\rm H_2}$}}
\newcommand{\mmsdf}{\mbox{$\Sigma^{\rm FWHM}_{\rm H_2}$}}
\newcommand{\mcl}{\mbox{$M^{\rm Total}_{\rm H_2}$}}

\newcommand{\tff}{\mbox{$t_{\rm ff}$}}
\newcommand{\lonrange}{\mbox{$10^\circ < \ell < 65^\circ$}}

\newcommand{\amm}{\mbox{{\rm NH}$_3$}}
\newcommand{\coo}{$^{13}$CO}
\newcommand{\co}{CO} % carbon monoxide

\newcommand{\hcop}{\mbox{${\rm HCO}^+$}} % HCO+
\newcommand{\metho}{\mbox{${\rm CH_3OH}$}} % methanol, CH3OH
\newcommand{\nnhp}{\mbox{${\rm N_2H^+}$}} % N2H+
\newcommand{\water}{\mbox{${\rm H_2O}$}} % H2O

\input{epsf}

\newcommand{\ovsample}{4683}  % num. BGPS clumps in overlap
\newcommand{\ovbcdpdfs}{1640}  % num. dpdfs in overlap with broadcasting
\newcommand{\newgbt}{1215}  % num. total unique GBT NH3/H2O obs from this work
\newcommand{\totmasers}{472}  % num. total unique with H2O maser detections
\newcommand{\ovproto}{2460}  % num. total protostellar in overlap
\newcommand{\ovscc}{2223}  % num. total starless in overlap (w/ AGB, HG70 3's)

\begin{document}

% affiliation marks
\newcommand{\affarizona}{1}
\newcommand{\affnrao}{2}
\newcommand{\affcfa}{3}
\newcommand{\affuab}{4}
\newcommand{\affeso}{5}
\newcommand{\affcasa}{6}
\newcommand{\affugu}{7}
\newcommand{\affyale}{8}
\newcommand{\affut}{9}

\title{The Bolocam Galactic Plane Survey. XIV. \\
Physical Properties of Massive Starless and Star Forming Clumps
}
\shorttitle{BGPS. XIV.}
\shortauthors{Svoboda et al.}
\email[$^{\dagger}$email: ]{svobodb@email.arizona.edu}

\author{
	Brian E. Svoboda\altaffilmark{\affarizona}$^{,\dagger}$,
	Yancy L. Shirley\altaffilmark{\affarizona,\affnrao},
	Cara Battersby\altaffilmark{\affcfa},
	Erik W. Rosolowsky\altaffilmark{\affuab}, \\
	Adam G. Ginsburg\altaffilmark{\affeso,\affcasa},
	Timothy P. Ellsworth-Bowers\altaffilmark{\affcasa},
	Michele R. Pestalozzi\altaffilmark{\affugu}, \\
	Miranda K. Dunham\altaffilmark{\affyale},
	Neal J.~Evans II\altaffilmark{\affut},
	John Bally\altaffilmark{\affcasa},
	Jason Glenn\altaffilmark{\affcasa}
}

\altaffiltext{\affarizona}{
	Steward Observatory,
	University of Arizona,
	933 North Cherry Avenue,
	Tucson,
	AZ 85721,
	USA
}

\altaffiltext{\affnrao}{
	Adjunct Astronomer,
	National Radio Astronomy Observatory
}

\altaffiltext{\affuab}{
	Department of Physics,
	4-181 CCIS,
	University of Alberta,
	Edmonton AB T6G 2E1,
	Canada
}

\altaffiltext{\affcasa}{
	CASA,
	University of Colorado,
	389-UCB,
	Boulder,
	CO 80309,
	USA
}

\altaffiltext{\affeso}{
	European Southern Observatory,
	Karl-Schwarzschild-Strasse 2,
	D-85748 Garching bei M\"unchen,
	Germany
}

\altaffiltext{\affugu}{
	University of Gothenburg,
	SE-412 96 Gothenburg,
    Sweden
}

\altaffiltext{\affcfa}{
	Harvard-Smithsonian Center for Astrophysics,
	60 Garden Street,
	Cambridge,
	MA 02138,
    USA
}

\altaffiltext{\affyale}{
	Department of Astronomy,
	Yale University,
	P.O. Box 208101,
	New Haven,
	CT 06520,
	USA
}

\altaffiltext{\affut}{
	Department of Astronomy,
	The University of Texas at Austin,
	2515 Speedway,
	Stop C1400,
	Austin,
	TX 78712-1205,
	USA
}

\begin{abstract}
We sort $4683$ molecular clouds between $10^\circ< \ell <65^\circ$ from the Bolocam Galactic Plane Survey based on observational diagnostics of star formation activity: compact $70$ $\mu$m sources, mid-IR color-selected YSOs, H$_2$O and CH$_3$OH masers, and UCHII regions.
We also present a combined NH$_3$-derived gas kinetic temperature and H$_2$O maser catalog for $1788$ clumps from our own GBT 100m observations and from the literature. We identify a subsample of $2223$ ($47.5\%$) starless clump candidates, the largest and most robust sample identified from a blind survey to date.
Distributions of flux density, flux concentration, solid angle, kinetic temperature, column density, radius, and mass show strong ($>1$ dex) progressions when sorted by star formation indicator.
The median starless clump candidate is marginally sub-virial ($\alpha \sim 0.7$) with $>75\%$ of clumps with known distance being gravitationally bound ($\alpha < 2$). These samples show a statistically significant increase in the median clump mass of $\Delta M \sim 170-370$ M$_\odot$ from the starless candidates to clumps associated with protostars.
This trend could be due to (i) mass growth of the clumps at $\dot{M}\sim200-440$ Msun Myr$^{-1}$ for an average free-fall $0.8$ Myr time-scale, (ii) a systematic factor of two increase in dust opacity from starless to protostellar phases, (iii) and/or a variation in the ratio of starless to protostellar clump lifetime that scales as $\sim M^{-0.4}$.
By comparing to the observed number of CH$_3$OH maser containing clumps we estimate the phase-lifetime of massive ($M>10^3$ M$_\odot$) starless clumps to be $0.37 \pm 0.08 \ {\rm Myr} \ (M/10^3 \ {\rm M}_\odot)^{-1}$; the majority ($M<450$ M$_\odot$) have phase-lifetimes longer than their average free-fall time.
\end{abstract}
\keywords{
    ISM: clouds, molecules, surveys ---
    stars: formation ---
    submillimeter: ISM ---
    masers
}
\maketitle

%%%%%%%%%%%%%%%%%%%%%%%%%%%%%%%%%%%%%%%%%%%%%%%%%%%%%%%%%%%%%%%%%%%%%%%%%%%%%%%%
%			                      Begin Body Text
%%%%%%%%%%%%%%%%%%%%%%%%%%%%%%%%%%%%%%%%%%%%%%%%%%%%%%%%%%%%%%%%%%%%%%%%%%%%%%%%

\section{Introduction}\label{sec:Introduction}
Massive stars ($M > 8 \ \msol$) strongly influence the evolution of galaxies and the ISM, yet their formation remains an open problem in contemporary astrophysics \citep[see reviews by][]{mckee07,tan14}.
A major bottleneck in the study of massive star formation is the difficulty in identifying and systematically analyzing the properties of the incipient phases of intermediate- and high-mass (protocluster) star formation, high-mass starless clumps.
These objects fragment into dense starless cores that subsequently contract to form individual or bound systems of protostars \citep{tan14}.
Starless cores (and gravitationally bound prestellar cores) are cold ($T_{\rm d} < 20$ K), dense ($n({\rm H_2}) > 10^4$ cm$^{-3}$), and centrally concentrated objects that lack an embedded protostar \citep{difrancesco07}.
The study of low-mass ($< 10 \ M_{\odot}$) starless cores has received considerable attention due to their proximity and the systematic mapping of nearby molecular clouds in the Gould's Belt \citep{wardthompson94,tafalla04,andre07,andre13}.
In contrast, the study of intermediate and high-mass starless cores and clumps is still in a nascent state due in part to their greater distances and few systematic observations to identify them \citep{wilcock12,tackenberg12, ragan13, traficante15b}. 
Our understanding of how cluster formation is initiated and ensuing protocluster evolution both ultimately depend on identifying and constraining the physical properties of representative samples of high-mass starless clumps.
This situation is beginning to change with the publication of a catalogs of 170 and 667 starless clumps identified with the \textit{Herschel Space Observatory} towards Infrared Dark Clouds \citep[IRDC;][]{wilcock12,traficante15b}.
In this paper, we analyze the star formation activity of clumps observed in the Bolocam Galactic Plane Survey to identify a population of over $2000$ massive starless clump candidates in a $84.4$ square degree region of the first quadrant of the Galaxy.

Recent blind surveys of the dust continuum emission at (sub-)millimeter wavelengths are able to detect embedded star forming regions in a wide range of evolutionary states, including the starless phases \citep[e.g.,][]{tackenberg12,traficante15b}, throughout the Galaxy.
There are three primary surveys that have mapped far-infrared through millimeter wavelength emission in the Milky Way.
The \bgpsfull\footnote{See {\tt http://irsa.ipac.caltech.edu/data/BOLOCAM\_GPS/}} \citep[\bgps;][]{aguirre11,rosolowsky10,ginsburg13} at $\lambda = 1.1$ mm surveyed from $-10\degree < \ell < 90\degree$ and targeted regions in the second and third quadrants with $|b| < 0.5\degree$ (expands to $|b| < 1.5$ at selected $\ell$) at 33$^{\prime\prime}$ resolution.
ATLASGAL \citep{schuller09,contreras13,csengeri14} surveyed at $\lambda = 870 \um$ between $-60\degree < \ell < 60\degree$ and $|b| < 1.5\degree$ with an extension covering $-80\degree < \ell < -60\degree$ and $-2.0 < b < +1.0$ at 22$^{\prime\prime}$ resolution.
In the far-infrared and submillimeter, the \herschel\ Infrared Galactic Plane Survey \citep[\higal;][]{molinari10} surveyed the entire Galactic plane in the five PACS and SPIRE bands from $\lambda = 60 - 500 \um$.
In this paper, we shall focus on studying the physical properties of starless and star forming clumps identified in the \bgps.

The \bgps\ survey identified $8294$ clumps with 1.1 mm flux densities above 100 mJy and $8594$ clumps cataloged in total \citep{rosolowsky10, aguirre11, ginsburg13}.
The second data release (v2.0) of the Bolocam images and source catalog \citep{ginsburg13} used improved analysis of the time-stream data to improve angular flux recovery with full recovery out to $\sim80\arcsec$ and partial recovery out to $\sim300\arcsec$.
The v2.0 data also resolves systematic flux and pointing calibration offsets and includes new fields in the second quadrant. 
In this work we exclusively use the \bgps\ v2.0 data products, and all clumps are referenced to the \bgps\ v2.0  maps and \bolocat\ clumps.

Extensive followup observations of \bgps\ clumps have been carried out to determine unique velocities, kinematic distances, and gas kinetic temperatures to use in calculations of physical properties.
Followup spectroscopic observations of the \bgps\ catalog in dense molecular gas tracers \hcop\ $3-2$ and \nnhp\ $3-2$ with the Arizona Radio Observatory's Heinrich Hertz Submillimeter Telescope \citep[HHSMT;][]{schlingman11,shirley13}  detected about half of the objects in the \bgps\ catalog ($>97\%$ of those clumps have a single velocity component) from which kinematic distances may be derived.
A novel Bayesian technique has been developed to calculate a continuous function of the probability that a source lies at a given heliocentric distance \citep{ellsworthbowers13,ellsworthbowers14}.
The posterior Distance Probability Density Function (DPDF) is constructed by multiplying the bimodal likelihood function (with equal probability of being located at the near or far kinematic distance) with prior distributions that account for the source positions relative to the known H$_2$ distribution in the Galaxy and the coincidence and contrast of $8 \um$ absorption features \citep{ellsworthbowers13}.
Since the dense molecular gas lines are only detected toward approximately half of BGPS clumps, \cite{ellsworthbowers14} extended the number of sources with bimodal likelihood functions by developing a morphological matching technique between $^{13}$CO $1-0$ from the Galactic Ring Survey \citep{jackson06} and the \bgps\ 1.1 mm emission.
This technique assigns a clump the dominant $^{13}$CO velocity component that matches the 1.1 mm morphology.
The DPDF formalism is a powerful tool that permits proper accounting and propagation of the distance uncertainty in calculations of distance dependent physical properties (i.e., size, mass, and luminosity).

\begin{figure}
\centering
    \includegraphics[width=0.47\textwidth]{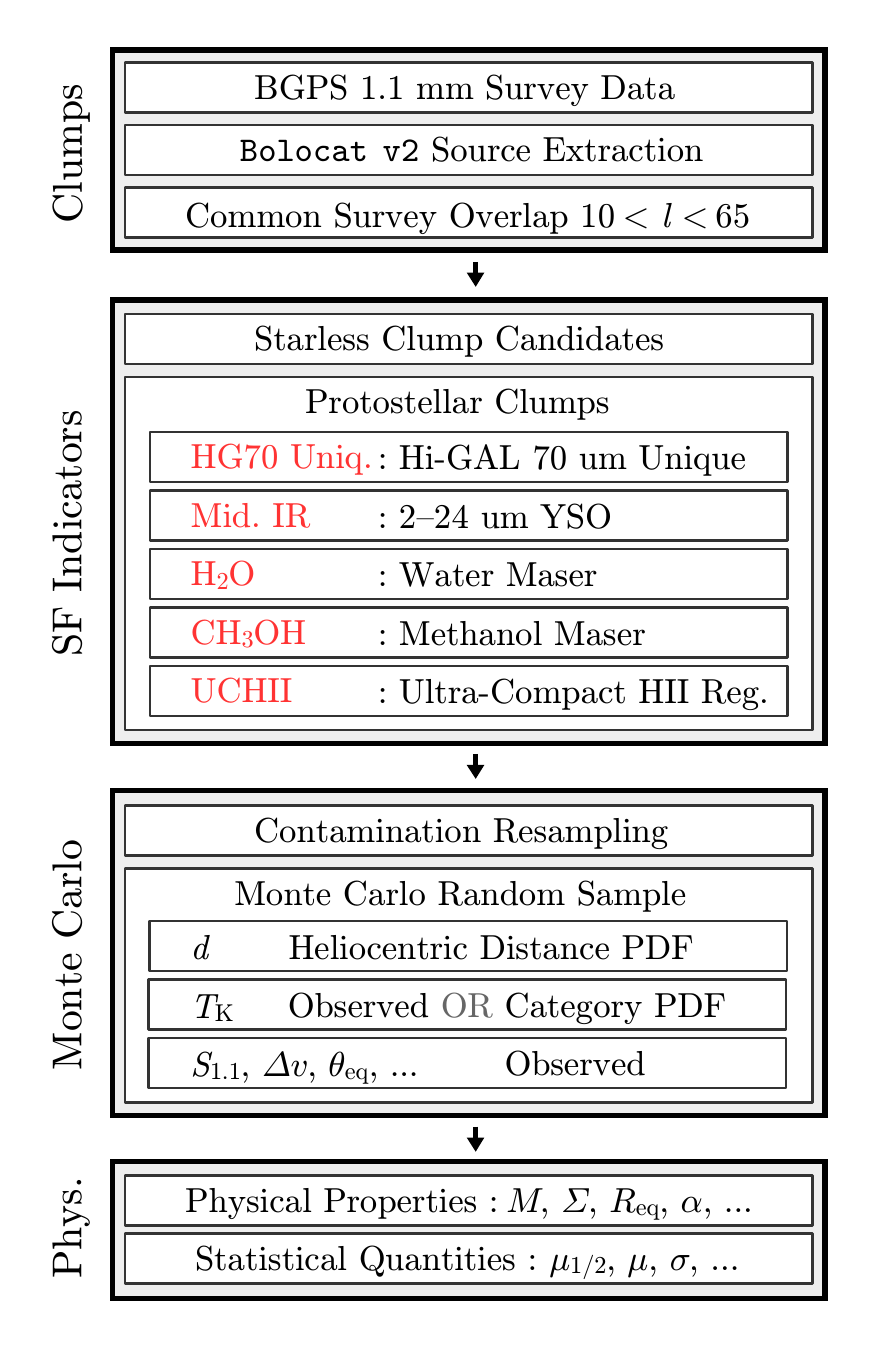}
\caption{
Flow chart describing the \bgps\ catalog, sorting by star formation
indicator, Monte Carlo random sampling, and property calculation.
}
\label{fig:Flowchart}
\end{figure}

Early BGPS papers have characterized clump physical properties on modest subsets of the \bgps.
\cite{schlingman11} determined the physical properties of 529 sources drawn uniformly from the \bgps\ 1.1 mm flux distribution, finding that BGPS sources have median physical radii of 0.75 pc, median total masses of $330$ M$_{\odot}$, and non-thermal linewidths that are $10$ times the thermal linewidth on median.
\cite{dunham11b} performed an initial evolutionary analysis on a subsample of 456 BGPS sources towards which NH$_3$ was detected, finding a median gas kinetic temperature of $16$ K and identifying a subset of starless clump candidates (SCCs) that composes $\sim1/3$ of their sample.
While \bgps\ sources are commonly referred to generically as clumps, \cite{dunham11b} pointed out that \bgps\ objects are part of a continuum of hierarchical structure which sample cores, clumps, and clouds on different scales and at different distances.
Indeed, higher resolution observations at $350 \um$ indicate that \bgps\ clumps typically break up into collections of smaller substructures \citep{merello15}.
All of these early studies made crude assumptions to resolve the kinematic distance ambiguity (i.e., all sources associated with $8 \um$ absorption are at the near distance or using the presence or lack of \hi\ self-absorption to distinguish between near and far kinematic distance; see \S4.3).
\cite{ellsworthbowers15} is the first study to utilize the full power of the DPDF formalism in which they characterize the clump mass distribution for the subset of $1710$ clumps with well-constrained DPDFs.
They find that the clump mass distribution is best fit with a lognormal distribution with an approximately power-law distribution for high-mass sources and a power-law index that is intermediate between that observed for giant molecular clouds and that observed for the stellar initial mass function.

A significant remaining challenge with the \bgps\ data set is to identify the star formation activity of all of the \bgps\ clumps and, in particular, identify the earliest phase of protocluster formation: starless clumps.
The extensive characterization and followup of the \bgps\ presents an excellent sample drawn from a blind survey to rigorously calculate population statistics and physical properties.
Measuring the physical properties of starless clumps will not only constrain the initial conditions of protocluster evolution but starless clumps will also be the proving grounds between theories of high-mass star formation, such as virialized turbulent core accretion \citep[TCA;][]{mckee03} and a sub-virial competitive accretion \citep[CA;][]{bonnell01,bonnell07}.
Resolving the current tension between these two theories is a major objective in the near-term study of high-mass star formation,
but requires the identification of a representative sample of massive starless clumps for further high resolution study.
In this paper we address questions related to how starless clumps, as the host environments of future intermediate- and high-mass stars, evolve through the starless phase into the protostellar phase.
How long does the starless phase last and how does the starless clump lifetime depend on the the mass of the clump?
Do clumps undergo significant evolution in their fundamental physical properties and do they undergo significant interaction with the surrounding molecular cloud throughout their lifetimes?
We study these questions by analyzing large, statistical subsamples ($N \sim 10^2-10^3$) of \bgps\ clumps.

We present a catalog of \ovsample\ \bgps\ clumps sorted by observational indicators of star formation activity from a comparison to other Galactic plane surveys in a common overlap range between \lonrange.
We also present \newgbt\ targeted \amm\ and \water\ maser observations with the 100 m Robert C. Byrd Green Bank Telescope (GBT) in \S\ref{sec:GbtObs}.
Schematically, we follow the procedure shown in Figure \ref{fig:Flowchart}. 
In \S\ref{sec:DevCatalog} we describe the survey data sets and methods in developing a \bgps\ star formation catalog.
In \S\ref{sec:Methodology} we describe the methods used to extend the sample of Distance Probability Density Functions to additional sources with a position-position-velocity (PPV) clustering algorithm.
We use Monte Carlo sampling of the DPDFs, re-sampling techniques to account for contamination of evolved stars, and sampling of uncertainties on observed quantities to calculate and analyze the physical properties of the \bgps\ clumps associated to star formation indicators in \S\ref{sec:Analysis}.
We discuss the properties of the starless clump candidate sample and possible evidence for clump mass growth during this phase in \S\ref{sec:Discussion}.

\section{\amm\ and \water\ Maser Survey of \bgps\ Sources}\label{sec:GbtObs}

\subsection{New GBT NH$_3$ Observations}\label{ssec:GbtObsAmm}
We have conducted targeted, spectroscopic observations of the \amm\ $(1,1)$, $(2,2)$, and $(3,3)$ inversion transitions as well as the $J_{\rm K_+, K_-} = 6_{1,6}-5_{2,3}$ $22$ GHz \water\ maser line with the GBT\footnote{The GBT is operated by the The National Radio Astronomy Observatory is a facility of the National Science Foundation operated under cooperative agreement by Associated Universities, Inc.} toward \newgbt\ new \bgps\ sources.
\cite{dunham11b} observed 631 \bgps\ sources within narrow ranges in Galactic longitude, while the new data presented in this work observed \bgps\ clumps that were detected in dense gas tracer \hcop\ $3-2$ and had $\tpk(\hcop) > 0.3$ K from \cite{schlingman11} and \cite{shirley13}.
All new observations were targeted with the center pixel of the K-band Focal Plane Array (KFPA) towards the $S_{\rm 1.1 mm}$ flux density peak in the \bgps\ v2.0 maps \citep{ginsburg13}.
We measure \amm\ properties using the slab model previously adopted in \cite{dunham11b}.
This model assumes a uniform-temperature, beam-filling slab of \amm\ with column density $N({\mathrm{NH_3}})$, intrinsic velocity dispersion $\sigma_{\rm v}$, kinetic temperature $T_{\rm K}$, and line-of-sight velocity $v_{\rm LSR}$.
The model assumes equal column densities of ortho- and para-NH$_3$ and that the levels are populated in thermodynamic equilibrium.
Given these assumptions, we use a non-linear least square regression to determine the optimal emission model incorporating the hyperfine structure and opacity of each component that matches the observed data \citep{rosolowsky08}.
As in previous pointed spectroscopy studies, the \amm\ fitting provides an excellent reproduction of the observed data with the derived parameters representing emission-weighted averages of the physical conditions within the GBT beam.
The observational setup, data reduction methods, and \amm\ fitting techniques are described in detail in \cite{dunham10,dunham11b}.

Previous KFPA observations based on the \bgps\ v1.0 are updated to match the v2.0 based on the coincidence of the pointing within the \bgps\ label-mask (discussed in \S\ref{ssec:CrossMatch} describing the catalog cross matching method).
Because of the changes in the \bgps\ catalog between data releases, 178 pointings are duplicates within the same v2.0 clumps and 245 pointings are not within a half-beam of a v2.0 clump.
We use the observation closest to the peak flux density $S_{\rm 1.1 mm}$ position when there are multiple \amm\ observations within the clump.
Columns are included in Table \ref{tab:ammprops} for the \bolocat\ v2.0 source and coordinate offset from the peak flux position.
Table \ref{tab:ammprops} also includes the observations from \cite{dunham11b} for consistency with the transition to \bgps\ v2.0.
We find an overall detection rate in \amm\ (1,1) at $75\%$.
For accurate estimates of the kinetic temperature we require a source to have $>3\sigma$ detections in both the (1,1) and (2,2) transitions, resulting in 1663 clumps with suitable \tk\ fits.
Table \ref{tab:ammprops} lists the observed \amm\ properties from the fits and $3\sigma$ upper limits for non-detections.
Unique \tk\ measurements are added for 48 \bgps\ clumps in \lonrange\ from observations in \cite{wienen12}.

\subsection{New GBT H$_2$O Maser Observations}\label{ssec:GbtObsWater}
The $22$ GHz \water\ maser line is a well known tracer of protostellar activity and was simultaneously observed in the fourth IF alongside the \amm\ $(1,1)-(3,3)$ inversion transitions.
A separate reduction in \gbtidl\ is used for the \water\ maser data because of the complex structure in the spectra from multiple overlapping maser spots in the beam, strong baseline ripples, and a telluric feature at the topocentric zero velocity.
The observations are dual-frequency switched with a 5 MHz ($69$ \kms) throw.
The telluric feature is pressure broadened by $\sim 2$ MHz ($\sim 30$ \kms) and adds significant structure to the baseline when frequency switched.
We baseline fit the spectra with first to third order polynomials to accommodate the complex baseline morphology.
Spectra are required to have a $>5\sigma$ detection in a single channel to qualify as detections.
Table \ref{tab:waterprops} lists the observed \water\ maser properties and $3\sigma$ upper limits for non-detections.
These values include the R.A.~and Dec.~of the GBT pointing, offset from the \bgps\ peak position $\theta_{\rm offset}$, velocity of the peak component $v_{\rm LSR}$, FWZI spread between the left- and right-most velocity components $v_{\rm spread}$, main-beam peak intensity $T_{\rm mb}$, main-beam integrated intensity $W$, and number of unblended velocity components $N_{\rm lines}$.
We observe \totmasers\ detections, $343$ of which are unique associations not contained in the published surveys by the Mopra and Arcetri observatories discussed in \S\ref{sssec:DescripWaterMasers}.
No significant deviation was seen between the center velocity defined by the first moment and that of the strongest peak.
The number of unblended line peaks are identified through visual inspection, but because the maser components can be heavily blended towards strong sources, this number represents a lower limit to the number of maser components towards a \bgps\ source.
For extreme \water\ maser emission, the weaker telluric feature is covered and adds a $\sim 0.5$ K km s$^{-1}$ uncertainty to the calculated \water\ maser integrated intensity.
This overlap introduces a $< 1 \%$ uncertainty in the $> 100$ K km s$^{-1}$ \water\ maser spectra where this occurs.

\section{Developing a Star Formation Indicator Catalog}\label{sec:DevCatalog}
Recently completed blind Galactic plane surveys with wavelength coverage from the near infrared to radio provide a wide range of indicators of star formation activity.
We develop an evolutionary catalog for clumps in the \bgps\ based on the indicators of star formation that can be associated with the clumps.
These evolutionary indicators are then grouped into the categories so that their observed and physical properties can be compared.
From this catalog, we identify a subsample of starless clump candidates that lack indicators of protostellar activity.
Below we describe the surveys and data sets used to define the flags of star formation indicators.
The published surveys include: the Red MSX Survey \citep{lumsden02}, Glimpse Red Source Catalog \citep{robitaille08}, Hi-GAL $70$ $\mu$m images \citep{molinari10}, H$_2$O Southern Galactic Plane Survey \citep{walsh11}, Methanol Multibeam Maser Survey (Breen et al.~2015), and Coordinated Radio and Infrared Sky Survey for High-mass Star Formation \citep{hoare12} in addition to the GBT H$_2$O maser and NH$_3$ observations presented in this paper.
We aim for the most uniform coverage in the surveys to accurately compare star formation indicator populations.
The common overlap region among the surveys described below is between \lonrange.

\subsection{Catalog Cross-Matching}\label{ssec:CrossMatch}
All published catalogs are cross-matched to the \bgps\ v2.0 catalog based on line-of-sight coincidence on the sky.
The \bolocat\ modified seeded watershed algorithm extracts the clump catalog from the Bolocam data and defines the spatial extent of each clump \citep{rosolowsky10}.
In this context the ``label-map'' or ``label-mask'' refers to the pixels in the Bolocam 1.1 mm flux image associated with a clump by the watershed.
For a source from another catalog to be matched with a clump we require that it must be contained within the clump label-mask.
In the common overlap region between \lonrange\ there are \ovsample\ BGPS clumps and 2203 clumps with either \hcop\ or \amm\ dense gas detections.
Table \ref{tab:mmprops} lists the observed 1.1 mm properties for the \ovsample\ BGPS clumps within the common overlap region.
Table \ref{tab:flagstats} shows the detection statistics for the cross-matched catalogs described in \S\ref{ssec:DescripMsx}-\S\ref{ssec:DescripCornish} to the \bgps.
The table lists the number of sources in the overlap region, the number of sources matched with \bgps\ clumps, and the number of \bgps\ clumps containing at least one matched source.

Without accurate independent distances for both a \bgps\ clump and a catalog source for a star formation tracer, we must rely on spatial coincidence on the sky to cross-match sources.
To assess the accuracy of the cross-matched associations, \cite{dunham11a} estimate the rate of chance alignments statistically.
\cite{dunham11a} compare the fraction of cross-matched sources in a given catalog to the average fraction of sources cross-matched when the coordinates have been randomized on area much greater than a single \bgps\ clump's environment (see Table 1).
Randomizing the coordinates of the catalog preserves the large-scale projected spatial distribution of sources throughout the Galaxy.
If the sources in a catalog are quasi-uniformly distributed on the sky, this randomized overlap fraction would simply be the area of the sky covered by \bgps\ clumps divided by the survey coverage.
However, because of Galactic structure and line of sight projection effects, the distribution of most catalog sources (e.g., red GLIMPSE sources) will not be uniform, and this will be reflected in a higher count of randomized cross-matches towards, e.g., the inner Galaxy.
The randomized cross-matching yields typically between $7-9\%$ of catalog sources being associated to \bgps\ clumps.
In this work we use the \bgps\ v2.0 whereas \cite{dunham11a} used the \bgps\ v1.0.
We find similar results for the RMS, EGO, and R08 sources matched to clumps (\S\ref{ssec:InfraredSurveys}) and the number of clumps with associated sources within a few percent of the total fraction of matched sources.
Catalogs that are new additions in this work such as MMB and CORNISH also have high-association rates, $>80\%$, so we do not suspect that the sources in these catalogs are highly contaminated by chance alignments.
Furthermore, other datasets in this work such as the \higal\ visual inspection and the GBT \water\ observations are targeted and the coordinates can not be randomized.

\subsection{Infrared Surveys}\label{ssec:InfraredSurveys}
\subsubsection{Red \msx\ Survey}\label{ssec:DescripMsx}
The Red MSX Survey \citep[RMS;][]{lumsden02,lumsden13} with observations at 8, 12, 14, and 21 $\micron$ defines a sample of massive YSOs complete above $L_{\rm bol} > 10^4 \ \lsun$ and $d_\odot < 10$ kpc, approximately a B0 type main sequence star.
The RMS catalog region has complete overlap with the \bgps\ for $\ell > 10\degree$, where confusion does not dominate the \msx\ data.
Extensive followup observations of RMS sources yield detailed classifications: proto-planetary nebula, planetary nebula, evolved star, \hii\ region, YSO, OH/IR star, young/old star, \hii/YSO, carbon star, and a miscellaneous ``other''.
We use the RMS classifications associated with YSOs (YSO, \hii/YSO, young/old star) to classify \bgps\ sources in the mid-IR category.
We do not associate \bgps\ sources with star formation activity for the other categories because contamination between the RMS classifications are low.
This is further evidenced by the minimal overlap in Table \ref{tab:flagstats}.
The RMS sources provide a well-vetted sample of massive YSOs, but for spectral types later than B0, or YSOs that are at distances $> 10$ kpc, the survey is incomplete.
Recent Galactic plane surveys by \spitzer\ and \herschel\ offer sensitive datasets through the mid- and far-IR to probe the star formation activity in \bgps\ clumps and complement the RMS data.

\subsubsection{\glimpse\ Red Source Catalog}\label{ssec:DescripGlimpse}
The \glimpse\ survey is a \spitzer/\irac\ Legacy survey of the Galactic mid-plane, at 3.6, 4.5, 5.8, $8.0 \um$ \citep{benjamin03,churchwell09}.
The \glimpse\ I and II fields provide complete overlap coverage with the \bgps\ for \lonrange.
\cite{robitaille08} (hereafter R08) use the \glimpse\ Catalog to a select a total number of $18{,}949$ intrinsically red ($[4.5]-[8.0] \geqslant 1$) sources that also meet photometric quality criteria.
MIPS $24\um$ magnitudes are also calculated at the same positions for $86.9\%$ of the sources from the MIPSGAL survey \citep{carey09}.
Because \glimpse\ has better spatial resolution and sensitivity than the \msx-survey, lower luminosity YSOs can be detected.
R08 estimate that a 100 $L_\odot$ Stage I YSO should be detected at distances between $0.2-10$ kpc and a $10^4$ $L_\odot$ YSO between $2-30$ kpc.
Intrinsically red sources are further classified by color-magnitude and color-color selection criteria \citep[see eq. 7-9][]{robitaille08} into candidate YSOs, ``Standard'' C- and O-rich AGB stars (sAGB), and ``Extreme'' AGB stars (xAGB).
The color criteria selects sources with spectral index $\alpha \geqslant -1.2$, which following the ``Class'' system of \cite{greene94} selects all Class I and a significant number of Class II YSOs.
Contamination from planetary nebula and background galaxies is expected to be between $2-3\%$.
However, because AGB and YSO sources have similar mid-IR colors, the YSO/AGB discrimination is uncertain based on 4.5, 8.0, and $24 \um$ photometry alone, with up to $50\%$ contamination between the two categories.
\cite{dunham11b} estimate a $\sim40\%$ contamination fraction of YSOs in the R08 candidate AGB sources.
We shall account for this contamination in the Monte Carlo sampling in \S\ref{sec:Analysis}.
The number of R08 candidate YSO or AGB sources matched to BGPS clumps is listed in Table \ref{tab:flagstats}.

An additional class of sources called Extended Green Objects (EGOs) have been identified from the \spitzer\ \glimpse\ data \citep{cyganowski08,chen13}.
EGOs are sources of extended $4.5 \um$ emission from shocked H$_2$ driven by strong outflows from massive star formation.
72 \bgps\ clumps match EGOs, and all except one clump (G044.008-00.026) contain RMS YSOs, R08 YSOs, or $70 \um$ compact sources (see \S\ref{ssec:DescripHerschel}).

\begin{figure*}
\centering
    \includegraphics[width=6in]{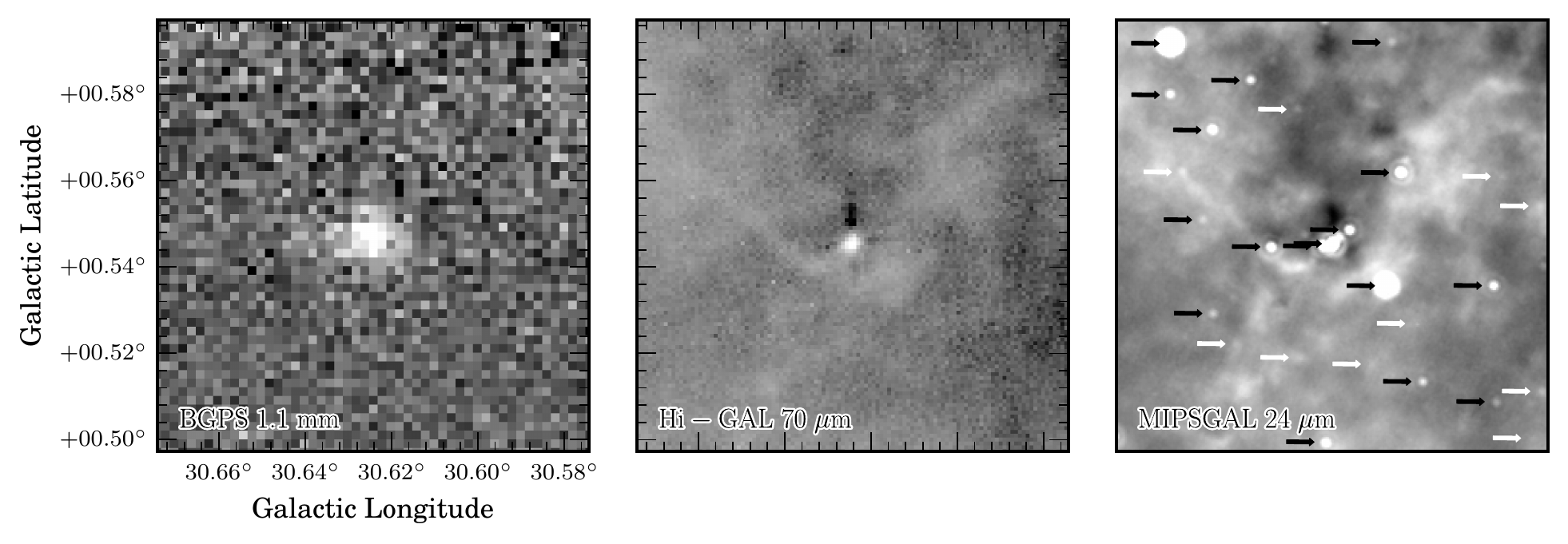}
\caption{
Comparison between BGPS 1.1 mm (Left), \higal\ $70 \um$ (Center), and MIPSGAL $24 \um$ (Right) with point sources overlaid.
The BGPS shows the location of G030.624+00.547 with a clear association to a \higal\ compact source where MIPSGAL shows numerous unassociated sources.
The black arrows point to the MIPSGAL high-quality Catalog and the white arrows point to the more complete Archive \citep{gutermuth15}.
}
\label{fig:Mipsgal}
\end{figure*}

\subsubsection{\mipsgal\ $24 \um$}\label{ssec:DescripMipsgal}
The Version 3.0 data products from the \spitzer\ MIPSGAL survey cover $-69\degree < \ell < 68\degree$ and $|b| < 1\degree$ at $24 \um$ \citep{carey09}.
The MIPS $24 \um$ band is sensitive to the hot dust found around YSOs and \hii\ regions, but also to envelopes of evolved stars.
\cite{gutermuth15} have created a comprehensive point-source catalog from the MIPSGAL $24 \um$ maps.
Two catalogs are presented: a high-quality catalog and more complete archive.
Combined the catalogs total $268{,}430$ point sources in the fields of the BGPS overlap region, with $4.9\%$ of MIPSGAL sources associated to clumps and $80.9\%$ of clumps associated to MIPSGAL point sources.
While some subset of MIPSGAL sources do represent observational indicators of embedded protostellar activity, this low match fraction of $4.9\%$ is consistent with the $6.8\% \pm 2.2\%$ randomized match fraction calculated in \cite{dunham11b} for RMS Evolved Stars (i.e., sources with no physical association).
Figure \ref{fig:Mipsgal} illustrates this poor correspondence to BGPS clumps.
Two maps at $70 \um$ and $24 \um$ are shown of the same field centered on a single, isolated clump.
Whereas the $24 \um$ image shows identified point sources from both the Catalog and Archive, only a few point sources have direct association to the clump 1.1 mm emission.
In contrast, the $70 \um$ image uniquely associates a single compact source to the clump.
The $70 \um$ dust continuum emission is more sensitive to the deeply embedded star formation activity that is better positionally correlated to BGPS clumps.
We conclude that line of sight association of $24 \um$ point sources from the MIPSGAL catalog are not a sufficient sole indicator of star formation activity because of the high contamination from evolved stars.
Thus while the MIPSGAL flags are not used in any calculations of physical properties, for completeness we present the flags in Table \ref{tab:flagstats} (see \S\ref{ssec:EvoCatSumm}).
These contamination issues also strongly affect the lower resolution WISE $22$ $\mu$m observations.
To further refine our set of star formation indicators, rather than pursue color- and quality-selection criteria to identify a less contaminated subsample from the MIPSGAL catalog, we associate clumps to $70 \um$ compact sources.

\subsubsection{\herschel\ \higal\ 70$\mu$m}\label{ssec:DescripHerschel}
The \herschel\ Infrared Galactic Plane Survey (\higal) is a far-IR survey of the Galaxy with the 70 and $170 \um$ PACS bands and the 250, 350, and $500 \um$ SPIRE bands \citep{molinari10}.
The \higal\ survey completely covers the \bgps\ fields that we discuss.
Previous survey data sets in the $100 \um$ regime from IRAS and \akari\ are not suitable for detecting faint point sources due to their poor angular resolution ($>1\arcmin$).
The \herschel\ \higal\ $10.2\arcsec$ resolution \citep{traficante11} at $70 \um$ is essential to disentangling the far-IR emission reprocessed by dust around embedded YSOs and their often active environment.
Most important however is the less severe contamination in the \higal\ $70 \um$ images from evolved stars compared to the MIPS $24 \um$ or WISE $22 \mu$m images.
Whereas $>80\%$ clumps have at least one associated MIPSGAL source, only $46\%$ of clumps have an associated $70 \um$ compact source from visual identification with clearly correlated positions to the 1.1 mm maps.

Due to complicated background structure, identifying compact sources algorithmically is difficult.
The \cutex\ algorithm \citep{molinari11} was developed specifically for identifying sources in highly spatially variable backgrounds such as those found in the \higal\ images.
Careful visual inspection is necessary because of false negatives where \cutex\ misses faint sources and false positives where \cutex\ mis-identifies sources in regions of strong emission.
Cuts on \cutex\ fit parameters were trained on publicly available data in the $\ell = 30\degree$ field \citep{bally10} and then applied to all cutout images to compare with the results of the visual inspection classifications.
The comparisons show that $77.7\%$ of the classifications agree, and $22.3\%$ disagree.
$81\%$ of the disagreements originate from sources flagged with fainter $70 \um$ sources from the visual inspection but missed by \cutex.

\begin{figure*}
\centering
    \includegraphics[width=6in]{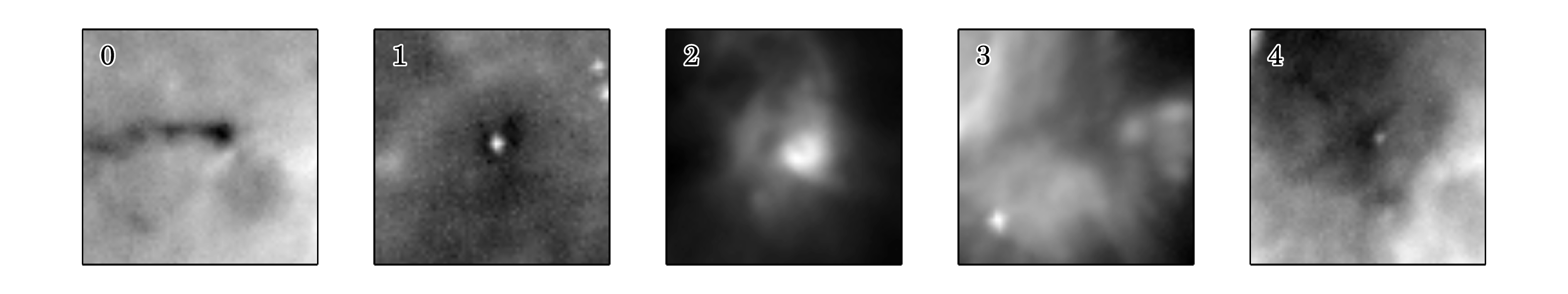}
\caption{
\higal\ visual inspection flag examples.
Flag 0: quiescent background with no source identified.
Flag 1: compact source.
Flag 2: diffuse source.
Flag 3: bright or complex background with no source identified.
Flag 4: faint or low confidence source.
The cutouts are $5\arcmin\times5\arcmin$.
}
\label{fig:HigalFlagCutouts}
\end{figure*}

We visually inspect $5\arcmin \times 5\arcmin$ cutouts of the reduced \higal\ $70 \um$ band images \citep{molinari11} for comparison against the \bgps\ data.
For consistency with the catalog matching scheme used for the other surveys in this study, we classify clumps as containing a \higal\ $70 \um$ source if it is within the \bolocat\ label-mask.
Visual inspection was carried out only on the $70 \um$ images alone, without simultaneous comparison to GLIMPSE or MIPSGAL images to prevent systematic bias in the classifications.
We assign \bgps\ clumps one of five flags in the visual inspection process.
Multiple flags are necessary to qualify the image background level and source characteristics.
Flags 1, 2, and 4 represent clumps identified with a compact source within the label-mask while flags 0 and 3 lack a clearly identifiable source.
Representative $70 \um$ cutouts of these flags are shown in Figure \ref{fig:HigalFlagCutouts}.
Flag 1's show a high confidence compact source on the scale of the \herschel\ $10.2\arcsec$ PSF.
Flag 2's are more diffuse than the PSF but identifiable as compact sources due to strong, localized emission above the background level.
These are common for extremely bright $70\um$ sources with $\geqslant 80\%$ of uniquely high-mass indicators, RMS sources, \metho\ masers, and \uchii\ regions, assigned flag 2's.
We identify 944 flag 1 and 1044 flag 2 clumps, or $42.5\%$ of clumps.
Flag 4's show a weak, PSF-like source whose classification is lower confidence.
Adding the 182 flag 4's increases the total to $46.3\%$ of clumps classified with a \higal\ source.
The brightness and complexity of the $70 \um$ background can make the visual source identification difficult.
Because of this we assign $1847$ flag 0's to clumps with quiescent, smooth, and low-level backgrounds and $655$ flag 3's to clumps with bright and more complex backgrounds.
The far-IR background of the Milky Way changes locally and as a function of Galactic longitude (generally $~10^2-10^3$ MJy sr$^{-1}$), and thus the flux threshold for source detection changes as well.

\begin{figure}
\centering
    \includegraphics[width=0.47\textwidth]{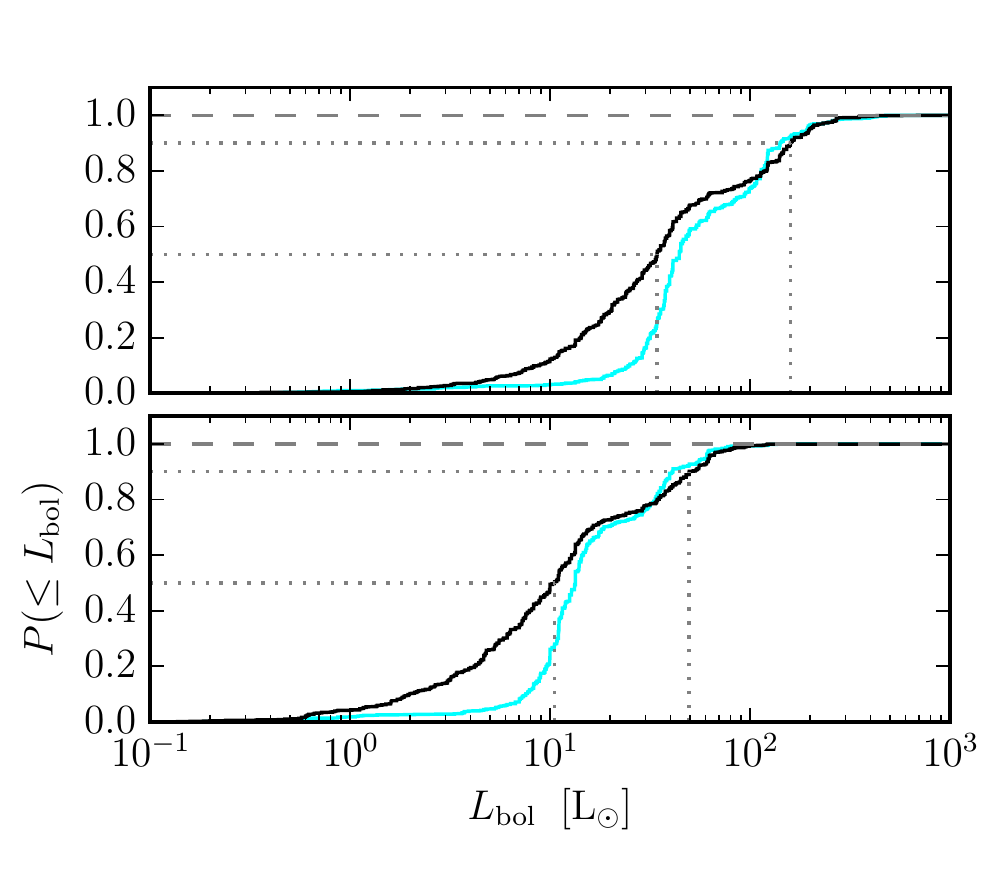}
\caption{
Completeness function for the bolometric luminosity, $L_{\rm bol}$, of \higal\ $70 \um$ sources associated to \bgps\ clumps in the Distance Catalog (black line).
Models from \cite{robitaille06} are used to compute $L_{\rm bol}$.
Top: Completeness in high background ($2000-3000$ MJy sr$^{-1}$) regions calculated with source detection threshold of $1$ Jy.
Bottom: Completeness in low background ($500-1000$ MJy sr$^{-1}$) regions calculated with source detection threshold of $0.3$ Jy.
Dotted lines show the $50\%$ and $90\%$ completeness limits and the cyan line shows the completeness function using only DPDFs between $\ell = 28-31^\circ$.
}
\label{fig:HigalComplFunc}
\end{figure}

The high rate of overlap between $70 \um$ sources and more extreme indicators along with the large number of unique $70 \um$ associations suggest that the \higal\ data is broadly probing a lower YSO bolometric luminosity, $L_{\rm bol}$, than the other indicators.
To assess the low-luminosity limit of the associations, we compute $L_{\rm bol}$ completeness functions for a representative sample of background levels given the sample of DPDFs in the Distance Catalog.
The $70 \um$ background level and complexity is the limiting factor in the identification of compact sources.
A detailed analysis of the background level distribution for the full \higal\ $70 \um$ dataset with the identification threshold quantified through fake source injection is beyond the scope of this paper.
Instead, we bracket the completeness function with a sample of $30$ clumps associated to the least confident $70 \um$ identifications (classified as \higal\ flag 4), evenly split between low background levels of $500-1000$ MJy sr$^{-1}$ and high background levels of $2000-3000$ MJy sr$^{-1}$.
Values are calculated from the cutouts described above.
Aperture photometry of the $70 \um$ sources in these selected regions yields sources identified with flux densities $\lesssim 0.3$ Jy for the low backgrounds and $\lesssim 1$ Jy for the high backgrounds.
Using these two detection thresholds, we compare to fluxes derived from the $20{,}000$ axisymmetric YSO radiative transfer models computed in \cite{robitaille06,robitaille07}.
We use the PACS $70 \um$ fluxes from the largest physical aperture, $100{,}000$ AU, which corresponds to the $20\arcsec$ diameter aperture at the median clump distance of $5$ kpc (see \S\ref{ssec:DpdfBiases}).
We then draw from Monte Carlo simulations for each clump distance probability density function (described in \S\ref{ssec:MonteCarloSampling}) to scale the \cite{robitaille06} PACS $70 \um$ model flux and apply the detection threshold cut.
Figure \ref{fig:HigalComplFunc} shows the completeness function for all clumps with well-constrained distances and the cyan line shows the same but for the subset of clumps between $28\degree < \ell < 31\degree$ for which publicly available \higal\ data exists \citep{bally10}.
The dotted lines show the $50\%$ and $90\%$ limits.
We find in the low background environments the $70 \um$ visual flags are $50\%$ complete to YSOs with $L_{\rm bol} > 10$ L$_\odot$ and $90\%$ complete to YSOs with $L_{\rm bol} > 50$ L$_\odot$.
For the high background environments, we find the $50\%$ completeness limit to YSOs at $L_{\rm bol} > 45$ L$_\odot$ and $90\%$ complete to YSOs with $L_{\rm bol} > 140$ L$_\odot$.
To put these luminosities in perspective, a $\sim1$ M$_\odot$ and $\sim5$ R$_\odot$ YSO accreting at $\sim10^{-5}$ M$_\odot$ yr$^{-1}$ corresponds an accretion luminosity of $\sim 30$ L$_\odot$ (N.B. episodic accretion rates result in large variations in the accretion luminosity at a given epoch; see \citeauthor{dunham14} \citeyear{dunham14}).
While the true sample completeness limits likely lie between these two values, it should be kept in mind that these limits are below the $L_{\rm bol}$ of the other star formation indicators by several orders of magnitude.

Among the positive \higal\ classifications, $1027/2170$ ($47.3\%$) have one or more additional star formation indicator and $1143/2170$ ($52.7\%$) are \higal\ $70 \um$ unique, representative a sample of deeply embedded and early protostellar candidates.
In the following analysis, this sample of clumps uniquely identified with star formation activity from $70 \um$ compact sources is treated separately as its own category.  It should be noted that the $70 \um$ unique category sources may have a MIPS $24 \um$ counterpart but does not contain any other star formation indicator (unique is defined in terms of our set of star formation indicators).

\subsection{Masers}
\subsubsection{\water\ Masers}\label{sssec:DescripWaterMasers}
We complement the mid- and far-IR star formation indicators with radio observations of masers.
The outflows driven by young protostars shock dense regions of the ISM and collisionally pump \water\ that creates a population inversion which drives the maser emission.
The typical physical conditions for a masing regions are a spatial density $n \sim 10^7 - 10^9$ cm$^{-3}$, kinetic temperature $\tk \sim 400 \kelvin$, spatial extent of $d \sim 0.66$ AU, and an aspect ratio $\sim50:1$ \citep{hollenbach13}.
\water\ masers are time variable on scales of months to years.
The beaming effect of \water\ masers make them a sensitive probe at large distances but because of the necessary narrow geometrical alignment, the non-detection of a \water\ maser does not imply the lack of star formation activity.
In addition to YSO molecular outflows, \water\ masers are also observed in the shells around AGB stars \citep{imai02}.
From high resolution followup of the \water\ Southern Galactic Plane Survey \citep[HOPS;][]{walsh11} and association to catalogs of star formation indicators \cite{walsh14} find that $69\%$ of water maser sites are associated with star formation, $19\%$ with evolved stars, and $12\%$ with unknown identification.
We expect limited contamination of AGB stars coincident with our GBT pointings because the observations are targeted towards the clump peak $S_{1.1 {\rm mm}}$ and $N({\rm H}_2)$ positions ($\theta_{\rm HPBW} \approx 30^{\prime\prime}$) where a YSO is likely to be located, but a chance alignment by a field AGB star is unlikely.
Thus we use the detection of a \water\ maser as an observational indicator of star formation activity.

Large observational \water\ maser survey programs have been carried out by the Arcetri survey \citep{valdettaro01}, HOPS \citep{walsh11}, and the RMS \water\ Maser Survey \citep{urquhart11}.
The variability of \water\ masers make multiple catalogs at different epochs complementary for systems that drop below a surveys sensitivity limits \citep{furuya03}.
The Arcetri catalog aggregates observations of all known literature \water\ water masers prior to 2001.
HOPS is a blind Galactic plane survey with overlap coverage with the \bgps\ extending to $\ell < 20\degree$.
The HOPS team discovered a number of new \water\ masers not found in the Arcetri catalog, most in the southern sky that is inaccessible to the Arcetri survey.
While the survey overlap between HOPS and the BGPS-GBT \water\ maser survey is limited to $10\degree < \ell < 20\degree$, it provides valuable context to evaluate the targeted GBT observations.
In this range, the GBT observations have a $3.4$ times higher detection rate towards \bgps\ clumps.
This is likely due to the tighter beam coupling of the GBT compared to Mopra and the lower baseline RMS for the targeted, single-pointing observing strategy.
The GBT observations primarily detect weaker \water\ masers with median integrated intensity $\sim 5$ Jy \kms\ compared to HOPS with $\sim 5\times10^2$ Jy \kms.

\subsubsection{\metho\ Masers}\label{ssec:DescripMethoMasers}
The 6.7 GHz Class II methanol (\metho) maser is thought to be exclusively associated with high-mass star formation because of the environments necessary for IR pumping \citep{breen13}.
A comparison between $6.7$ GHz \metho\ maser emission from the MMB and ATLASGAL $870$ $\mu$m dust emission finds that 99\% of MMB sources are associated with ATLASGAL emission indicating that methanol maser emission is almost ubiquitously associated with massive protostellar clumps \citep{urquhart15}. 
We associate \bgps\ clumps to \metho\ maser indicators based on available literature data.
We use data from the Arecibo \metho\ survey \citep{pandian07,pandian11}, the Methanol Multibeam Survey \citep[MMB;][2015]{breen12a,breen12b,breen14}, and aggregated literature data in \cite{pestalozzi05}.
Within the overlap region, MMB data extends from $10\degree < \ell < 60\degree$, Arecibo data extends from $35\degree < \ell < 54\degree$, and the data in \cite{pestalozzi05} are all-sky targeted observations.
Thus there is a small gap in the \metho\ maser survey data from $60\degree < \ell < 65\degree$, but there are only 15 BGPS clumps in this range, so there should not exist a strong sampling bias.
While the MMB is the most uniform and complete of the surveys, multiple overlapping datasets are useful because the intensity of \metho\ masers is time variable.
This can be important for masers that drop below the sensitivity limits of a single survey.
With multiple surveys at different epochs we create a more complete catalog of \metho\ masers.

\begin{figure}
\centering
    \includegraphics[width=0.47\textwidth]{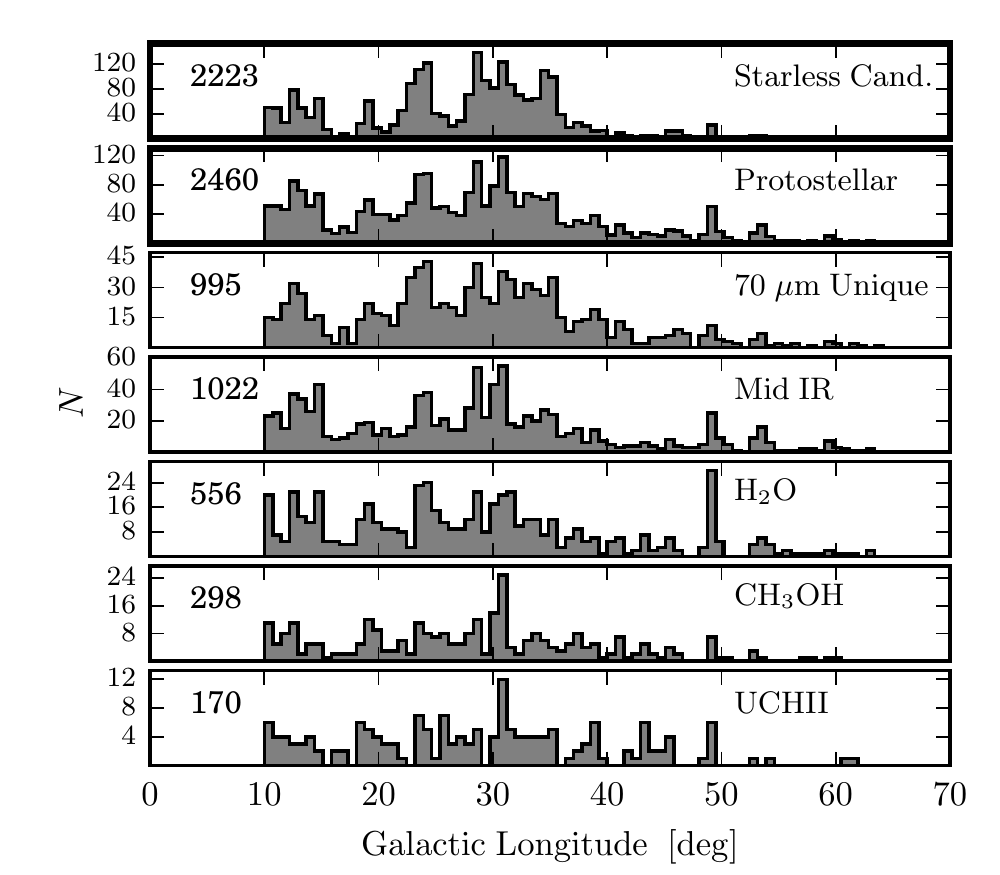}
\caption{
Observed distributions in Galactic longitude for different star formation indicators.
The top two panels show the distributions for starless clump candidates and protostar-containing clumps for all BGPS clumps in the overlap region.
The bottom five panels are ordered by star formation indicator, ranked by the approximate luminosity of a single protostar that produces the indicator.
The number of BGPS clumps in each subset is shown in the upper left.
Note that the y-axis scaling is different for each panel.
}
\label{fig:StagesHistLongitude}
\end{figure}

\subsection{Radio Continuum -- UCH{\small II} Regions}\label{ssec:DescripCornish}
\uchii\ regions are an unambiguous indicator of high-mass star formation because an OB star must be present to internally photoionize the clump.
We compare the \bgps\ with the Coordinated Radio and Infrared Sky Survey for High-Mass Star Formation \citep[CORNISH;][]{hoare12,purcell13} observed with the JVLA in 5 GHz continuum from \lonrange.
We compare all $241$ sources classified as \uchii\ region candidates in \cite{purcell13}, excluding $43$ sources classified as either ``\hii-Region'', ``Diffuse \hii-Region'', or ``Dark \hii-Region'', representative of later stages of evolution.
We find $219/243$ ($90.1\%$) CORNISH \uchii\ candidates associated with 170 \bgps\ clumps.
No \bgps\ clumps are exclusively identified with star formation activity from the CORNISH \uchii\ region sources.
All such clumps are associated with an additional evolutionary indicator: 169 \higal\ $70 \um$, 69 mid-IR, 103 \water\ maser, and 70 \metho\ maser.
\cite{urquhart13b} compare CORNISH with ATLASGAL and find $213/243$ are matched to $170$ ATLASGAL clumps.
The $24$ CORNISH sources classified as \uchii\ region candidates that are not associated with millimeter continuum emission in the \bgps\ could potentially be \bgps\ non-detections, contamination/mis-identification in the CORNISH survey, or more evolved systems that have disrupted most of the surrounding clump.

\subsection{Star Formation Indicator Catalog Summary}\label{ssec:EvoCatSumm}

\subsubsection{Star Formation Indicator Statistics}\label{sssec:IndStat}
We combine these catalogs of observational indicators of star formation activity into a series of flags to sort clumps into groups for joint analysis of their physical properties.
The most basic distinction is between clumps that are associated with no star formation indicators, the ``Starless Clump Candidates'' (SSC), and ``Protostellar'' Clumps that are associated to one or more indicator.
The star formation indicator groups are then ranked by the approximate luminosity of a single protostar that produces the indicator (see Figure \ref{fig:StagesHistLongitude}).
These rankings are a heuristic motivated by proposed linear clump evolutionary sequences \citep{battersby10,battersby11,chambers09}, but the ordering is settled upon because it consistently yields monotonic trends for clump properties.
``$70 \um$ Unique'' are clumps that only contain a \higal\ compact source, and thus are candidates for deeply embedded ($\approx$ Class 0/I) protostellar activity.
``Mid-IR'' clumps are associated to one or more R08 YSO-flagged source, RMS protostellar source, and/or EGO.
We group R08, RMS, and EGOs together because they are frequently coincident with the same protostellar sources and they use similar wavelengths to identify activity.
``\water'' clumps are associated to one or more \water\ maser from the \bgps\ GBT survey, Arcetri, and/or HOPS.
Similarly, ``\metho'' clumps are associated to one or more \metho\ masers from \cite{pestalozzi05}, Arecibo, and/or MMB.
``\uchii'' clumps are associated to the CORNISH \uchii\ regions.
In section \S\ref{sec:Analysis} we will use these categories to compare physical properties.

Table \ref{tab:flagstats} lists the detection statistics for sources matched to the \bgps\ in the overlap region and number of clumps matched to sources.
We find that in the common overlap region $\ovscc / \ovsample$ ($47.5\%$) are starless candidates and $\ovproto / \ovsample$ ($52.5\%$) show some indicator of protostellar activity.
This fraction of starless clump candidates is lower than the upper limit of $67\%$ estimated from the ATLASGAL survey by 
\cite{csengeri14} which compare ATLASGAL clumps to WISE $22 \um$ and MSX $21 \um$ point source catalogs.  
The smaller fraction of starless clump candidates in our analysis is likely due to the inclusion of protostellar indicators with lower luminosity sensitivity such as \textit{Herschel} 70$\mu$m sources in our analysis.
The \textit{Herschel} $70$ $\mu$m sources accounts for the largest fraction of star formation indicators detected toward Protostellar clumps ($2168/2460$ or $88\%$ of Protostellar clumps contain a $70$ $\mu$m source flag).
No strong variation in the Galactic longitude distribution is observed between Starless Clump Candidates and Protostellar Clumps (Figure \ref{fig:StagesHistLongitude}).
Individual star formation indicator flags for each clump are listed in Table \ref{tab:SfFlag}.

\begin{figure*}
\centering
    \includegraphics[width=0.7\textwidth]{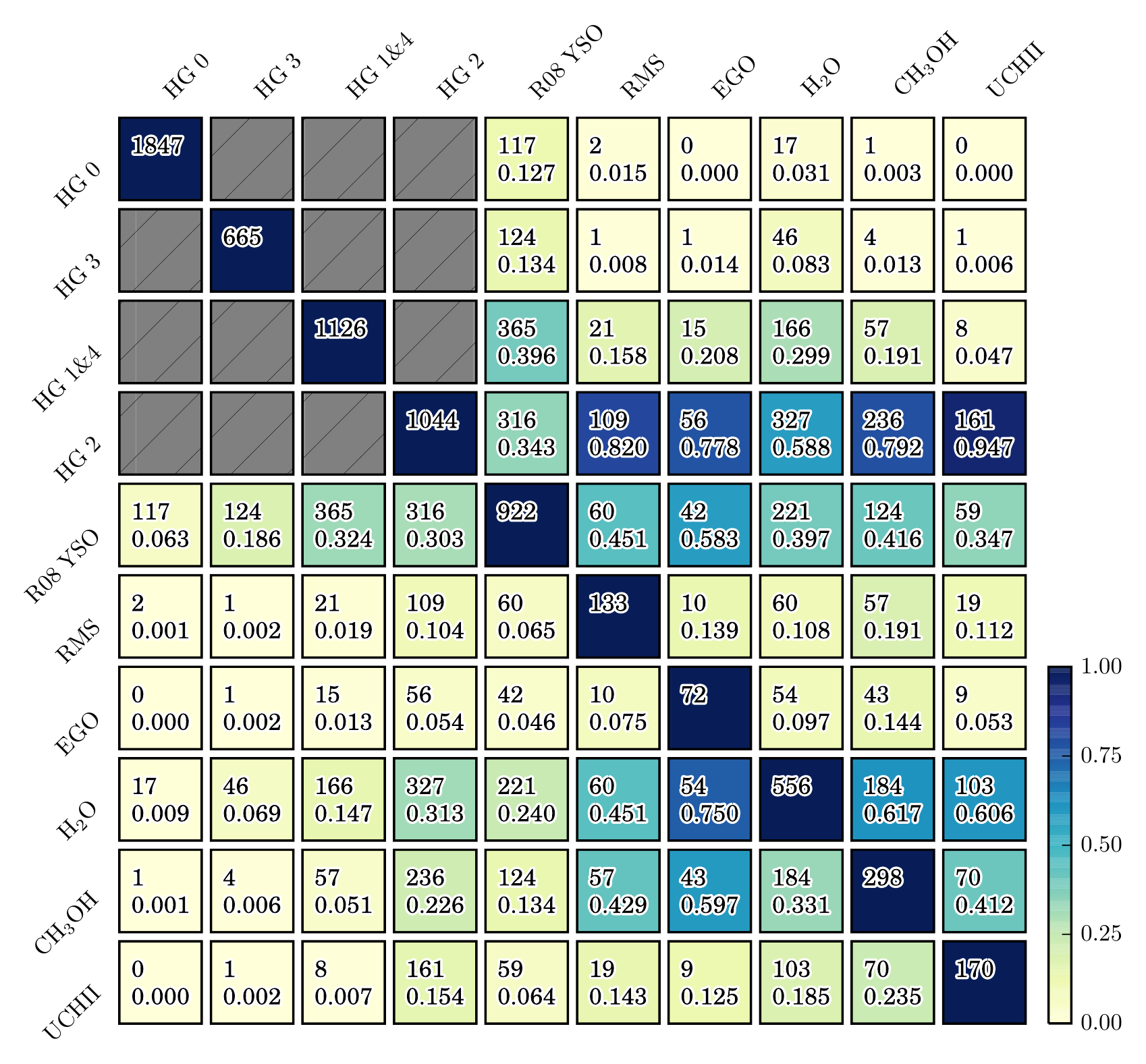}
\caption{
Intersection between pairs of star formation indicators.
Totals for each indicator are shown along the diagonal.
The top value in each square shows the number of overlapping indicators and the bottom value shows the overlap fraction calculated with respect to the total of the column.
The overlap fraction is color-coded by the scale on the right.
To read, select the first indicator along the top axis, and drop down to the row of the second indicator along the left axis.
The \higal\ flags are mutually exclusive for each clump and are grayed out.
}
\label{fig:overlap}
\end{figure*}

Except for $70 \um$ Unique, the \bgps\ indicator groups are not mutually exclusive.
Indeed, $70/170$ ($41\%$) of \uchii\ clumps are also associated to a \metho\ maser and $103/170$ ($61\%$) are associated to a \water\ maser.
There are no BGPS clumps uniquely associated with \water , \metho , or \uchii\ flags; those clumps are always associated with another protostellar indicator.
This fact indicates that it is unlikely that a population of H$_2$O unique clumps (meaning H$_2$O masers are the only indicator of star formation activity) exist toward clumps that have been targeted by our GBT observations (\S\ref{ssec:GbtObsWater} and \S\ref{sssec:DescripWaterMasers}).
Figure \ref{fig:overlap} shows the intersection between indicators with the fraction of the total calculated for each.
For example, $70$ clumps are associated to both \metho\ masers and \uchii\ regions: $0.412$ ($70/170$) of those with \uchii\ regions are also associated to \metho\ masers, alternatively $0.235$ ($70/298$) of those with \metho\ masers are also associated to \uchii\ regions.

Among these indicators the \metho\ and \uchii\ clumps are the most consistent indicators of high-mass protostellar activity.
However, it is important to note that more extreme indicators of star formation activity do not necessarily represent a later stage in a linear clump evolutionary sequence.
For one, not all clumps will have sufficient mass to form an OB star to drive an observable \metho\ maser or \uchii\ region.
The sequence of star formation indicators outlined in \cite{battersby10} is consistent with the expected sequence of observational indicators for a high-mass protostellar {\it core}, but \bgps\ {\it clumps} are hierarchical constructs of density, likely composed of multiple cores \citep{merello15}.
The surveys also do not have high enough resolution to identify single protostellar sources at large distances.
For example, what appears to be a \higal\ compact source could be a protocluster at several kpc.
In this case, the only certain true evolutionary transition is from the starless clump phase to the protostellar phase.

\subsubsection{Comparison with Published Starless Clump Catalogs}
We compare our protostellar indicator flags to the previously published starless clump catalog of \cite{tackenberg12} which identified $210$ starless candidates with peak column density ${\rm N(H_2)} > 10^{23}$ cm$^{-2}$ from ATLASGAL $870 \um$ images in the $10\degree < l < 20\degree$ range.
\cite{tackenberg12} use a combination of mid-IR colors from \spitzer\ imaging plus visual analysis of $24 \um$ point sources to identify starless candidates.
Within this longitude range, there are $52$ \bgps\ catalog sources with peak positions that lie within the \bgps\ angular resolution ($33\arcsec$) of the \cite{tackenberg12} starless candidate position.
We find that $30$ out of those $52$ clumps ($58\%$) contain an indicator of star formation activity in our catalog.
Nearly all of those associations result from identification of a $70 \um$ source in our catalog.
This result highlights the necessity of far-IR imaging to identify deeply embedded and lower luminosity YSOs.

We also compare to the \cite{traficante15b} catalog of $1684$ \higal\ clumps associated to IRDCs in the $10\degree < \ell < 55\degree$ and $|b| \leq 1\degree$ range.
Clumps are extracted from the IRDC catalog of \cite{peretto09}.
A novel algorithm \citep[\texttt{Hyper};][]{traficante15a} is used for clump extraction and photometry, and counterparts at $70$ $\mu$m are used for protostellar identification.
When compared by positions, $963$ clumps in \cite{traficante15b} are associated to $692$ BGPS clumps ($14.8\%$ of the BGPS clumps in the overlap region), whereas the remaining $721$ clumps are not associated.
Among the $692$ BGPS clumps with associations, $550$ ($80.3\%$) star forming categorizations agree and $142$ ($19.7\%$) conflict.
The conflicts are dominated ($109$ of $142$ conflicts) by cases of a smaller ``sub-component'' starless clump in the \cite{traficante15b} catalog associated to a larger ``parent'' BGPS clump with a protostellar indicator (N.B. the \textit{Herschel} \higal\ resolution at 160 $\mu$m is $13.6^{\prime\prime}$ which is $2.4$ times higher resolution than the BGPS at 1.1 mm).
A minority of $33$ ($4.8\%$) BGPS clumps remain that we have classified as starless candidates that contain $70$ $\mu$m counterparts from the \cite{traficante15b} catalog.
In these cases we have conservatively assigned \higal\ flag 3s ($28/33$) for ambiguous high-backgrounds in the visual inspection.
%In total $692/4683$ ($14.8\%$) BGPS clumps in the overlap region have clump counterparts in the \cite{traficante15b} catalog. 
%The BGPS has a larger number of clumps because it is drawn from a blind dust continuum survey.

\section{Determining Heliocentric Distance}\label{sec:Methodology}
\subsection{Distance Probability Density Functions}
\label{ssec:DpdfSampling}
A radial velocity measurement for Galactic positions within the Solar Circle yields a kinematic ambiguity between a near and far distance.
Unique radial velocities are determined from observations of dense gas tracers (HCO$^+$, N$_2$H$^+$, or NH$_3$; \citeauthor{shirley13} \citeyear{shirley13}, \citeauthor{dunham11b} \citeyear{dunham11b}) 
and from morphological matching the \bgps\ clumps to the \coo\ spectra from the Galactic Ring Survey (GRS; see 
\citeauthor{ellsworthbowers14} \citeyear{ellsworthbowers14}).
A Bayesian statistical methodology of formulating clump distance probability density functions (DPDFs) based on a range of prior distributions has been carried out and applied to the \bgps\ by \cite{ellsworthbowers13} and \cite{ellsworthbowers14}.
Further, to accurately propagate our uncertainty in heliocentric distance, we shall use Monte Carlo random sampling of the DPDFs when calculating the derived physical properties.
The DPDF is calculated by multiplying the likelihood function, derived from the measured  $v_{\rm LSR}$ with a $7$ km/s uncertainty to account for typical GMC dispersion and the \cite{reid14} rotation curve of the Galaxy, with prior distributions
\begin{equation}
{\rm DPDF}(d_{\odot}) = \mathcal{L}(v_{\rm LSR},l,b;d_{\odot})\, \prod_i P_i(d_{\odot},l,b) \;\; .
\end{equation}
The priors are calculated from morphological matching to $8 \um$ absorption features, latitude offsets compared to the \cite{wolfire03} model of the H$_2$ distribution in the Galaxy, maser parallax measurements, and proximity to an \hii\ region of known distance.
The well-constrained DPDFs are defined to have full-width of the $68\%$ maximum-likelihood error bar ${\rm FW}_{68} \leq 2.3$ kpc \citep{ellsworthbowers13}.
The total number of well-constrained DPDFs is $1710$ in the BGPS survey.

\subsection{Distance Resolution Broadcasting}\label{ssec:KdarBroadcasting}
% Motivation for expanding DPDFs & KDARs
The well-constrained \bgps\ distance sample contains $1414$ clump DPDFs representing $43\%$ of the \bgps\ velocity catalog in \lonrange.
Clumps show an easily distinguishable tendency to cluster in velocity coherent groups.
Indeed, $75.4\%$ of clumps in the overlap region have at least one adjacent neighbor (i.e., pixels touching) in the \bolocat\ label-masks.
Clumps that are within spatial and kinematic proximity are more likely to be cospatial than chance alignments of non-cospatial near/far kinematic distances.
In order to increase the size of the Distance Sample, we apply a modified Friends-of-Friends algorithm \citep{huchra82} to associate or ``broadcast'' the known DPDFs based on whether clumps are nearby in $(\ell, b, \vlsr)$ or position-position-velocity (PPV) space.
Because this technique uses only proximity in PPV-space, the number of new associations does not have a strong selection effect from evolutionary stage or physical properties.
To evaluate the optimal search angle-velocity search parameters we use the Distance Sample as a training set, and minimize the fraction of near/far misassociations.

We assign a clump to a ``PPV-group'' or ``group'' by searching a distance $\rho$ around the clump in angular separation, $\theta_s$, and velocity, $v_s$:
\begin{equation}
    \rho = \sqrt{ \Delta \theta^2 + \left( \frac{\theta_s}{v_s} \Delta v \right)^2} < \sqrt{2} \theta_s
\end{equation}
Clumps within this neighborhood are assigned to a parent group and the group is iteratively expanded by searching around the neighbors.
This algorithm effectively creates groups out of a minimum density in PPV-space.
We calculate group assignments over a $30\times30$ grid of angular search and velocity search radii from $\theta_s = 1.8 - 10.5\arcmin$ and $v_s = 1 - 6 \kms$.
We use the known maximum likelihood distances as a training set to evaluate the accuracy of the search parameters.
We denote the conflict ratio, $R_{\rm conf}$, as the number of clumps in groups with disagreeing distance resolutions divided by the number of clumps in groups with at least two distance resolutions.
With $R_{\rm conf}$ as a figure of merit, we select the search parameters that maximizes the number of new distance resolutions for fixed $R_{\rm conf}$.
It is important to note however that the DPDFs are not simply Boolean near/far kinematic distance resolutions.
Thus, $R_{\rm conf}$ can only be approximately as low as the mean weight between the near/far probabilities in the DPDFs.
\citep{ellsworthbowers13} find that $8\%$ of the DPDFs disagree with the \coo\ Galactic Ring Survey kinematic distance ambiguity resolutions (KDARs) from \cite{romanduval09} using \hi\ self-absorption (\hi SA).
We fix the conflict ratio to be reasonably consistent with this figure at $R_{\rm conf} = 0.12$ and select the search parameters $(\theta_s,\ v_s) = (3.9\arcmin,\ 4.0 \kms)$ that maximize new associations.
In total, we associate $446$ new DPDFs over the full survey, and $226$ in the survey overlap region.
The KDA resolution broadcasting method thus increases the Distance Sample used in this study by $226$ ($16\%$) to $1640$ clumps in the overlap region.

\begin{figure}
\centering
    \includegraphics[width=0.47\textwidth]{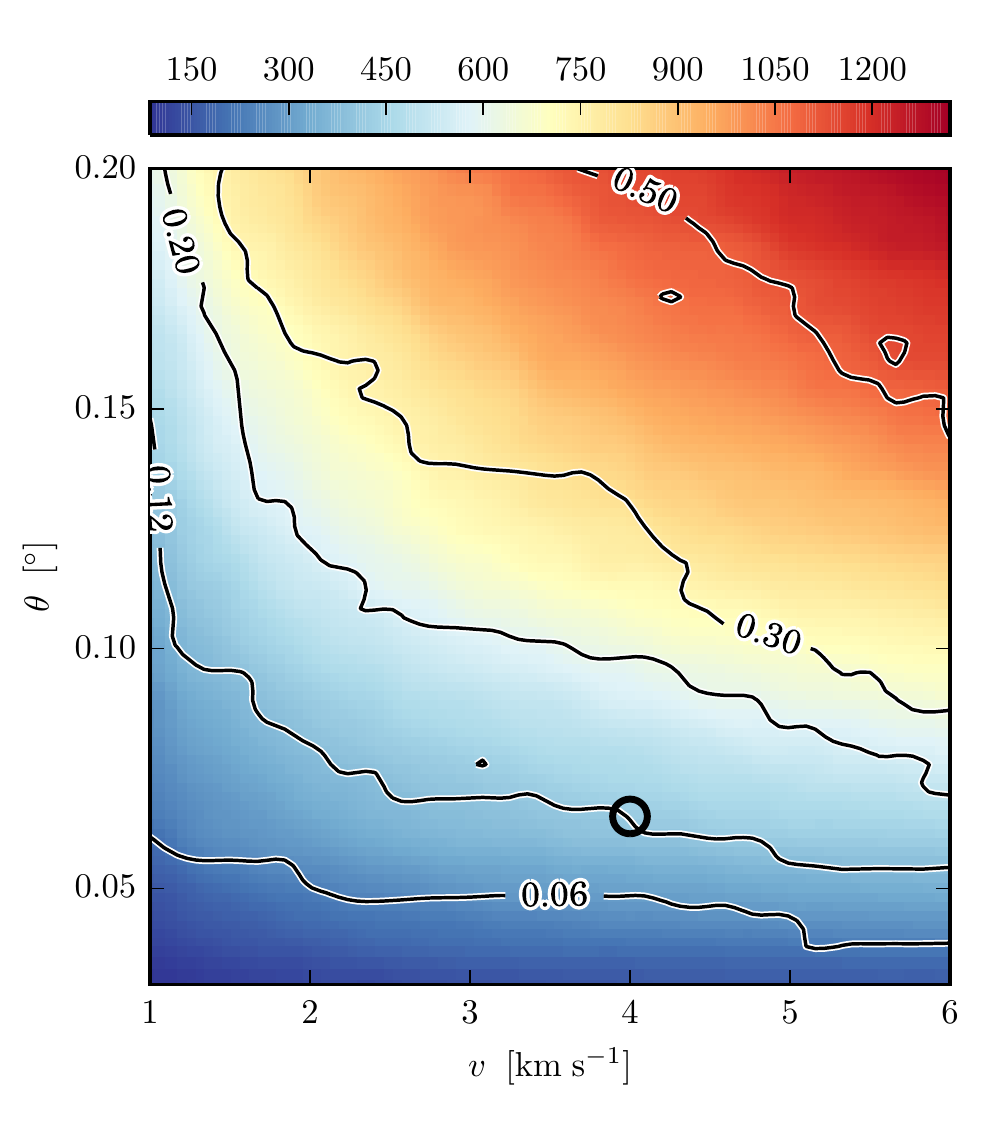}
\caption{
Optimization of the velocity and angular search parameters for the number of new clump KDA resolutions (color-scale).
The new KDA resolutions are in PPV-groups with one or more resolved KDA member that have a well-constrained DPDF.
Black contours show the conflict ratios $R_{\rm conf}$, or the ratio of clumps in PPV-groups with conflicting Near/Far KDA resolutions to the total number of clumps in PPV-groups with two or more distance-resolved members.
Maximizing the number of new KDAR associations for a fixed $R_{\rm conf} = 0.12$ gives $446$ new distance resolutions at $\theta = 0.065\degree$ (3.9\arcmin) and $v = 4.0 \kms$ (black circle).
}
\label{fig:PPVModelNewAssoc}
\end{figure}

The DPDFs added to the Distance Sample are constructed from the weighted priors of the DPDFs within the group.
The optimized search parameters are used to associate the neighboring clumps in the group to a common kinematic near/far distance.
To create a new posterior DPDF for a clump in a group, we apply a weighted average of all priors from the clumps in the group with a DPDF to the the new group-member clump. 
The priors are given Gaussian weights by the PPV-distance $\rho$ with FWHM values set to the optimized $\theta_s$ and $v_s$ above.
This weighting ensures that the closest clump priors in the group are most strongly weighted.
To propagate our uncertainty in broadcasting the distance resolution of an individual clump, we apply a weighted error function (as a step-function analogue) to the new posterior DPDF to account for $R_{\rm conf} = 0.12$.
Down-weighting the distance resolutions by the conflict fraction accounts for the mean uncertainty in the group associations.

\subsection{Distance Comparisons}
\cite{dunham11b} used a combination of visually identified $8 \um$ absorption features and \hi\ self-absorption (\hisa) to resolve the kinematic distance ambiguity (KDA) towards a sample of \bgps\ clumps.
Because these sources are also in our sample, we compare the distance resolutions in \cite{dunham11b} with the resolutions of the well-constrained DPDFs in \cite{ellsworthbowers13} and \cite{ellsworthbowers14}.
The results of \cite{dunham11b} are based on the \bgps\ v1.0 data release and are assigned to \bgps\ v2.0 sources by mapping the peak flux positions of the v1.0 sources onto the v2.0 source masks.
Of the $456$ sources with resolved distances in \cite{dunham11b}, $403$ match to v2.0 sources and $100$ have well-resolved KDAs in our sample.
For sources flagged as either ``near'' or ``tangent'' in \cite{dunham11b}, we find that $69 / 78 \approx 88\%$ agree.
For sources flagged as ``far'' that had no incidence of an $8 \um$ absorption feature nor \hisa, we find that only $11 / 20 \approx 53\%$ agree.

The lack of an \hisa\ feature is commonly used to associate GMCs to the far kinematic distance \citep{romanduval10}, where the superposition against the background \hi\ distribution produces absorption at the velocity coincident with the high-column density present in the GMC.
The clumps found in the \bgps\ are both on smaller spatial scales and at higher spatial densities than GMCs.
The disagreement for far distance resolutions for clumps is because the \hi\ distribution does not follow that of the clump dust continuum, and thus the lack of a \hisa\ feature can not be used to accurately resolve the kinematic distance ambiguity for BGPS clumps.

\cite{wienen15} use associations of $^{13}$CO GRS isocontours in PPV-space and the associated \hi\ absorption with those GRS clouds to resolve $44\%$ of ATLASGAL clumps to the far kinematic distance.  
This is twice the $22\%$ fraction found in \cite{ellsworthbowers14}.
Comparison of the \cite{wienen15} KDA resolutions with \cite{ellsworthbowers14} DPDFs indicate a $30$\% conflict fraction. 
The origin of the larger conflict fraction is the larger PPV spaces used to associate objects in the \cite{wienen15} paper.
Figure \ref{fig:PPVModelNewAssoc} shows that the conflict fraction increases with angular separation and velocity for larger PPV groups.
This result combined with the larger number of far kinematic resolutions will systematically skew distributions of physical properties such as $M$, $R$, $n$, and $\langle t_{\rm ff} \rangle$ to larger values for the \cite{wienen15} sample.

\section{Analysis of Physical Properties with Star Formation Indicators}\label{sec:Analysis}
In the following subsections, we shall compare and analyze observable and physical quantities for both the starless clump candidates and protostellar clumps.
For the protostellar clumps, we sort clumps by their star formation indicators into subsamples described in \S\ref{ssec:EvoCatSumm}.
As noted in \S\ref{sec:DevCatalog}, this ordering is not an evolutionary sequence but is roughly ordered by the typical luminosity required to produce each indicator.
In the following figures, the top two panels indicate the starless clump candidate and protostellar candidate subsamples while subsequent panels indicate various star formation indicator subsamples.
For the physical properties and sampling techniques described below, Table \ref{tab:PhysProps} lists the derived physical properties and uncertainties for each clump in the Distance Sample, and Table \ref{tab:StagesStats} lists the statistical properties and uncertainties for subsamples of clumps sorted by star formation indicator.

\subsection{Monte Carlo Sampling}\label{ssec:MonteCarloSampling}
For each subsample we compute Monte Carlo (MC) simulations to randomly sample all of the input parameters, both measured quantities with statistical uncertainties (e.g.,~$S_\nu$, \tk) and DPDFs. 
By calculating descriptive statistical quantities (e.g.,~$\mu_{1/2}$, $\rho_{\rm spear}$) from suites of MC simulations, we can estimate the uncertainty and significance of comparisons between samples of clumps associated with different star formation indicators.
The suites of MC simulations also more accurately propagate the uncertainties when compared to simply creating a single histogram of the maximum likelihood values, because some quantities such as heliocentric distance can have highly asymmetric and/or bimodal PDFs.
Thus, while a single clump may have large fractional uncertainties in any given derived physical property, we can make robust and accurate calculations for large ($\sim10^2-10^3$), statistically significant subsamples of clumps. 

\subsubsection{AGB Contamination Re-sampling}
Contamination from AGB stars is a significant source of uncertainty when the only indicator of star formation activity within a clump is from the mid-infrared colors of a point source.  In order to propagate this uncertainty due to AGB contamination, we re-sample flags discriminating between R08 YSOs and AGBs sources in the MC simulations. We use conservative estimates of the contamination fractions between the R08 YSO contaminants in the AGB classification and AGB contaminants in the YSO classification.
As described in \S\ref{ssec:DescripGlimpse}, we apply contamination fractions of  $40\%$ for AGB sources and $50\%$ for YSO sources.
All R08 sources within clumps are re-sampled this way.
If a clump is associated to another star formation indicator besides R08, then it is still flagged as protostellar and the MC flag re-sampling will not affect the clump's overall designation.
In the overlap region, only 280 clumps ($6.0\%$) are uniquely associated with R08 sources, and 122 clumps ($7.4\%$) in the Distance Sample.  Re-sampling propagates our uncertainty in the R08 flags, but because of significant coincidence with other star formation indicators, it introduces only a minor effect in discriminating between starless clump candidates and protostellar clumps.  In the following analysis, AGB contamination re-sampling is used in all MC calculations.

\subsubsection{DPDF Sampling and Distance Biases}\label{ssec:DpdfBiases}
For the calculation of quantities which are distance dependent (e.g., $R$, $M$), we draw the heliocentric distance from the source DPDF.
Each MC simulation consists of $10^4$ random draws.
Since the output distributions of the physical properties may be skewed, we use robust statistical indicators such as the median ($\mu_{1/2}$) and median absolute deviation ($\sigma_{1/2}$)\footnote{The median absolute deviation is defined by $\sigma_{1/2} = \mu_{1/2,i}(|X_i - \mu_{1/2,j}(X_j)|)$ for a univariate dataset $X$. For normally distributed data, $\sigma_{1/2}$ is related to the standard deviation by $\sigma \approx 1.48260 \sigma_{1/2}$.} as conservative characterizations of the distribution of properties for a particular subsample. 

\begin{figure*}[tbh]
\centering
    \includegraphics[width=0.47\textwidth]{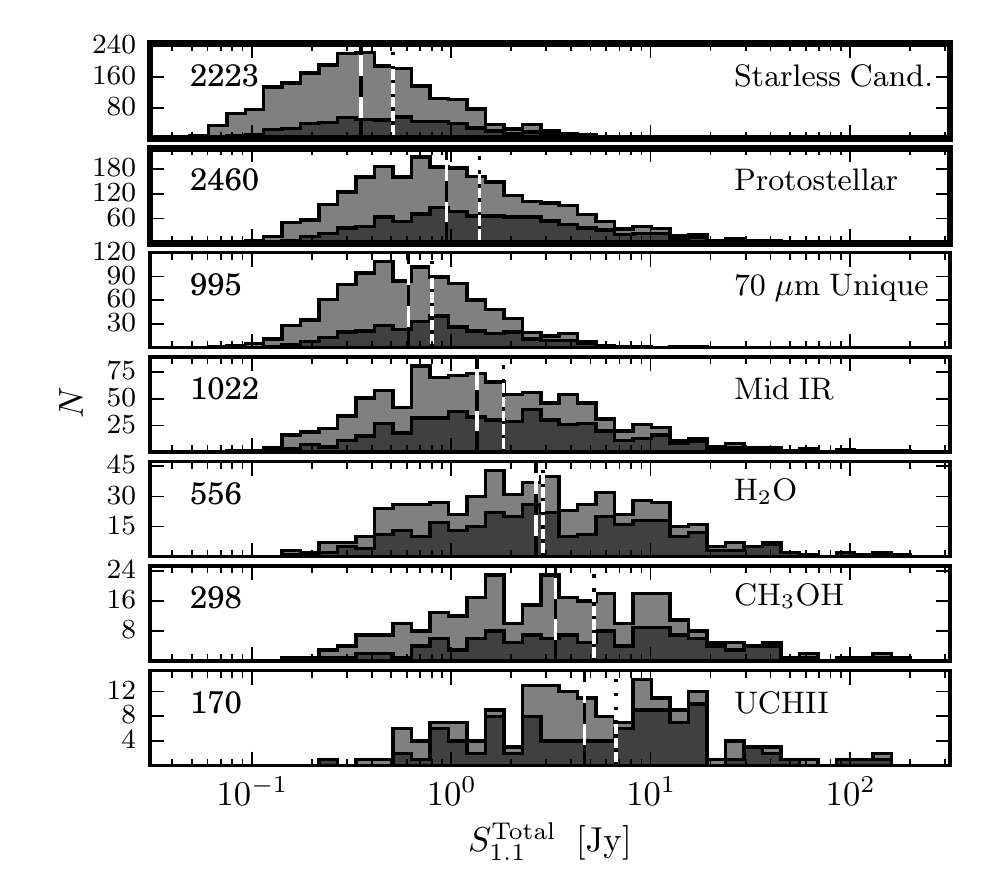}
    \includegraphics[width=0.47\textwidth]{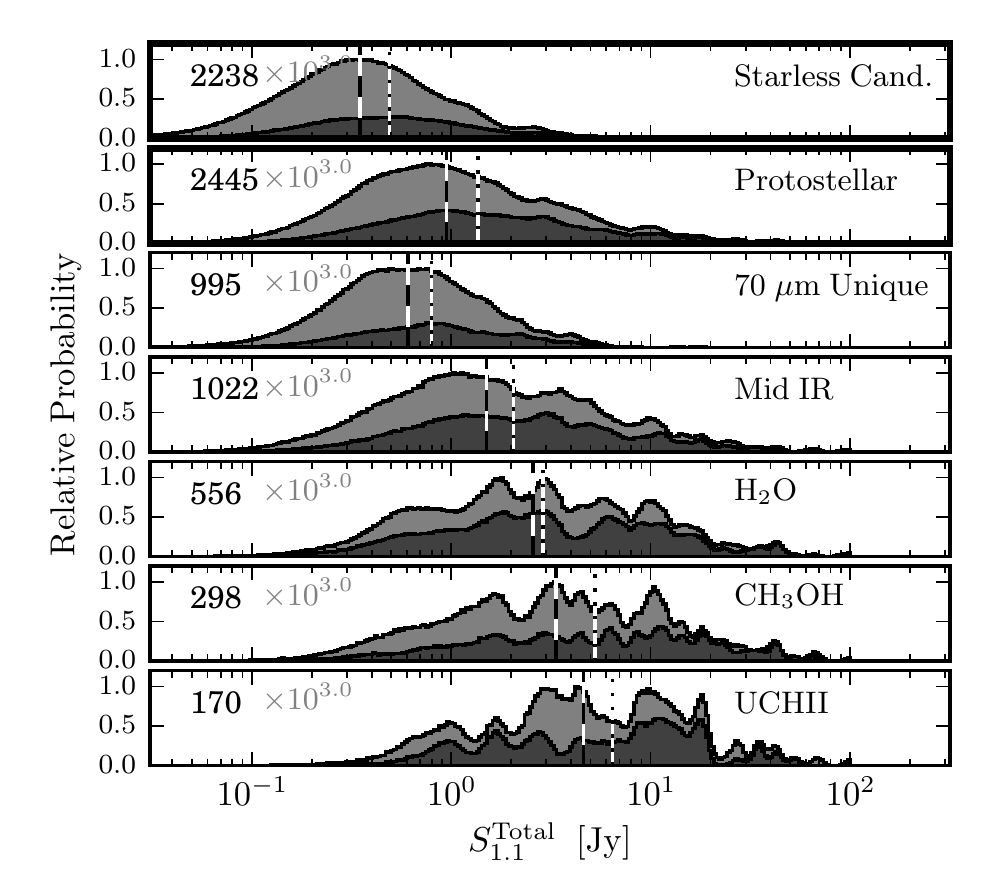} \\
    \includegraphics[width=0.47\textwidth]{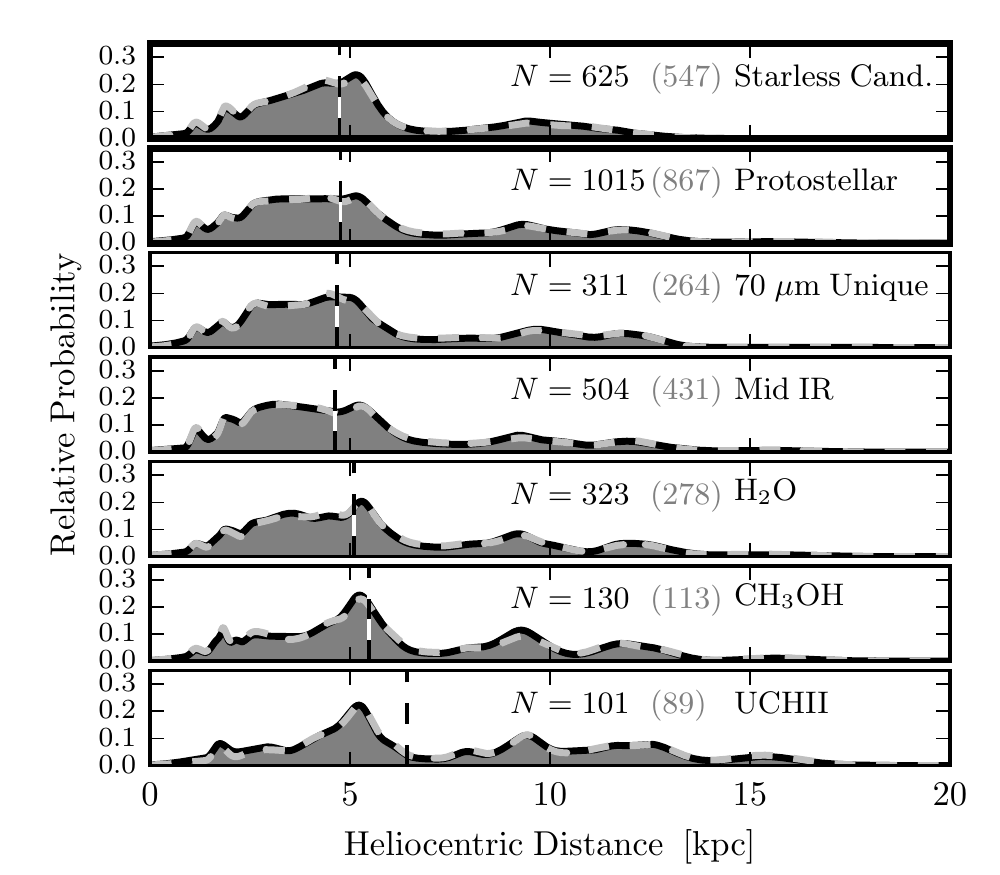}
    \includegraphics[width=0.47\textwidth]{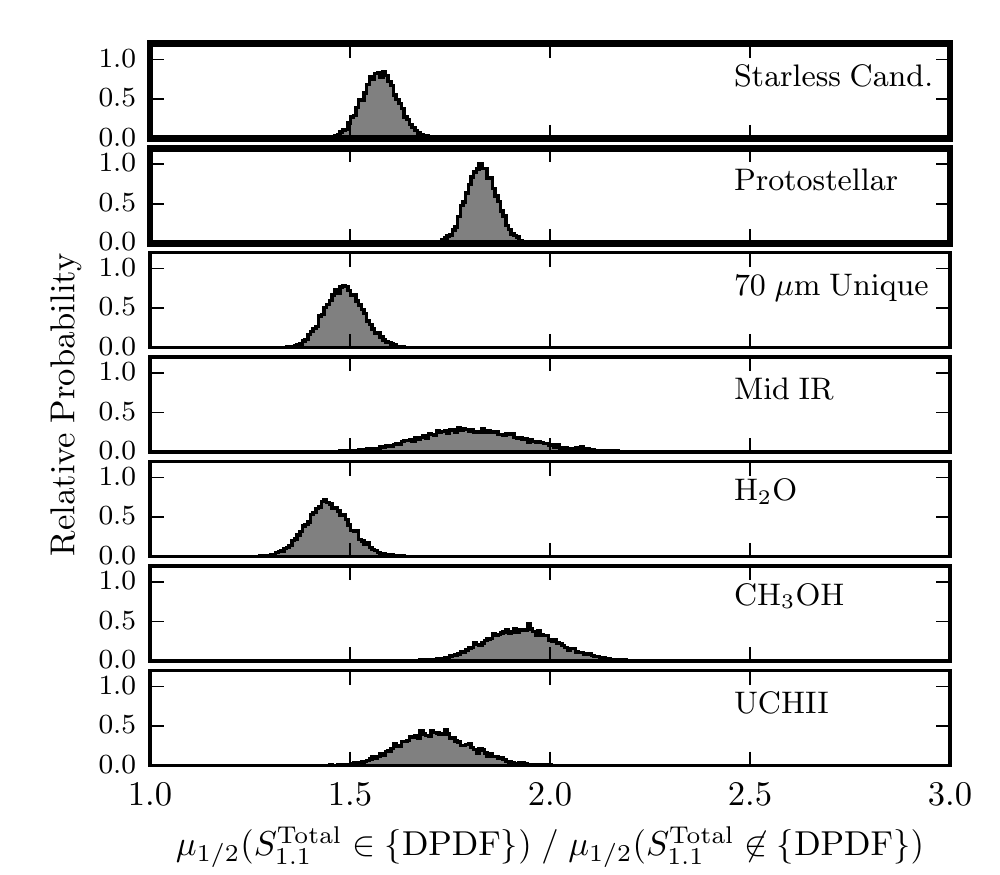}
\caption{
Distributions of total flux density for clumps in the full sample (light gray) and Distance Sample (dark gray) by star formation indicator.
Top left: flux density histogram of observed values.
Top right: flux density distribution drawn from MC simulation.
Bottom left: average DPDF for the expanded DPDF sample (black line) and original DPDF sample (dashed gray line). The number $N$ of the clumps are shown in black and gray, respectively.
Bottom right: ratio of median total flux density of clumps with a DPDF to clumps without a DPDF.
Clumps in the Distance Sample have a factor of $\sim 1.5-2.0$ higher median flux density than clumps not in the Distance Sample.
The star formation indicator is shown in the upper-right of each panel.
}
\label{fig:MonteCarloFlux}
\end{figure*}

Before we compare physical properties among clumps with different star formation indicators (\S\ref{ssec:FluxAndConcentration}-\ref{ssec:MassAndSurfDens}), we compare the distribution of DPDFs for each star formation indicator subsample defined in \S\ref{ssec:EvoCatSumm}.
Figure \ref{fig:MonteCarloFlux}c shows the mean distribution or average DPDF for each indicator.
The indicator categories show similar distributions of heliocentric distance.
While the number of clumps associated with each indicator vary, the range and distribution of $d$ are similar, with medians between $4.4-5.1$ kpc.
As a result, physical properties which depend linearly or quadratically on distance will not be strongly biased by the underlying distance distribution for each indicator subsample.
Sources with associated \metho\ masers and/or \uchii\ regions show a relative deficit of distances from $2-4$ kpc.
This could be an effect from the method of resolving distances with the $8 \um$ absorption prior, where better spatially resolved clumps at the near distance are dominated by mid-IR emission from a high-mass YSO.
These regions may not have the necessary contrast in the GLIMPSE $8 \um$ maps to result in a well constrained distance estimate, whereas further, and thus larger, clumps could have sufficient surrounding $8 \um$ absorption.
In addition, the PPV method of associating nearby clumps with known distances for a fixed radius in angle and a fixed velocity better associates regions at larger distances where the angular separation between clumps is smaller.

\begin{figure}
\centering
    \includegraphics[width=0.47\textwidth]{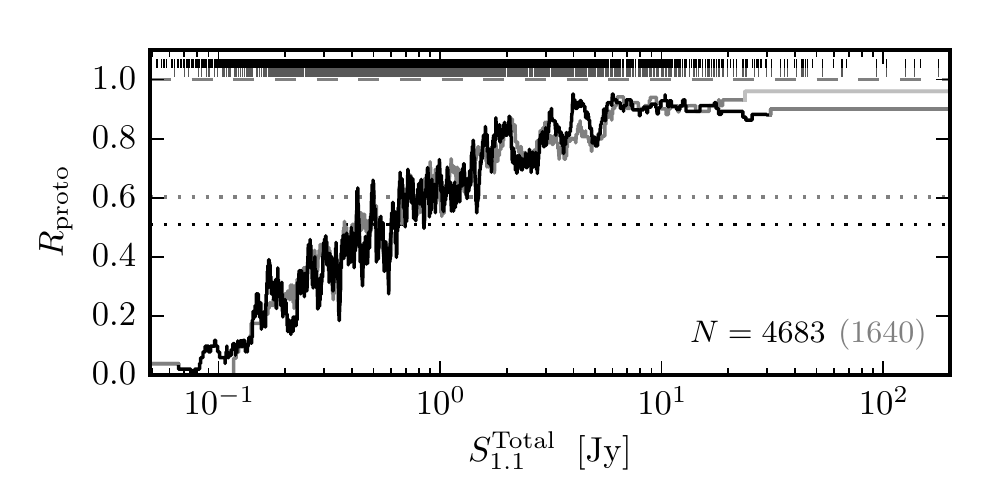} \\
    \vspace{-3mm}\includegraphics[width=0.47\textwidth]{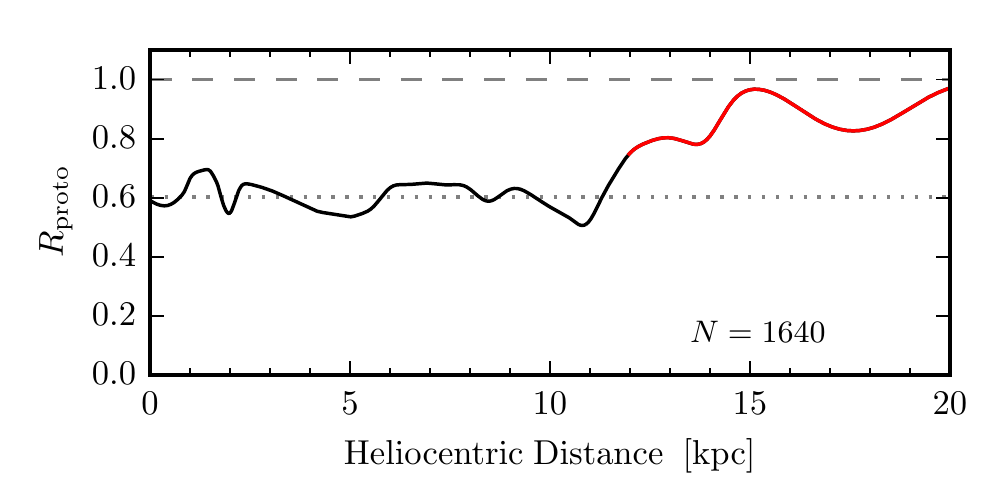}
\caption{
Protostellar fraction distributions, computed using the MC simulations.
{\it Top:} Total flux density.
The black line shows $R_{\rm proto}$ for a moving window of 50 sources over the full overlap sample and the Distance Sample is shown in gray.
Vertical lines show the clump $S^{\rm Total}_{1.1}$ values.
{\it Bottom:} Heliocentric distance.
The solid line shows $R_{\rm proto}$ computed as a weighted sum of the DPDFs.
$R_{\rm proto}$ is approximately constant at $< 12$ kpc, suggesting that the star formation indicators are mostly complete for BGPS clumps.
In red are the 58 clumps with $d_{\rm ML} > 12$ kpc that compose $3.5\%$ of the Distance Sample.
Sample protostellar fractions are shown by horizontal dotted lines.
}
\label{fig:SfFraction}
\end{figure}

When comparing distance independent quantities between starless candidate and protostellar categories, the total number of sources is nearly equal (\ovscc\ SCCs vs. \ovproto\ protostar-containing clumps); however, when comparing distance dependent physical properties, it should be noted that in the Distance Sample there are $1.675$ times as many protostellar-containing clumps than starless clump candidates ($613$ SCCs vs. $1027$ protostellar clumps).
The fraction of clumps in the Distance Sample that contain at least one star formation indicator is shown as a function of heliocentric distance in Figure \ref{fig:SfFraction}.
The median protostellar clump fraction in the Distance Sample is $0.60$, and there is no significant trend in this ratio out to heliocentric distances of $12$ kpc.
Since only $58$ clumps ($3.5\%$) in this study have maximum likelihood distances greater than $12$ kpc, this result indicates that differences in the properties of starless candidates compared to protostellar clumps are not driven by a bias in the protostellar fraction by distance.

We can only calculate distance-dependent quantities from the subset of clumps in each category of the Distance Sample.
We run MC simulations for both the subsamples that have a well-constrained DPDF and those without to test how this selection criteria may bias our distributions.
Figure \ref{fig:MonteCarloFlux}a shows the observed total flux density $S^{\rm Total}_{1.1}$ for each indicator, with the total sample (light gray) and Distance Sample (dark gray).
Figure \ref{fig:MonteCarloFlux}b shows the cumulative results of $10^4$ draws from the Monte Carlo simulation of the 1.1 mm flux density Gaussian-deviates and R08 AGB contamination re-sampling.
The subsamples with DPDFs have greater median $S_{1.1}$ when compared to full samples.
Similarly, figure \ref{fig:MonteCarloFlux}d shows the distribution of the ratio of medians for each MC simulation.
All indicator categories show a factor of $1.5-2.2$ higher median flux for clumps with DPDFs; however, there is not a consistent trend in the ratio between star formation indicators.  The ratio of median flux densities for protostar-containing clumps 
is only a factor of $1.1 - 1.2$ times larger than the corresponding ratio for starless clump candidates.
If we make the simple assumption that sources without well constrained DPDFs follow the average DPDF distributions for the Distance Sample, then this result implies that physical quantities that depend linearly on flux density (i.e., mass) are biased by factors of $1.5-2.2$ larger in the Distance Sample versus the total sample.

\subsection{Flux Density, Flux Concentration, and Size}\label{ssec:FluxAndConcentration}
The observed 1.1 mm flux densities increase towards more extreme indicators from starless clump candidates to \uchii\ associated clumps (Fig. \ref{fig:MonteCarloFlux}a).
The median total 1.1 mm flux density of starless clump canddiates is $0.356\pm0.005$ Jy and is a factor of 15 lower in total flux density compared to the median for \uchii\ associated clumps.
The starless candidate and \uchii\ associated clump distributions are distinct where 99\% of starless candidates have lower total 1.1 mm flux density than the median total flux of \uchii\ associated clumps.
Even the presence of a deeply embedded YSO observed as a compact \higal\ $70 \um$ source preferentially increases the median observed flux density by a factor of $\sim1.8$.
Figure \ref{fig:SfFraction} shows these trends in an alternative way as a strongly increasing protostellar clump fraction with increasing 1.1 mm flux density.  
The ratio crosses 0.5 around 500 mJy and approaches 1.0 above 5 Jy.
Without internal star formation activity, starless clumps would only be externally heated by the interstellar radiation field (ISRF; \citeauthor{evans02} \citeyear{evans02}, \citeauthor{jorgensen06} \citeyear{jorgensen06}, \citeauthor{wilcock12} \citeyear{wilcock12}).
As a result, clumps with weaker internal heating will have colder dust temperatures and lower observed 1.1 mm flux densities (see \S\ref{ssec:GasTempWidth}).

The deconvolved clump equivalent angular radius $\theta_{\rm eq}$ is calculated by subtracting the \bgps\ beam $\theta_{\rm HPBW}=33\arcsec$ ($\Omega_{\rm beam}=1234$ sq. arcsec) in quadrature from the observed equivalent angular radius $\theta_{\rm obs}=\Omega_{\rm obs}/\pi$ calculated directly from the sum of the $7.2\arcsec\times7.2\arcsec$ map pixels in the \bolocat\ label-mask frames
\begin{equation}
    \theta_{\rm eq} = \sqrt{\frac{\Omega_{\rm obs}}{\pi} - \frac{\Omega_{\rm beam}}{\pi}} \; \; ,
\end{equation}
where $\Omega_{\rm obs}$ is either the total solid angle defined by the Bolocat v2.0 seeded watershed algorithm or the FWHM solid angle defined by the contour at half-maximum flux density.
Quantities that are derived from the \bolocat\ label-masks of the \bgps\ maps are sensitive to the spatial filtering imposed by the principal component analysis performed on the Bolocam time stream data.
A clump size based on the FWHM definition is more robust to spatial filtering and is also more prevalent in the literature \citep[e.g.,][]{beuther02,shirley03}.
In the following analyses, we shall use both total size definition, i.e., the full label masks, as well as the FWHM size definition to calculate derived properties.

\begin{figure*}
\centering
    \includegraphics[width=0.47\textwidth]{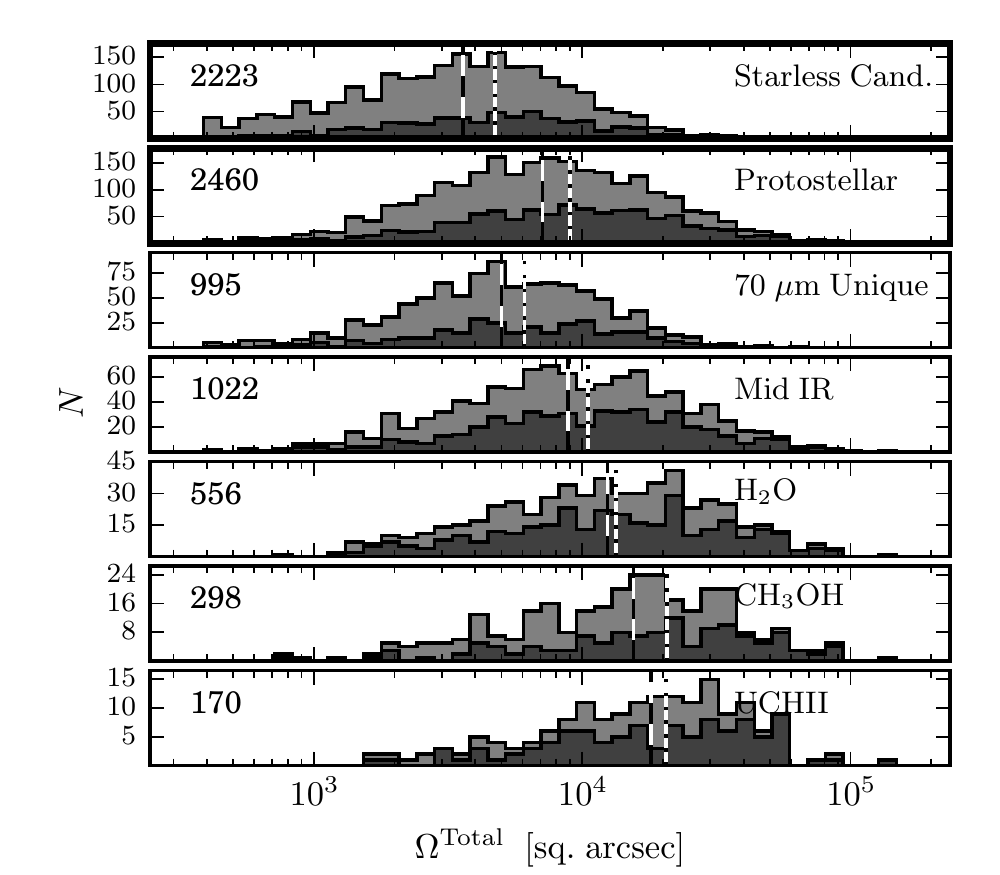}
    \includegraphics[width=0.47\textwidth]{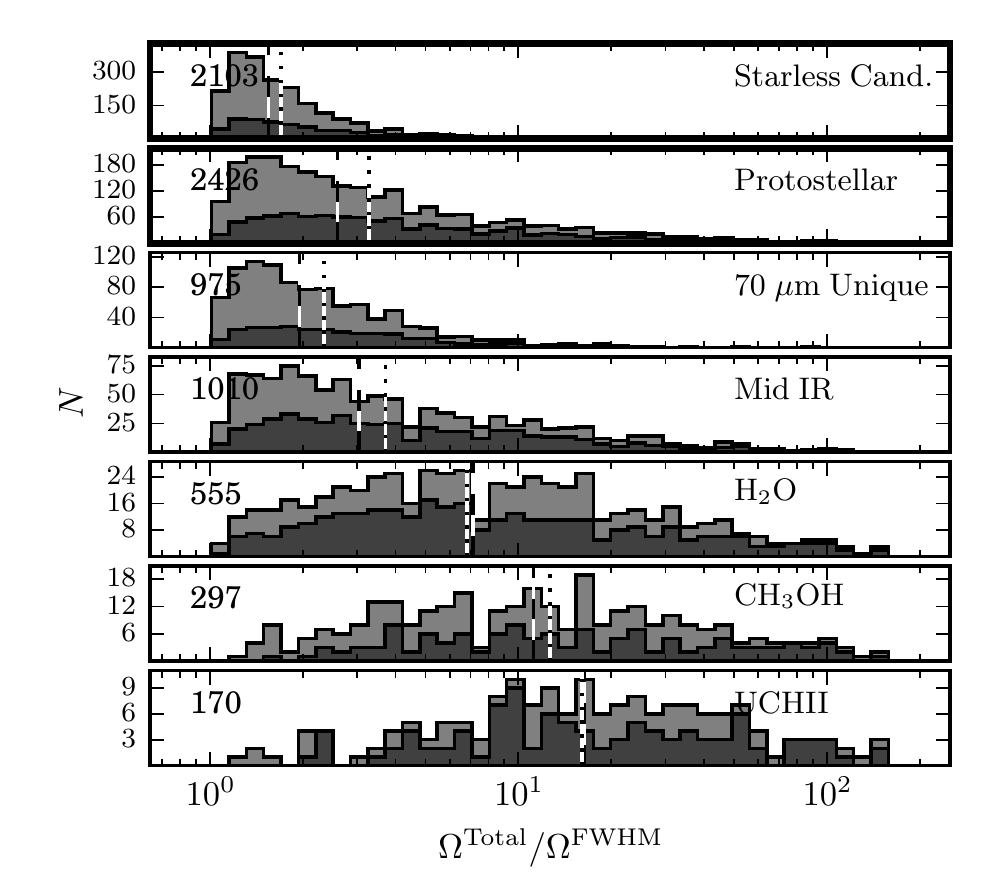}
\caption{
Left: Total solid angle calculated from the BGPS maps for the full sample (light gray) and Distance Sample (dark gray).
Right: Total to FWHM concentration.
The FWHM size is calculated for the extent of the emission above the half-max.
Starless candidate clumps tend to have flux that is less extended and less concentrated than more extreme indicators.
}
\label{fig:SolidAngle}
\end{figure*}

Figure \ref{fig:SolidAngle}a shows $\Omega^{\rm Total}$ for each indicator, with an increasing trend in the distributions with median $3.62\times10^3$ square arcsec ($1.01$ square arcmin) for the starless candidates to $1.81\times10^4$ square arcsec ($5.03$ square arcmin) for the \uchii\ associated clumps.
There is also a systematically increasing trend in the concentration of the 1.1 mm flux density distributions when sorted by indicator.
Figure \ref{fig:SolidAngle}b shows the ratio in deconvolved solid angles, or Total to FWHM ``concentration'', defined as $\Omega^{\rm Total}/\Omega^{\rm FWHM}$.
Clumps that are near uniform brightness and with a peak brightness less than twice the $1\sigma$ ($\sim 0.1-0.5$ Jy) cut-off threshold of the clump boundaries should have concentrations near unity, while clumps with highly concentrated brightness distributions will have large concentration values ($\gg 1$).
A strong increasing trend is seen from a median ratio $\Omega^{\rm Total} / \Omega^{\rm FWHM}$ for the starless candidate phase at $1.55$ to $16.38$ for \uchii\ clumps.
Clumps associated with \water\ masers, \metho\ masers, and \uchii\ regions are concentrated at similar levels and also more concentrated than those with less extreme indicators.

\begin{figure*}
\centering
    \includegraphics[width=0.47\textwidth]{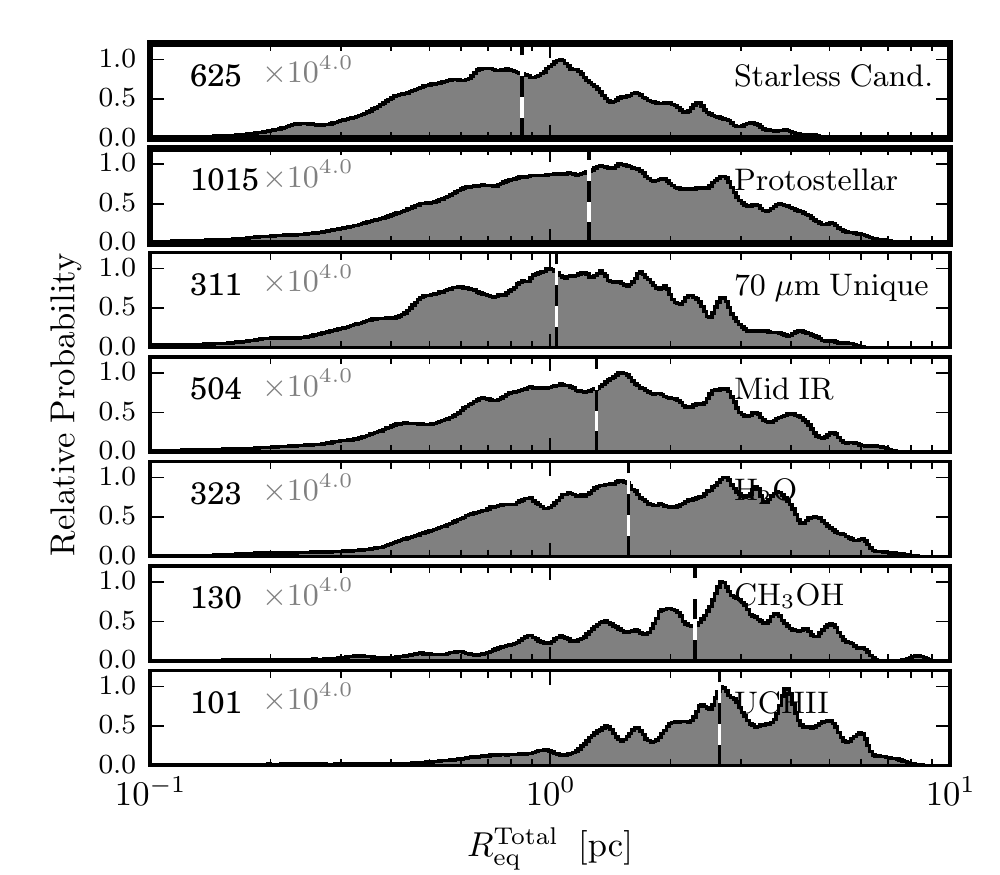}
    \includegraphics[width=0.47\textwidth]{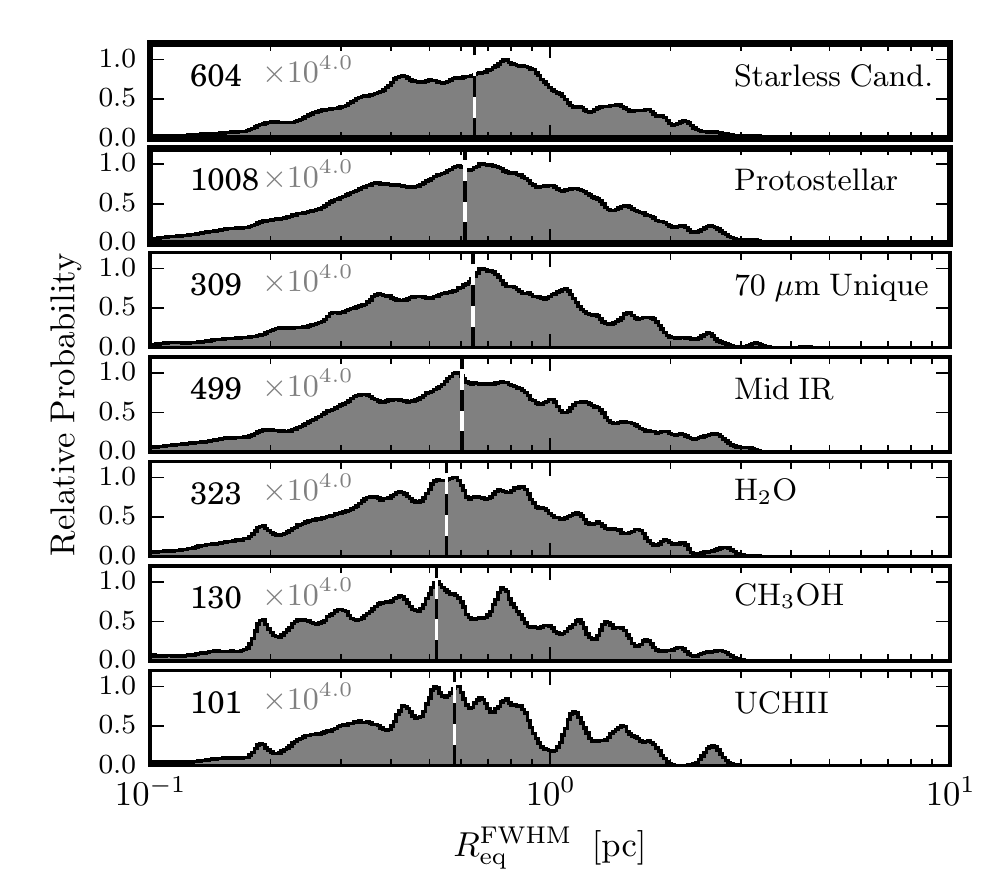}
\caption{
Physical equivalent radius histograms from the MC simulations for each star formation indicator sample.
Left: Physical total radius.
Right: Physical FWHM radius.
The dashed lines show the median value and the number of clumps in the subsample is shown in the upper left.
}
\label{fig:StagesHistArea}
\end{figure*}

The equivalent physical radius $R_{\rm eq}$ is calculated from the observed $\Omega$ when projected to a heliocentric distance $d_\odot$ drawn from the DPDFs
\begin{equation}
    R_{\rm eq} = 0.29 \; \left( \frac{\theta_{\rm eq}}{\rm arcmin} \right) \left( \frac{d_\odot}{\rm kpc} \right)
\end{equation}
Figure \ref{fig:StagesHistArea} shows the equivalent radius distributions for the Total (\ref{fig:StagesHistArea}a) and FWHM (\ref{fig:StagesHistArea}b) definitions.
The median $R^{\rm Total}_{\rm eq}$ increases from $0.842\pm0.017$ to $2.66\pm0.08$ pc from starless candidates to \uchii\ clumps; however, little separation among different indicators is observed using the FWHM definition with median $R^{\rm FWHM}_{\rm eq}\sim0.6$ pc.
The $R^{\rm Total}_{\rm eq}$ in this study are larger than those found by \cite{schlingman11} (median $R_{\rm eq}=0.75$ ) for a subset of known $529$ sources taken from the BGPS v1.0.
This discrepancy can be attributed to the recovery of flux at lower spatial frequencies in the \bgps\ v2.0 maps resulting in larger clumps \citep{ginsburg13}.
The different trends observed between the total source size and FWHM size are more striking and can be explained by offsetting factors.
Clumps containing indicators of more luminous protostars (i.e., CH$_3$OH masers, \uchii ) have the highest flux concentrations which partially offset their larger total solid angle resulting in their FWHM sizes remain nearly the same on median as starless clump candidates and clumps with lower luminosity protostellar indicators.  
This result highlights the need to explore multiple size definitions in the analysis of source physical properties.

\subsection{Gas Kinetic Temperature and Linewidth}\label{ssec:GasTempWidth}
\begin{figure*}
\centering
    \includegraphics[width=0.47\textwidth]{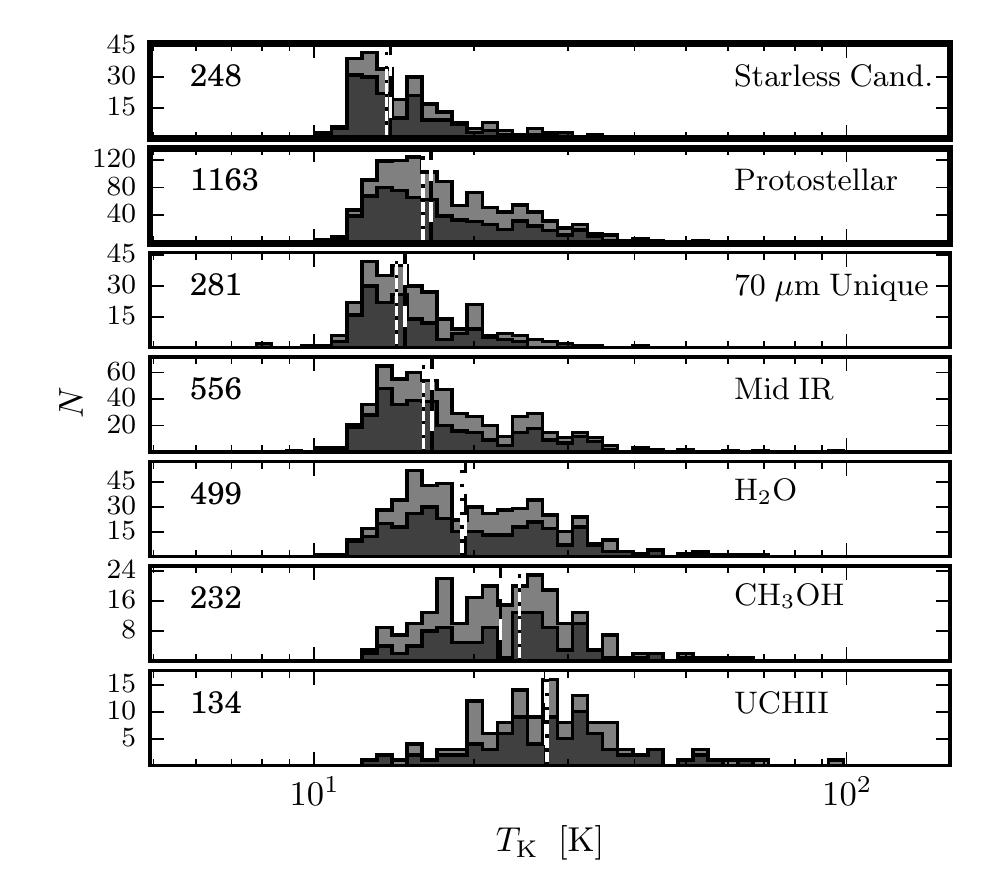}
    \includegraphics[width=0.47\textwidth]{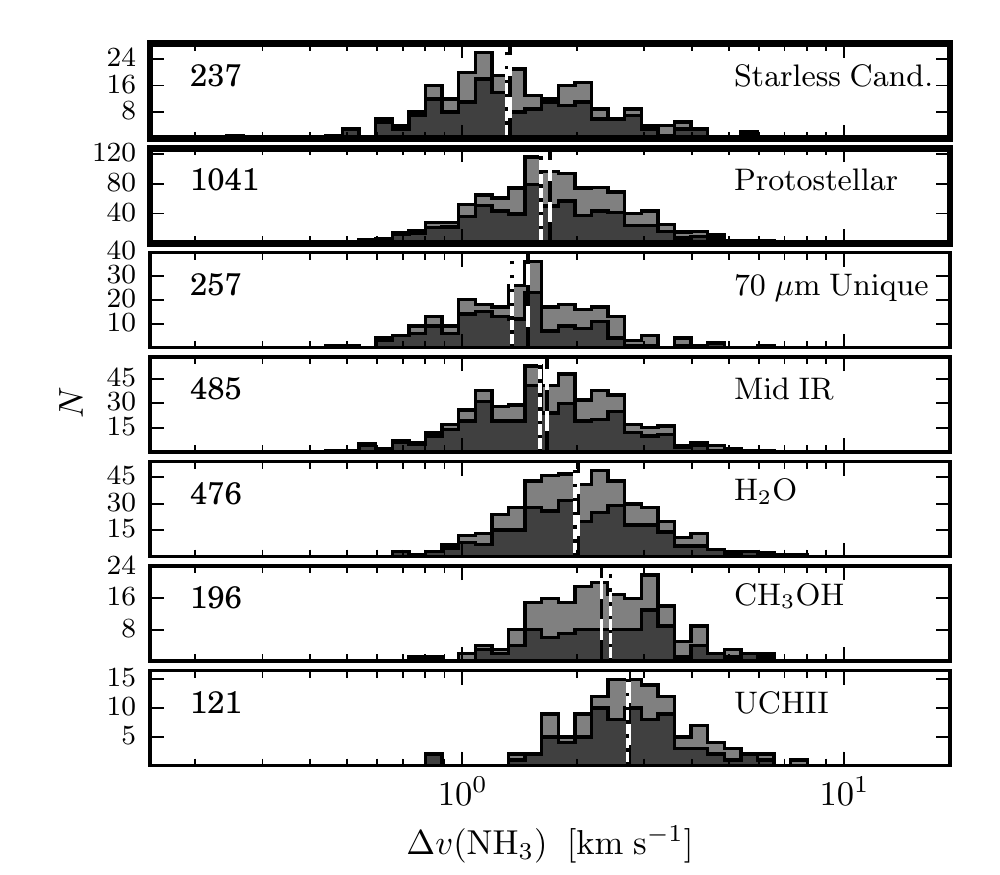}
\caption{
Observed distributions of \amm-derived kinetic temperature (Left) and linewidth (Right) by star formation indicator.
The dashed lines show the median value and the number of clumps in the subsample is shown in the upper left.
}
\label{fig:StagesHistAmmonia}
\end{figure*}

\amm\ is an excellent intermediate to dense gas tracer with an effective excitation of $\sim 10^3$ cm$^{-3}$ for the lowest 
pure inversion transition \citep{shirley15}.  Fitting the intensities of the lowest energy inversion transitions provide estimates of the gas kinetic temperature \citep{ho83}.  The rotation temperature of the (1,1) and (2,2) p-\amm\ inversion transitions saturates for $T_{\rm K} > 30$ K, but the simultaneously observed (3,3) transitions retain sensitivity to $T_{\rm K} < 100$ K if the o-\amm\ (3,3) transition is not masing \citep{danby88}.
Figure \ref{fig:StagesHistAmmonia} shows the distributions of \amm-derived gas kinetic temperature with values ranging between $T_{\rm K} \sim 10-100$ K.
The starless candidate phase has a median $T_{\rm K} = 13.96\pm0.10$ K with an increasing trend to $T_{\rm K} = 27.2\pm0.2$ K for \uchii\ clumps.
The clumps uniquely associated with compact \higal\ $70 \um$ sources, a sign of deeply embedded protostellar activity, show only a slight enhancement in the gas kinetic temperature.
As a clump evolves from a quiescent phase to one with active star formation, radiative heating from star formation is expected to raise the gas kinetic temperature.
With more extreme indicators of star formation activity suggesting higher luminosity YSOs, the CH$_3$OH and UCHII region containing clumps are subject to stronger radiative heating from massive star formation.
The starless clump candidate distribution does display a positive skew. This could be a result of undetected embedded sources or enhancement by the local radiation field from neighboring star forming regions.
The lower $S_{1.1}$ in the candidate starless clump stage shown in Figure \ref{fig:MonteCarloFlux} can, in part, be attributed to this observed trend in \tk.
The flux density of a modified blackbody is $4.9$ times higher at $T_{\rm K} = 30$ K than $10$ K at 1.1 mm.
We shall account for the temperature differences observed among the different star formation indicators in our calculations of the clump mass in \S\ref{ssec:MassAndSurfDens}.

The observed FWHM linewidths of \amm\ for clumps may be broadened due to a variety of factors including thermal broadening, bulk motion, optical depth, and turbulence.
In the following analysis we use the \amm\ rather than \hcop\ linewidth because the \amm\ observations have three times better spectral resolution, $0.3$ km s$^{-1}$, and \amm\ has lower optical depths, $\tau_{{\rm NH_3}(1,1)} \sim 1-5$ compared to $\tau_{{\rm HCO^+}(3-2)} \sim 10$ \citep[see][]{shirley13}.
Figures \ref{fig:StagesHistAmmonia}b shows the distributions of observed \amm\ linewidth, indicating a factor of two increase from $1.339\pm0.014$ \kms\ on median towards starless candidates to $2.736\pm0.03$ \kms\ on median towards \uchii\ associated clumps.
For gas temperatures between $10-30$ K the thermal linewidth is $\Delta v_{\rm therm}({\rm NH_3}) = 0.16 - 0.28$ \kms.
The observed \amm\ linewidths at $> 1$ \kms\ suggest supersonic turbulence exists on scales smaller than the $33\arcsec$ beam of the \bgps.
The quiescent starless candidate phase and the deeply embedded candidate phase of uniquely identified \higal\ $70 \um$ compact sources have $\Delta v_{\rm turb} / \Delta v_{\rm therm} \sim 10$.
The observed increase in linewidth with more extreme star formation indicators could be due to an increase in turbulent feedback in more luminous protostellar clumps.
More luminous protostars will drive more powerful jets and outflows \citep[see][]{bontemps96} as well as provide more radiative feedback on the surrounding clump that can drive turbulence \citep[see][]{krumholz12}.
Alternatively, the larger linewidths towards clumps with more extreme indicators could be due to larger bulk-flow motions that are unresolved at the GBT angular resolution or a systematic tendency for higher mass stars to form in more turbulent regions.
Without higher resolution observations of the clumps, it is difficult to determine which scenario drives the observed trend.

\begin{figure}
\centering
    \includegraphics[width=0.47\textwidth]{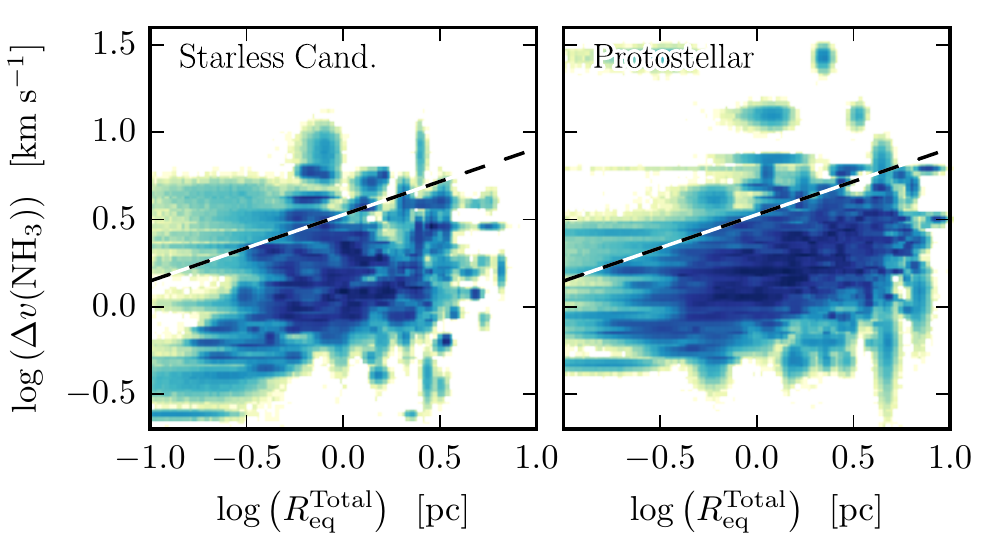}
\caption{
Linewidth-size relationship plotted for starless clump candidates (Left) and protostellar clumps (Right) drawn from MC simulations of each quantity.
The protostellar clumps have a higher Spearman rank correlation coefficient $\rho_{\rm sp}=0.50$ than the starless clump candidates $\rho_{\rm sp}=0.24$.
The dashed line indicates the converted relationship $\Delta v = 3.37 \, {\rm km \; s^{-1}}(R/{\rm pc})^{0.38}$ observed in \cite{larson81} for molecular clouds traced by \co\ $1-0$.
}
\label{fig:LinewidthSize}
\end{figure}

Prior observations of a significantly smaller subsample of BGPS clumps have indicated a breakdown in the linewidth-size relationship \citep{schlingman11}.
We plot the linewidth-size relationship for SCCs and protostellar clumps observed in NH$_3$, shown in Figure \ref{fig:LinewidthSize}.
The Spearman's rank correlation coefficients for SCCs and protostellar clumps are $\rho_{\rm sp}=0.24$ and $\rho_{\rm sp}=0.50$ respectively from the MC simulations.
The starless clumps appear uncorrelated while protostellar clumps have a weak correlation that is only marginally better than the \cite{schlingman11} correlation ($\rho_{\rm sp}=0.40$).
SCCs, as traced by NH$_3$ emission, seems to have decoupled from the supersonic scaling relationship observed toward molecular clouds more so than protostellar clumps.
Indeed the observed NH$_3$ model-fit velocity dispersion (expressed as FWHM linewidth) are below the extrapolated relationship observed by \cite{larson81} for molecular clouds traced by CO $1-0$.
While these clumps themselves are still turbulent structures, their level of turbulence appears dissipated compared to the expected relationship in clouds when observed on clump spatial scales.

\subsection{Mass Calculations}\label{ssec:MassAndSurfDens}
\subsubsection{Mass Surface Density}
The mean mass surface density is calculated using:
\begin{equation}
\Sigma_{\rm H_2} = \frac{S_{1.1}}{B_{1.1}(T_{\rm dust}) \kappa_{1.1} \zeta}\frac{1}{\Omega}
\end{equation}
where $B_{1.1}$ is the Planck function, $\kappa_{1.1}$ is the dust opacity, $\zeta$ is the dust-to-gas ratio, here assumed to be $1/100$, and $\Omega$ is the source solid angle, and $S_{1.1}$ is the source integrated flux density.
Both the Total and FWHM size definitions are used for $\Omega$ or $S_{1.1}$.
We assume the 1.1 mm emission is optically thin and use Gaussian-deviates to sample the statistical uncertainty.
We also assume OH5 dust opacities calculated for coagulated grains with thin ice mantles \citep{ossenkopf94}.
For the Monte Carlo calculation, we draw Gaussian-deviates from the observational uncertainty on \tk , and assume that the dust temperature is equal to the gas kinetic temperature.
Because $47\%$ ($772/1640$) of clumps with a distance lack \tk\ measurements, for those clumps we draw \tk\ from that indicator's observed distribution of \tk\ (Fig. \ref{fig:StagesHistAmmonia}a).
This takes into account the trend of increasing gas \tk\ for clumps associated to more extreme star formation indicators.
For clumps associated with multiple star formation indicators, the indicator with the hottest median \tk\ is used.

Figure \ref{fig:StagesHistMassSurfDens} shows \mmsd\ with an increasing trend towards higher \mmsd\ for clumps associated to more extreme star formation indicators.
The median \mmsd\ increases from median values $\mmsd = 0.0145\pm0.0002$ g cm$^{-2}$ and $\mmsdf = 0.0165\pm0.0003$ g cm$^{-2}$ for the starless candidate clumps to $\mmsd = 0.0207\pm0.0005$ g cm$^{-2}$ and $\mmsdf = 0.076\pm0.002$ g cm$^{-2}$ for the \uchii\ associated clumps.
A similar trend in increasing column density from quiescent (potentially starless) clumps to protstellar clumps has also been observed by other (sub)millimeter Galactic plane surveys \cite[see][]{csengeri14, guzman15}.

The factor of four increase in \mmsdf\ is consistent with the greater flux densities and flux concentrations observed towards clumps with more extreme indicators.  
The median mass surface density for protostellar sources is close to the purported threshold for star formation of approximately 120 M$_{\odot}$ pc$^{-2}$ \citep[or 0.025 g cm\( ^{-2} \);][]{lada10,heiderman10}.
Since half of these clumps presumably contain protostellar sources despite being below this threshold, this result highlights that BGPS clumps themselves contain significant substructure which are at higher mass surface densities.

\subsubsection{Clump Mass}\label{ssec:ClumpMass}
We calculate total clump masses with:
\begin{align}
M_{\rm H_2} & = \frac{S_{1.1} \ d_{\rm ML}^2}{B_{1.1}(T_{\rm dust}) \kappa_{\rm dust, 1.1} \zeta} \\
            & \approx 13.1 \left( \frac{e^{13 {\rm K} / T_{\rm dust}} - 1}{e^{13 {\rm K} / 20 {\rm K}} - 1} \right) \left( \frac{S_{1.1}}{1 \ {\rm Jy}} \right) \left( \frac{d_\odot}{1 \ {\rm kpc}} \right)^2 M_\odot \nonumber
\end{align}
Figure \ref{fig:StagesHistMass} shows the distributions of \mcl\ when sorted by indicator from $10^4$ MC simulations that account for R08 AGB catalog contamination and uncertainties in $S_{1.1}$, \tk\ (or randomly drawn from the appropriate \tk\
distribution if no \tk\ observations exists), and the DPDFs.
The masses show a systematic increase from $\mu_{1/2}(M_{\rm H_2})=226\pm11$ M$_\odot$ for SCCs and $\mu_{1/2}(M_{\rm H_2})=600\pm20$ M$_\odot$ for protostellar clumps.
Among the clumps associated to star formation indicators, the masses increase from $\mu_{1/2}(M_{\rm H_2})=390\pm30$ M$_\odot$ for the $70\um$ Unique category to $\mu_{1/2}(M_{\rm H_2})=2600\pm200$ M$_\odot$ for the \uchii\ category.
The difference in median masses between the SCC and Protostellar categories of $\Delta\mu_{1/2}(M_{\rm H_2})=370\pm20$ M$_\odot$ and the SCC and $70 \um$ Unique category of $\Delta\mu_{1/2}(M_{\rm H_2})=170\pm30$ M$_\odot$.
The median mass difference is driven by more SCCs than Protostellar clumps below $470$ \msun\ and more Protostellar clumps than SCCs above that mass.
We statistically estimate the highest mass SCC by integrating the upper-tail of \mcl\ PDF to where it equals one part in the sample size, $1:625$, or the $99.84$\textsuperscript{th} percentile.
This sets the maximum observed mass of SCCs at $<1.4\times10^4$ \msun .
The observed increase in mass towards clumps associated to more extreme star formation indicators could be due to several physical and systematic explanations.
In the remainder of this section, we explore possible systematic effects to generate the observed mass difference.

\begin{figure*}
\centering
    \includegraphics[width=0.47\textwidth]{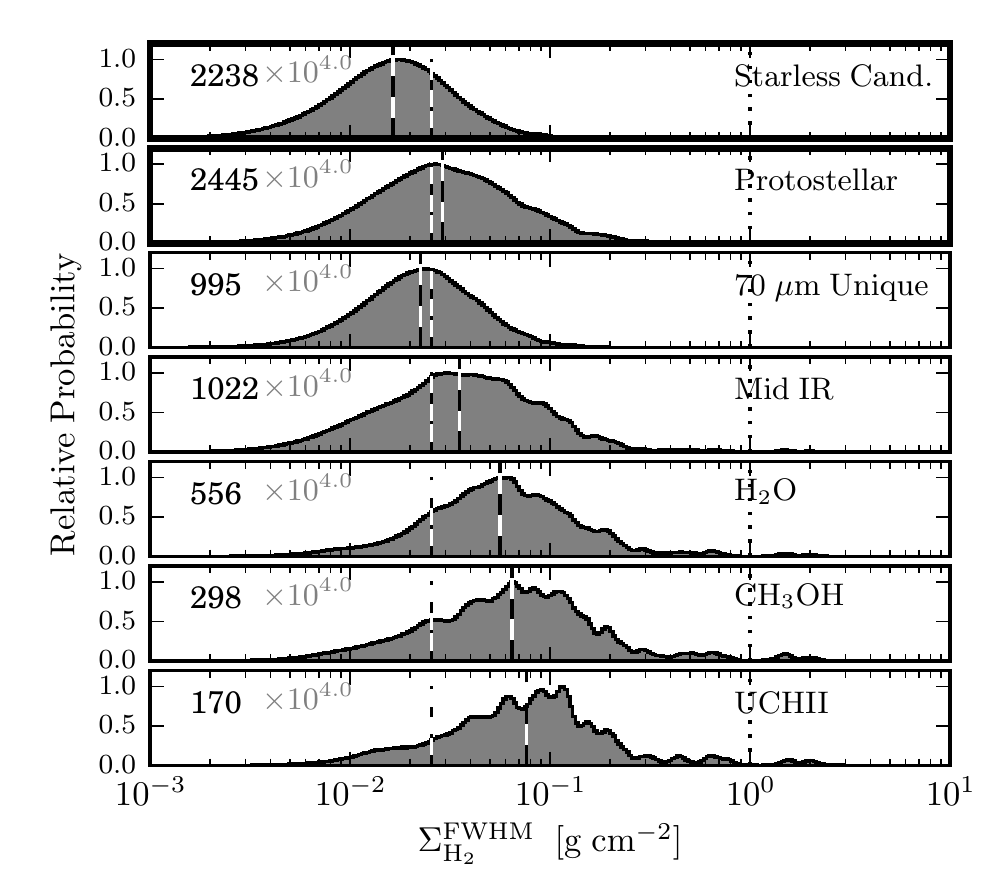}
    \includegraphics[width=0.47\textwidth]{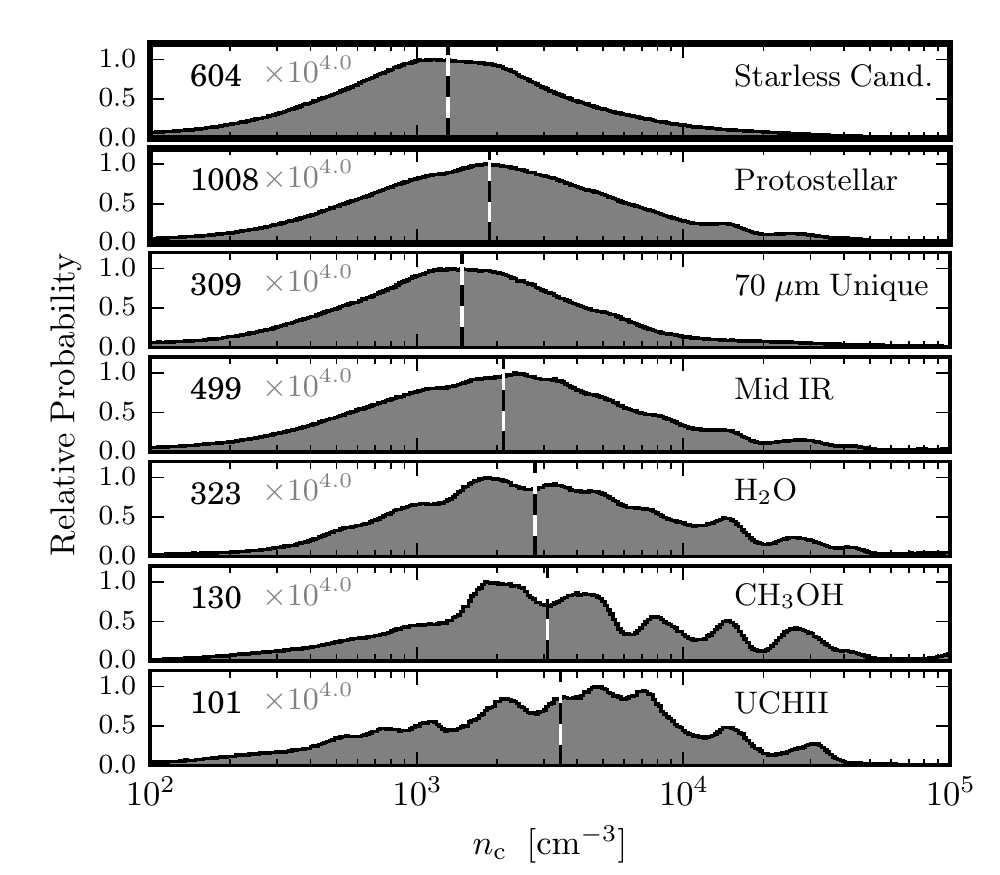}
\caption{
Distributions of FWHM mass surface (Left) density and central density (Right) from MC simulations by star formation indicator.
The dashed lines show the median value and the number of clumps in the subsample is shown in the upper left.
The dotted line shows the predicted 1 g cm$^{-3}$ threshold for the formation of a massive star from \cite{krumholz08}.
The dash-dotted line shows the 0.025 g cm$^{-3}$ threshold for star formation observed in local clouds \citep{lada10,heiderman10}.
}
\label{fig:StagesHistMassSurfDens}
\end{figure*}

For physical properties that depend on the dust temperature of the clumps, we assume an isothermal temperature with $T_{\rm dust} = T_{\rm K}$ in the clumps.
%The NH$_3$ gas kinetic temperature is an average along the line-of-sight that is only sensitive to gas above the effective excitation density \citep{shirley15}.
Nominally, these two temperatures are only expected to be well coupled by collisions at densities greater than $\sim 10^5$ cm$^{-3}$ \citep{goldsmith01,young04}.
\cite{battersby14a} find no correlation between \amm-derived \tk\ and \herschel-derived $T_{\rm dust}$ for a quiescent clump; however, agreement within $20\%$ is observed when the \herschel\ data is not background subtracted.
\cite{battersby14a} suggest that the gas and dust are weakly coupled at densities $10^4-10^5$ cm$^{-3}$.
Similarly, the $T_{\rm dust}$ derived from far-infrared SEDs in \cite{traficante15b} show a $20\%$ agreement to $T_{\rm K}$ for both starless candidate and protostellar clumps.
A $20\%$ decrease in $T_{\rm dust}$ (from $T_{\rm K} \sim 12$) will increase our mass and column density measurements for the starless candidate group by $30\%$.
This effect alone is not enough to explain the observed mass differences.
There is no reason to decrease only the dust temperatures of starless clump candidates.
Any simultaneous decrease in $T_{\rm dust}$ for protostellar clumps would mitigate the decrease of the median mass differences.  

In reality, there are dust temperature gradients within the BGPS beam due to heating from the interstellar radiation field and due to embedded protostars in protostar-containing clumps.
The general effect of line-of-sight dust temperature gradients on greybody SED fitting of dust continuum emission is to overestimate the dust temperature and underestimate the mass more severely for starless clumps than protostellar clumps \citep{malinen11}.
The corresponding effect on the line-of-sight average NH$_3$ $T_{\rm K}$ cannot explain the observed mass difference unless the temperature gradients cause an underestimate of the average line-of-sight $T_{\rm K}$ for the starless clump candidates from the observed median value of 12 K down to 8.8 K, a value that is lower than all NH$_3$-based $T_{\rm K}$ measurements in this paper and 95\%\ of SED-based $T_{\rm dust}$ measurements in \cite{traficante15b}.
Analysis of the dust temperature profiles from grids of radiative transfer models of low-mass starless cores \citep{shirley05, launhardt13} find that the mass-weighted average $T_{\rm dust}$ never drops below $10$ K for a grid of Bonner-Ebert spheres \citep{ebert55, bonnor56} with central densities spanning $10^4$ to $3 \times 10^6$ cm$^{-3}$ and strength of the interstellar radiation field equal to half the standard value \citep{habing68}.
These low-mass starless core models are not the exact analogues of massive starless clumps which are possibly fragmented and clumpy, have lower observed average density (Figure \ref{fig:StagesHistMassSurfDens}), and are likely subject to stronger heating from the interstellar radiation field than half the Habing value, effects which will increase the average $T_{\rm dust}$ compared to low-mass starless cores.
These results indicate it is unlikely that dust temperature gradients result in an average $T_{\rm dust}$ as low as $8.8$ K.
Radiative transfer modeling of the entire BGPS clump population is beyond the scope of the current paper and requires higher spatial resolution continuum and NH$_3$ images.
In the absence of reliable dust continuum temperatures for the full sample, a single component $T_{\rm dust} = T_{\rm K}$ assumption is the next best alternative.

The other assumption made in calculation of mass surface density and mass is that the dust opacities are well described by OH5 dust.
They are a popular set of opacities because grains are expected to coagulate and accrete ice mantle in dense environments and there are observational constraints that indicate OH5 opacities are a reasonable match, albeit toward nearby, low-mass cores \cite[see][]{shirley11}.
The OH5 model assumes that the grains have coagulated for a period of $10^6$ years at a density of $10^5$ cm$^{-3}$.  
Both this time period and the density used in the coagulation model are likely larger than the typical values observed toward BGPS clumps \cite[see][]{schlingman11,dunham11a}.
In order to explain the observed $\Delta \mcl$  the opacity ratio towards protostellar clumps versus starless clumps would have to be $1.7 - 2.6$ (for 70$\mu$m unique clumps or the full protostellar clump sample).
While it is certainly likely that there are opacity variations among clumps that deviate from the assumed OH5 opacities, the variation would have to be systematic between starless candidate and protostellar clumps.  There is currently no evidence for or against such a systematic variation in opacities.

The observed mass difference cannot be due to selection effects on the flux density $S_{1.1}$ of the Distance Sample alone (i.e., the subsample of clumps that have DPDFs compared to those without) because  SCCs and protostellar clumps have similar ratios in median $S_{1.1}$ (see \S\ref{sec:Analysis},  Fig.~\ref{fig:MonteCarloFlux}d).
At most this effect could account for $\sim10\%$, lowering the mass difference to $\Delta\mu_{1/2}(M_{\rm H_2}) \approx 330$ \msun .
In addition, as a flux limited survey, the \bgps\ is effected by Malmquist bias and mass incompleteness.
\cite{ellsworthbowers15} derive the mass completeness function for clumps in the Distance Sample by computing Monte Carlo simulations drawn from the DPDFs, where the sample is $50\%$ complete above $\mcl \approx 70$ \msun\ and $90\%$ above $\mcl \approx 400$ \msun.
The median SCC mass, $\mcl = 230$ \msun, is at the $80\%$ completeness level, and the $90\%$ completeness level is achieved at the $64$\textsuperscript{th} mass percentile.
The incompleteness in the SCC category should not, qualitatively, affect the observed median mass difference.
Because $R_{\rm proto}$ is smaller at low flux densities and masses, the population of undetected clumps with $M<400$ would disproportionately add to the SCC category and simply enhance the median mass difference further. 

One additional possibility is the spatial filtering in the BGPS survey has systematically affected the flux densities of sources.
\cite{ginsburg13} estimate the effects of spatial filtering by approximating the source brightness distribution as a power-law with varying index and finds that the flux density for sources with angular radii less than $40^{\prime\prime}$ is essentially recovered (see \cite{ginsburg13} Figure 8).
Depending on the exact value of the power-law index, the spatial filtering recovers $> 75 - 80$\%\ of the flux density for sources with angular radii $< 60^{\prime\prime}$.
This accounts for $77$\%\ (3606/4683) of the BGPS sources.
It is evident from Fig.~\ref{fig:SolidAngle}a that the sources more strongly affected by spatially filtering are protostellar stages with more luminous evolutionary indicators which tend to have larger angular radii.
Thus, accounting for spatial filtering should systematically increase the median mass difference between starless clump candidates and protostellar clumps.
A detailed analysis of the effects of spatial filtering on a source by source basis is beyond the scope of the present work.

\begin{figure*}
\centering
    \includegraphics[width=0.47\textwidth]{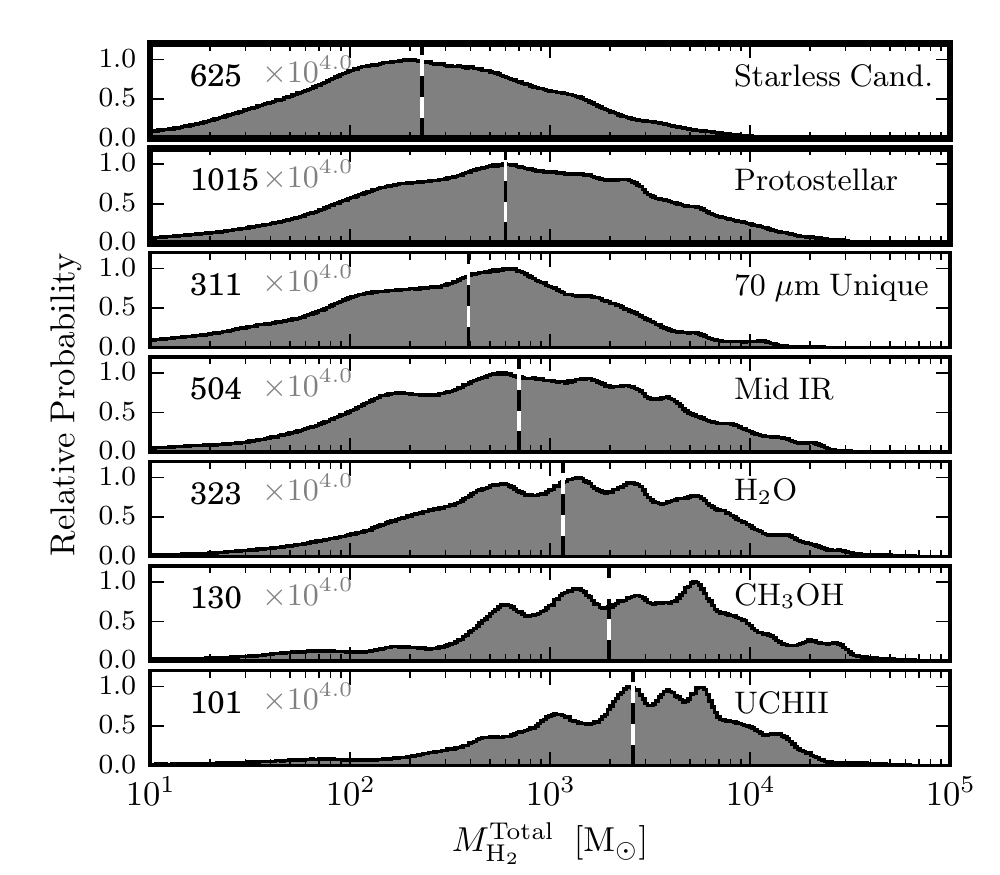}
    \includegraphics[width=0.47\textwidth]{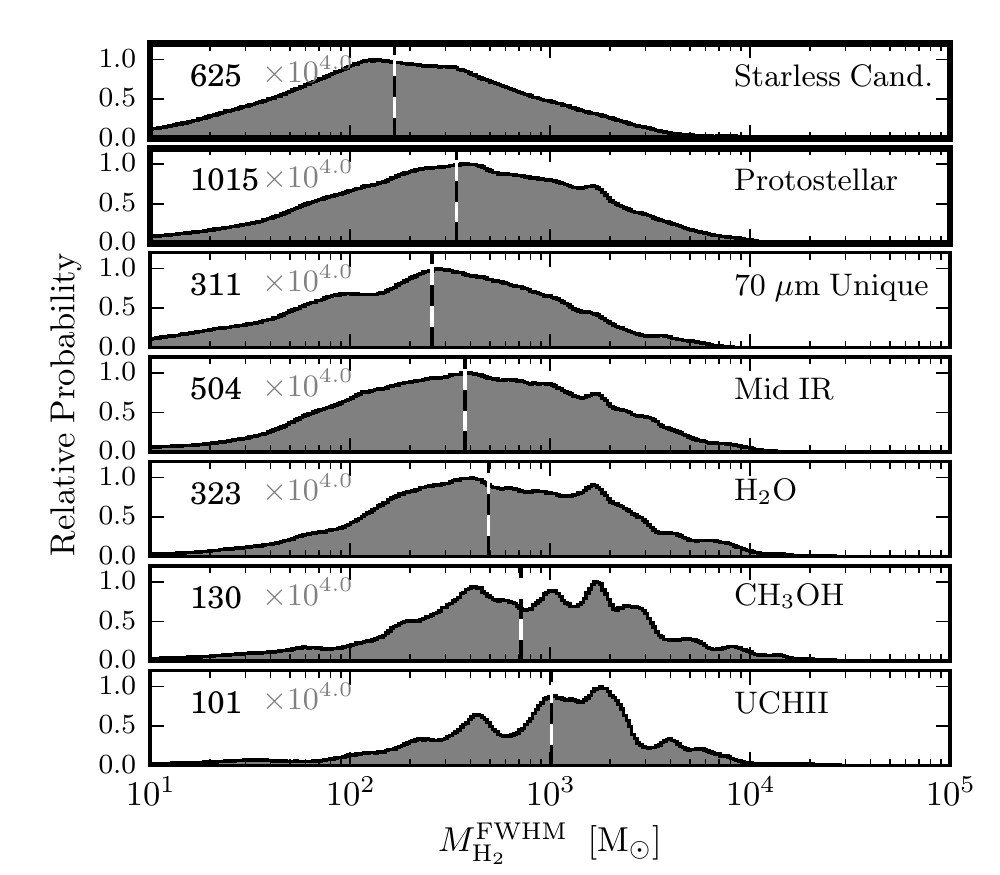}
\caption{
Distributions drawn from MC simulations of total mass (Left) and FWHM mass (Right) sorted by star formation indicator.
The number of clumps in a sample is shown in the upper left and a vertical dashed line shows the median value.
}
\label{fig:StagesHistMass}
\end{figure*}

In summary, the observed mass difference cannot be explained by biases between the Distance sample and the full sample, incompleteness, spatial filtering, or our assumptions in the mass calculation unless there is a factor $1.7 - 2.6$ systematic variation in dust opacities between starless clump candidates and protostellar clumps.
Interestingly, the BGPS sample is the second statistically significant sample to see this systematic offset in mass.
The \textit{Herschel} survey of \citep{traficante15b} toward IRDCs also find a modest offset of $\sim 80$ \msun\ between the average mass of starless clumps and protostellar clumps.
While the mass difference is smaller by a factor of two in their study than ours, the mass difference in their survey is significant.
Alternatively, \cite{hoq13} using data from the first year data release of the MALT90 molecular line mapping survey of several hundred ATLASGAL clumps in the fourth and first quadrants do not observe an increasing trend in total mass from an ``$8/24 \um$ quiescent'' phase to a ``PDR'' phase.
It should be noted that these two surveys are drawn from subsets of clumps in the Galactic plane; therefore interpretation of these results depend on how the properties of the subsets compare to the more complete clump populations observed in (sub)millimeter continuum Galactic plane surveys.
In \S\ref{ssec:OriginOfMassDifference} we explore possible physical explanations for the variation in the median mass between the starless clump candidates and protostellar candidates under the assumption that the observed mass difference is real.

\subsubsection{Virial Parameter}
\begin{figure}
\centering
    \includegraphics[width=0.47\textwidth]{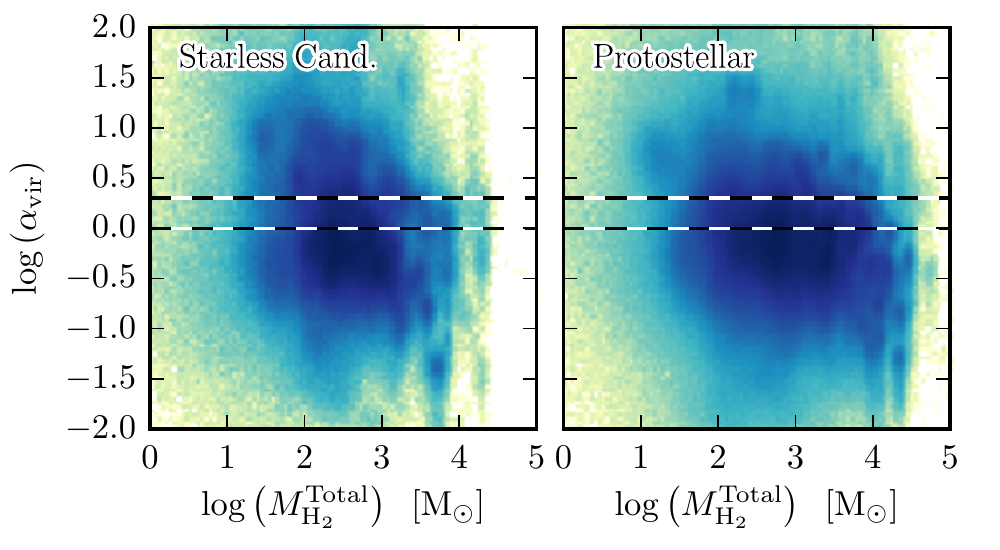}
\caption{
Comparison of virial parameter to clump total mass for SCCs (left) and protostellar clumps (right) computed using values drawn from the MC simulations.
Virial parameters are derived from the model-fit \amm\ velocity dispersion.
The dotted lines show $\alpha=1$ and $\alpha=2$.
Note the logarithmic scaling.
}
\label{fig:MassVir}
\end{figure}

The virial parameter, $\alpha = M_{\rm vir} / M$, expresses the relative importance of gravitational and kinetic energy, and can be used to assess whether clumps are gravitationally bound and/or stable to collapse.
We calculate the virial parameter for a homoeoidal ellipsoid (ellipsoid of revolution) via
\begin{align}
    \alpha & = \frac{5}{8 \ln 2} \frac{1}{a_1 a_2} \frac{\Delta v^2 R}{G M} \\
    & \approx 1\; \frac{1}{a_1} \left( \frac{\Delta v}{1 \ \rm km \ s^{-1}} \right)^2 \left( \frac{R}{1 \ {\rm pc}} \right) \left( \frac{M}{209 \ {\rm M}_\odot} \right)^{-1} \nonumber
\end{align}
where $a_1 = \frac{1 - p/3}{1 - 2 p/5}$ is the correction factor for a power-law density distribution with index $p$ and $a_2\sim1$ for sources with aspect ratios less than 2 \citep{bertoldi92}.
For a clump with negligible magnetic fields, $\alpha = 1$ is gravitational virial equilibrium and $\alpha \approx 2$ is marginally gravitationally bound \citep[see][]{kauffmann13}.
The \bgps\ does not have sufficient resolution to constrain $p$, so lacking appropriate data, we draw $p = 1.8 \pm 0.4$ Gaussian-deviates in the MC simulations, as constrained from dust continuum modelling of high-mass star forming regions \citep{mueller02}.
Here we use the total mass and radius, thus setting the outer boundary to calculate $\alpha$ at the extent of the $1.1$ mm emission.
Figure \ref{fig:MassVir} shows the virial parameter calculated from MC simulations for starless candidates and protostellar clumps with median values $\alpha = 0.73\pm0.06$ and $\alpha = 0.68\pm0.03$ respectively, suggesting that most clumps are in approximate virial equilibrium with more than half of clumps in the Distance Sample with \amm\ observations showing (sub)-virialized motions ($\alpha \lesssim 1$).
Further, $76\%$ of the starless candidates and $86\%$ of the protostellar clumps have $\alpha < 2$ implying that most BGPS clumps are gravitationally bound.
Although $24\%$ of the starless candidates are unbound by the criteria $\alpha > 2$, it is probable that these clumps host gravitationally bound higher density substructures.
These results suggest that the majority of BGPS clumps in the Distance Sample are self-gravitating or collapsing.

Observations of GMCs indicate virial parameters that are typically $\gtrsim 2$ \citep{larson81,solomon87,scoville87}.  
For instance, analysis of GMCs in the Galactic Ring Survey find a mean value of $\alpha = 1.9$ \citep{heyer09}.
For smaller clump scales there exist a range of virial parameter measurements that report samples with $\alpha \gtrsim 1$ or $\alpha \lesssim 1$ (see \citeauthor{kauffmann13} \citeyear{kauffmann13} for a summary of literature values).
We find our measurements of $\alpha \lesssim 1$ consistent with other Galactic plane surveys towards clumps from both \bgps\ and ATLASGAL with \amm-determined linewidths.
The most direct comparison are previous studies using GBT observations of \amm\ that have been carried out towards the BGPS sources \citep{dunham10,dunham11b}.
\cite{dunham11b} measured the virial parameter towards $456$ \bgps\ clumps with a median $\alpha = 0.74$. 
\cite{wienen12} also measure the virial parameter toward ATLASGAL clumps using \amm, and find a mean $\alpha \lesssim 1$ for clumps.

One important aspect that is often ignored in the standard virial parameter analysis is the importance of magnetic fields in virial balance.
Magnetic fields in dense gas are notoriously difficult to measure, and there have been no systematic observations toward BGPS clumps.
We can  estimate the B-field strength required for virial balance from the expressions derived in \cite{kauffmann13} (their Equations 6 and 16)
\begin{equation}
B \gtrsim 15 \ \mu{\rm G} \left(\frac{2}{\alpha} - 1\right) \left( \frac{\Delta v}{1 \ {\rm km \ s^{-1}}} \right)^2 \left( \frac{1 \ {\rm pc}}{R} \right) \;\;.
\end{equation}
Evaluating the above expression for the median values of the starless clump candidates indicate that the B-fields of only $50$ $\mu$G are needed to support typical starless clump candidates against collapse.
Magnetic fields of this magnitude and larger are indeed observed toward IRDCs.
For instance, \cite{pillai15} find a total B-field strength of $\sim 250$ $\mu$G toward the IRDC G11.1 from analysis of submillimeter continuum polarization.
The required B-field strength scales as $B \sim 1/\alpha$, implying that very sub-virial clumps with $\alpha \lesssim 0.1$ required $B \gtrsim 600$ $\mu$G.
In these more extreme cases, B-field support may prove difficult, but for the typical (median property) starless clump candidates, the required B-field strength for support is within a reasonable range of observed values \citep[for a summary of B-field measurements, see][]{crutcher12}.

\section{Discussion}\label{sec:Discussion}
For many of the observed properties and calculated physical properties analyzed in \S\ref{sec:Analysis}, there is a systematic trend in the property with the ordering of the star formation indicators from $70$ $\mu$m unique flags, mid-IR star formation flags, \water\ maser flags, \metho\ flags, to \uchii\ flags.
Protostellar clumps that contain the most extreme indicators of luminous protostars (CH$_3$OH maser and \uchii\ containing clumps) tend to have the highest flux densities, are the most centrally condensed, have the highest temperatures, are more turbulent and more massive on median than other clumps.
A partial evolutionary analysis of the ATLASGAL clumps associated to RMS sources, \metho\ maser, and \uchii\ regions, has been performed \citep{urquhart11,urquhart13a,urquhart13b,urquhart14}.
The BGPS results are consistent with the observed trends ($M$, $T_{\rm K}$) for the properties of ATLASGAL clumps that contain methanol masers and \hii\ regions \citep[see Table 4 of][]{urquhart14}.
In contrast, the observable and physical properties indicate that starless clump candidates have lower flux densities, are less centrally concentrated, have smaller sizes, are colder and less turbulent, and are less massive on median when compared to clumps with indications of protostellar activity.
In this section, we explore the potential for forming massive stars, the timescales for evolution of this newly discovered population of starless clumps, and explore possible explanations for the observed median mass difference between starless and protostellar clumps.

\subsection{Massive Star-forming Potential of Clumps}

\begin{figure}
\centering
    \includegraphics[width=0.47\textwidth]{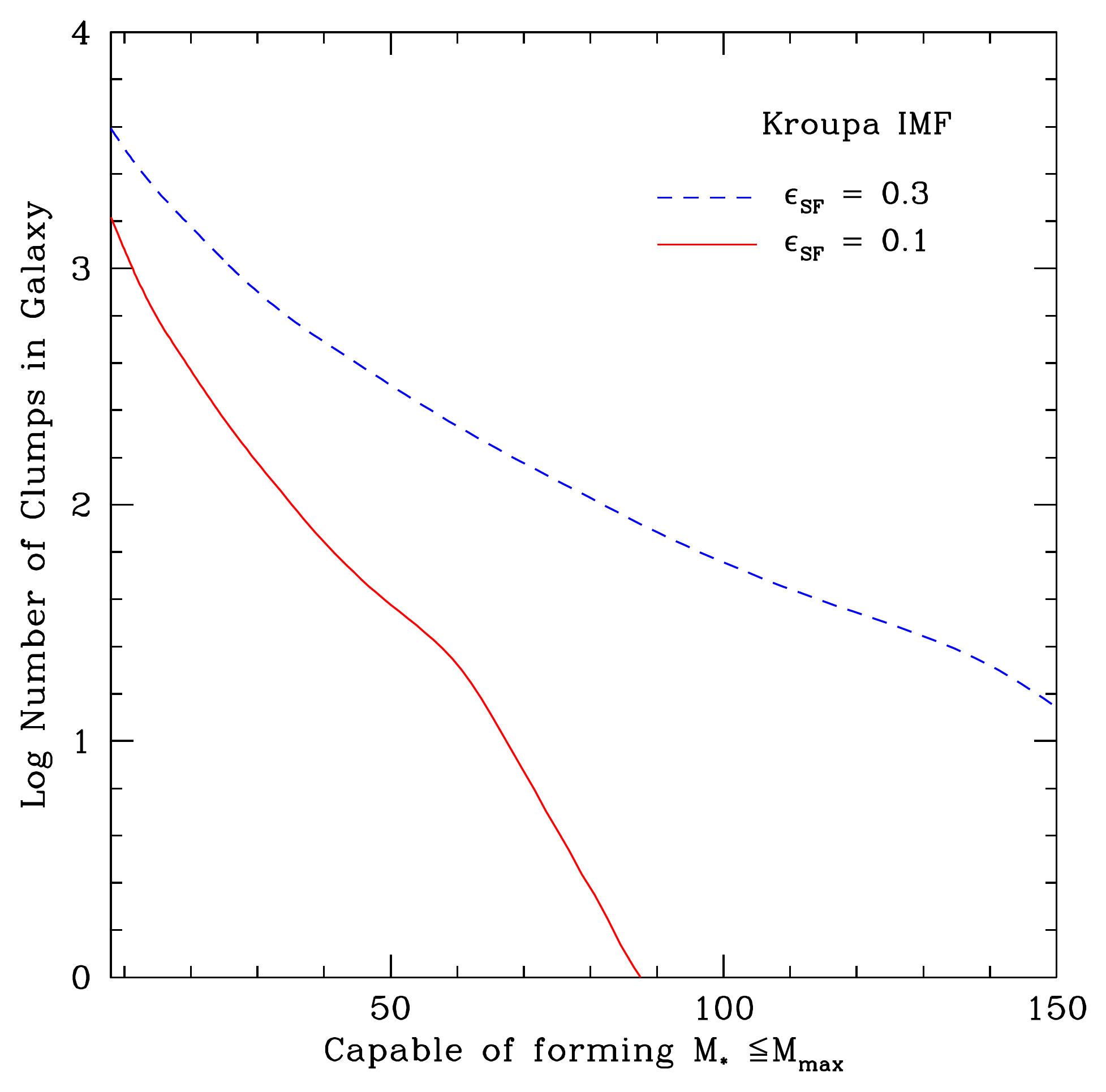}
\caption{
The number of clumps in the Milky Way capable of forming a star with a mass $\leq M_{\rm max}$ assuming a Kroupa IMF.  Two star formation efficiencies are plotted for the clumps ($0.1$ in red solid line and $0.3$ in blue dashed line).
}
\label{fig:NumGalaxy}
\end{figure}

With a robust sample of starless clump candidates, it is important to assess their potential to form high-mass stars.
We can estimate the mass of the most massive star formed by a starless clump candidate, M$_{\rm max}$, using the stellar Initial Mass Function (IMF) from \cite{kroupa01}\footnote{$N(M) \propto M^{-1.3}$ for $0.08 \ \msun \leq M \leq 0.5 \ \msun$, $N(M) \propto M^{-2.3}$ for $M > 0.5\ \msun$.  We assume a maximum stellar mass of 150 \msun\ in IMF calculations.}.
The total stellar cluster mass is equal to the mass of the progenitor clump times a star formation efficiency factor $\epsilon_{\rm SF}$.
We can derive a relationship between M$_{\rm max}$ and the mass of the clump from
\begin{eqnarray}
\epsilon_{\rm SF} M_{\rm clump} & = & \frac{\int_{0.08}^{150} N(M) M dM}{\int_{M_{\rm max}}^{150} N(M) dM} \\
M_{\rm max} & \approx & 20 \,\msun\ \left( \frac{\epsilon_{\rm SF}}{0.3} \frac{M_{\rm clump}}{1064 \, \msun } \right)^{1/1.3} \;\; ,\label{eq:IMF}
\end{eqnarray}
where the IMF is normalized such that at least $1$ star with $M \geq M_{\rm max}$ is formed.
The values of $\epsilon_{\rm SF}$ range from $0.05-0.5$ with a typical value of $0.3$ \citep[see][]{lada03,shirley03,krumholz07,kuiper10,bontemps10}. 
Using $8$ \msun\ as the definition of the minimum mass of a high-mass star, the corresponding BGPS clump mass is $320$ \msun .  
In the Distance Sample $42\%$ of SCCs ($264$ clumps) have masses above $320$ \msun\ in the MC simulations.
This corresponds to $603$ SCCs in the BGPS survey volume if we assume the SCCs without well-constrained DPDFs have a similar average DPDF and account for the different flux density distributions.  
If we use the \cite{wolfire03} model of the H$_2$ distribution in the Galaxy as an axisymmetric proxy for the massive clump spatial distribution\footnote{
The correction factor is calculated numerically by integrating the H$_2$ density distribution over the volume of the wedge observed by the BGPS divided by the integral of the H$_2$ density distribution over the total volume of the Galaxy.
The \cite{wolfire03} density distribution is given by $\rho(R_{\rm g}, z_{\rm g}) \propto \exp(-4\ln2(R_{\rm g} - 0.571 R_{\odot})^2 / (0.52 R_{\odot})^2) \exp(-|z_{\rm g}| / 0.059)$ for $R < 0.82 R_{\odot}$ and $\rho(R_{\rm g}, z_{\rm g}) \propto \exp(-R_{\rm g} / (0.34 R_{\odot})) \exp(-|z_{\rm g}| / 0.059)$ for $R \geq 0.82 R_{\odot}$ where $R_{\rm g}$ is the Galactocentric radius in kpc, $z_{\rm g}$ is the height above the Galactic mid-plane in kpc, and $R_{\odot} = 8.5$ kpc is the distance to the Galactic center.
Note that we account for the vertical offset of $25$ pc of the Sun above the Galactic plane in the calculation of the BGPS wedge integrals \citep[see Appendix C of][]{ellsworthbowers13}.}, then there are $\sim 3900$ SCCs capable of forming a $8$ \msun\ star in the entire Milky Way.  
Figure \ref{fig:NumGalaxy} plots the number of clumps in the Milky Way capable of forming stars with masses $ \leq M_{\rm max}$.
Note that the number of clumps is very sensitive to assumed star formation efficiency, especially for very massive stars.

\begin{figure}
\centering
    \includegraphics[width=0.47\textwidth]{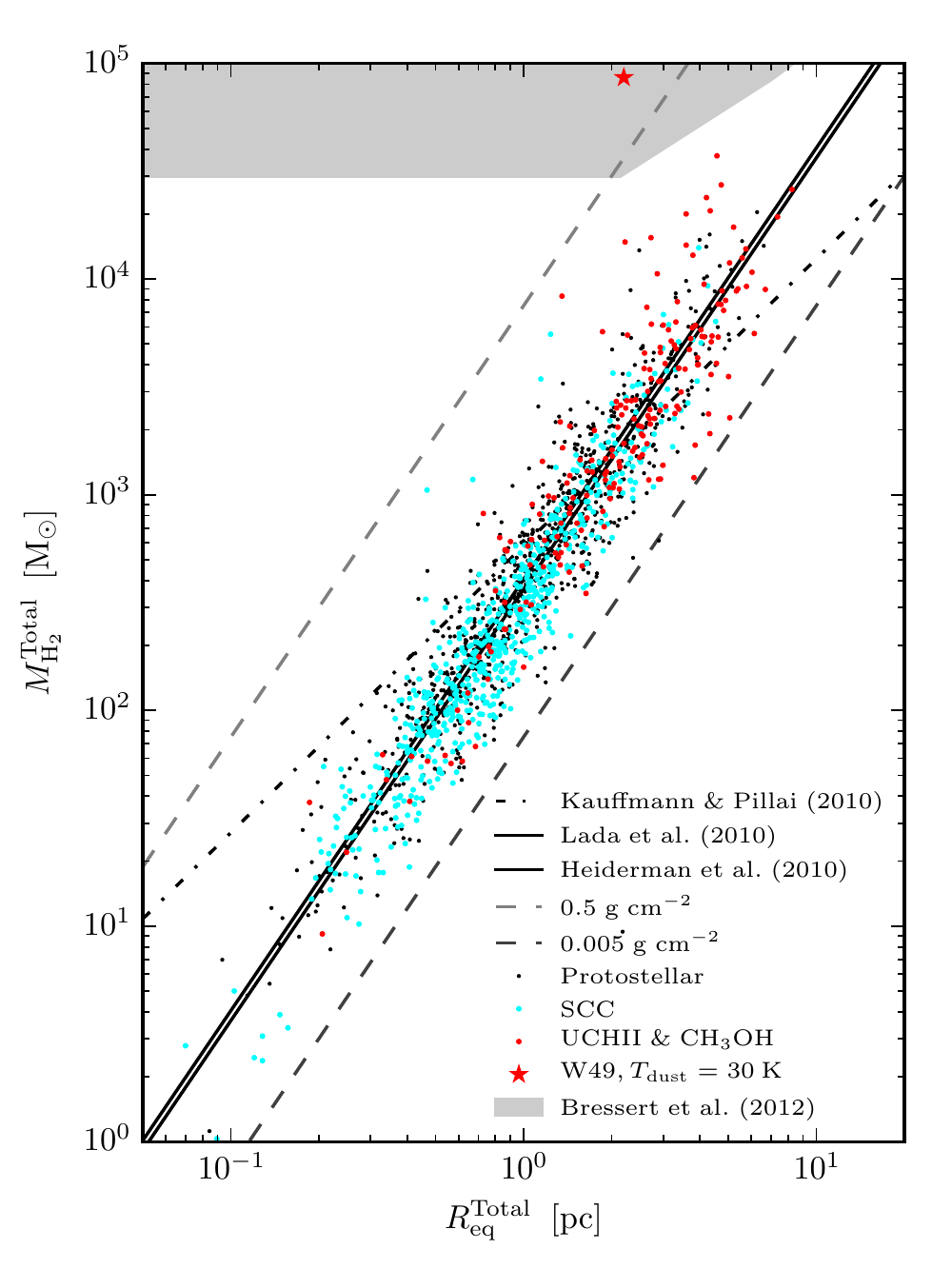}
\caption{
Equivalent radius compared to total mass.
Median values for each clump are computed from MC simulations.
Clumps are shown associated to \uchii\ or \metho\ masers (red), other protostellar (black), and SCCs (cyan).
The dash-dotted line shows the threshold for high-mass star formation $M(R) \geq 580 M_\odot (R \ {\rm pc}^{-1})^{1.33}$ from \cite{kauffmann10} when scaled to OH5 dust opacities.
Solid lines shows thresholds for ``effecient star formation'' of $116$ M$_\odot$ pc$^{-2}$ \citep{lada10} and $129$ M$_\odot$ pc$^{-2}$ \citep{heiderman10}.
The upper and lower dashed gray lines show $0.5$ g cm$^{-2}$ and $0.005$ g cm$^{-2}$ respectively.
The shaded gray region in the upper left indicates the part of parameter space for Young Massive Clusters as defined in \cite{bressert12}.
}
\label{fig:MassRadius}
\end{figure}

An alternative way to assess the massive star-forming potential of BGPS clumps is to compare $M_{\rm H_2}$ and $R_{\rm eq}$ (Figure \ref{fig:MassRadius}).
Values are represented as the median value drawn from the MC simulations per clump.
A robust increasing correlation is observed with $\rho_{\rm sp}=0.942$.
To show the differences in clumps by star formation activity we plot SCCs in yellow and clumps associated to \uchii\ or \metho\ masers as unambiguous detections of high-mass star formation activity in red.
The distributions of points follow the general trends in discussed in \S\ref{sec:Analysis}: SCCs are generally lower mass, smaller radius, and lower mass surface density.
Several thresholds for star formation determined from local clouds are over-plotted.
Fitting star formation rates to local molecular clouds, \cite{lada10} and \cite{heiderman10} find relationships for ``effecient'' star formation of $\approx116$ M$_\odot$ pc$^{-2}$ and $129\pm$ M$_\odot$ pc$^{-2}$, respectively, where above these thresholds the star formation rate is linearly proportional to mass surface density.
The thresholds from \cite{lada10} and \cite{heiderman10} approximately bisect the samples, with $50.6\%$ and $42.5\%$ of the full samples, $46.6\%$ and $39.9\%$ of the \uchii\ and \metho\ maser sample, and $41.4\%$ and $33.6\%$ of the SCC sample each exceeding the respective limits.
\cite{kauffmann10} derive the more restrictive criteria of $M \leq 580 {\rm M_\odot}(R_{\rm eq} / {\rm pc})^{1.33}$ from observations of nearby molecular clouds.
Note that \cite{kauffmann10} scale their OH5 dust opacities down by a factor of $1.5$, so for consistency to values in this work we use the leading factor of $580$ M$_\odot$ rather than the original $870$ M$_\odot$.
In the \bgps\ Distance Sample, $24.8\%$ of the full sample, $54.9\%$ of the \uchii\ and \metho\ maser sample, and $10.8\%$ of the SCC sample each exceed this criteria.
Because nearly half of \bgps\ clumps with an unambiguous detection of a high-mass YSO do not meet this criteria, this suggests that the \bgps\ cloud structures do not strictly follow the criteria derived in \cite{kauffmann10}.
While the majority of SCCs lie below this threshold, any systematic growth would increase clump total mass and radius, placing them in phase-space for a higher probability of high-mass star formation (see \S\ref{ssec:ClumpMassGrowth}).
In a similar blind sample from the ATLASGAL survey, \cite{wienen15} find that $92\%$ of ATLASGAL clumps meet this criteria and $100\%$ above the \cite{lada10} and \cite{heiderman10} lines. 
Notably, only 1 clump in the \bgps\ from $10^\circ < \ell < 65^\circ$ (W49, G043.167+00.011) meets the criteria for the progenitors of Young Massive Clusters \citep[YMCs;][]{longmore14} with clump masses $\gtrsim 10^5$ \msun , although W49 is not a SCC because it is actively forming stars.

\subsection{Clump Timescales}\label{ssec:Lifetime}
\subsubsection{Average Free-fall Time}
\begin{figure}
\centering
    \includegraphics[width=0.35\textwidth]{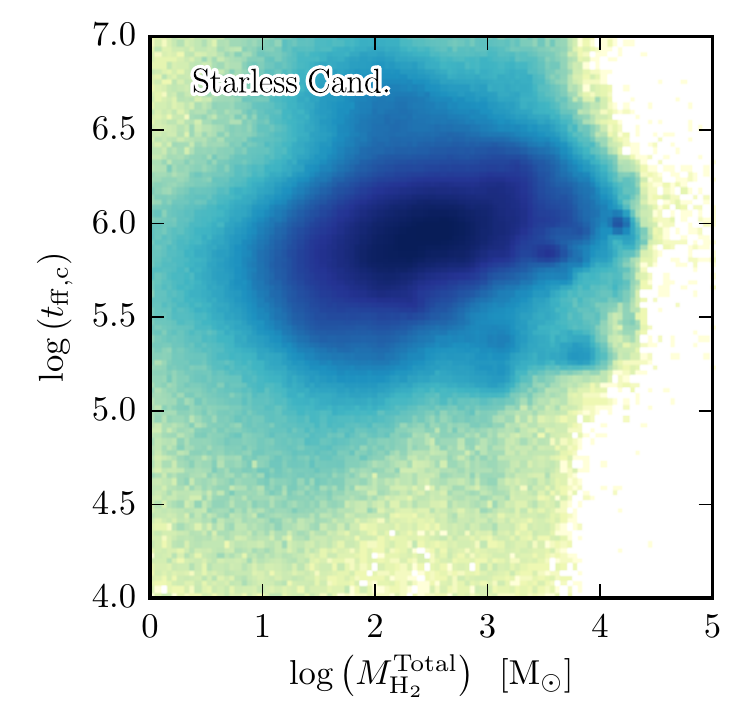} \\
\caption{
Comparison of clump free-fall time to clump total mass for starless clump candidates computed using values drawn from the MC simulations.
Free-fall time values calculated from central density $n_{\rm c}$.
}
\label{fig:FreeFallMass}
\end{figure}

For the SCC sample we estimate the average free-fall time, \tff, calculated with
\begin{align}
\tff  & = \sqrt{\frac{3\pi}{32 G \mu m_{\rm p} n_c}} \\
      & \approx 0.98 \left( \frac{n_{\rm c}}{10^3 \ {\rm cm^{-3}}} \right)^{-1/2} \ \ {\rm Myr}. \nonumber
\end{align}
where $n_{\rm c}$ is the clump central density.  
We estimate the clump central density $n_{\rm c}$ from the peak mass surface density and FWHM radius, assuming a spherically symmetric Gaussian density distribution \citep[cf.~Appendix C of ][]{pattle15}.
Integrating over the FWHM cylindrical aperture yields the expression
\begin{equation}
n_{\rm c} = \sqrt{\frac{\ln{2}}{\pi}} \frac{\Sigma^{\rm peak}_{\rm H_2}}{\mu m_{\rm H} R^{\rm FWHM}} \left[ {\rm erf} \left( \sqrt{\ln{2}} \frac{\theta^{\rm Total}}{\theta^{\rm FWHM}} \right) \right]^{-1}
\end{equation}
where $\theta$ is the deconvolved angular radius for the Total or FWHM definition (see \S\ref{ssec:FluxAndConcentration}) and $\mu$ is the mean molecular weight.
While a density distribution described by a singly peaked Gaussian is a crude assumption, it is more representative of higher density sub-structures within starless clumps than just the average density with the FWHM.
Figure \ref{fig:FreeFallMass} shows the average free fall time plotted versus the mass of the clumps of the MC simulation. 
There is no significant correlation between the average free-fall time with the mass of the clumps.
The median is $t_{\rm ff,c} = 0.84\pm0.02$ Myr for the starless clump candidates.

Since the \bgps\ clump FWHM sizes are typically a factor of two larger when compared to the Jeans length of $0.36 \, {\rm pc} \, (T_{\rm K} / 15 \: {\rm K})^{1/2} (10^3 \: {\rm cm}^{-3}/ n_{\rm c})^{1/2}$, these starless clumps should fragment into smaller structures.
Higher resolution observations of a subset of \bgps\ clumps show this fragmentation into objects with higher densities \citep{merello15}.
As a result, the median \tff\ is likely an upper limit to the free-fall time for a typical starless clump candidate to form a single detectable star.

\subsubsection{Clump Lifetimes from Number Statistics}\label{ssec:AverageLifetimes}
If the Galactic population of clumps is in steady state where the clump formation rate is equal to the destruction rate then the number fraction of clumps in a given state is proportional to the lifetime of that state.
We find nearly equal numbers of starless clump candidates ($47.5\%$) compared to protostellar clumps.
This result implies that the typical lifetime of a starless clump candidate is nearly equal to the typical lifetime of a protostellar clump.  
The starless clump lifetime is defined as the time it takes for a clump detectable by the BGPS to form a single star that is detectable in current infrared surveys of the Galactic plane.
The protostellar clump lifetime is determined by the time it takes a cluster of stars to form (not a single star) and dissipate the gas and dust to the point where the clump is no longer detectable in the BGPS.
This latter timescale is difficult to determine, but we can use the cluster formation timescale as a rough estimate.
This cluster formation timescale has a large range and is typically longer than $1$ Myr and up to a few Myrs (\citeauthor{rebull07} \citeyear{rebull07}, \citeauthor{chomiuk11} \citeyear{chomiuk11}, \citeauthor{morales13} \citeyear{morales13}; see review by \citeauthor{longmore14} \citeyear{longmore14}).
The total population statistics imply a long average timescale for starless clump candidates that is on the order of a few times the average free-fall time of the clumps and that is consistent with the effects of turbulence and or magnetic fields lengthening the starless evolution timescale.

A single timescale cannot properly describe the starless clump evolution timescale because clumps with different masses should evolve out of the starless phase at different rates. 
While a median mass starless clump candidate ($230$ \msun ) may have a lifetime over $1$ Myr, it seems highly unlikely that $10^4$ \msun\ starless clumps exist in a starless phase for that long.
The lack of very massive ($M > 1.4 \times 10^4$ \msun ) starless clump candidates argues for a shorter timescale for these objects to form a single detectable star and move to the Protostellar clump category. 
Current single timescale estimates (or upper limits) in the literature for massive starless clumps \citep{csengeri14,traficante15b} are problematic because they do not account for this variation with mass and because they also attempt to compare to a single timescale for the lifetime of the embedded massive star formation phase \citep[i.e.,][]{davies11,duartecabral13}.
For the remaining discussion, we analyze how starless clump lifetimes vary with the mass of the clumps.

We can estimate how the clump lifetime for massive clumps vary with mass by using a method pioneered in \citeauthor{battersby13t} (\citeyear{battersby13t}, Ph.D.~Thesis) by pinning the relative lifetimes of clumps to an estimated absolute lifetime of the Class II \metho\ maser $\tau_{\rm CH_3OH}=2.5-4.5\times10^4$ yr \citep{vanderwalt05}.
This timescale is a statistical estimate based on correcting the number of the observed \metho\ masers in \cite{pestalozzi05} for completeness to predict the total Galactic count of high-mass stars with mass $> 20$ M$_\odot$.
\cite{vanderwalt05} uses a ${\rm SFR}=4$ M$_\odot$ yr$^{-1}$ based on the ${\rm SFR}=3-6$ M$\odot$ yr$^{-1}$ range in \cite{boissier99}.
\cite{chomiuk11} re-analyze several Galactic SFR measurements (see Table 1 therein) to derive ${\rm SFR}=1.9\pm0.4$ M$_\odot$ yr$^{-1}$.
The \cite{chomiuk11} study is unique in that they apply the same IMF \citep{kroupa01} and SFR law to all measurements and use the Starburst99 code to self-consistently compare studies with substantially different methodologies (including free-free radio continuum observations as well as YSO counts).
Scaling the CH$_3$OH maser lifetime inversely by their revised SFR yields $\tau_{\rm CH_3OH}=6.1-9.3\times10^4$ yr.

The clump lifetime is calculated using the relative number fraction of sources compared to the \metho\ maser sample at a given mass via
\begin{equation}
    \tau_{\rm SCC} = \tau_{\rm CH_3OH} \left( \frac{N_{\rm SCC}}{N_{\rm CH_3OH}} \right) \; \; 
\end{equation}
where $\tau_{\rm CH_3OH}$ is the completeness corrected and the SFR corrected Class II \metho\ maser lifetime.
The ratio of $N_{\rm SCC}/N_{\rm CH_3OH}$ is corrected for the fraction of SCCs and clumps with CH$_3$OH that are in the Distance sample compared to the total sample.
The fundamental assumption of this method is that all clumps capable of forming a $20$ \msun\ star ($M_{\rm clump} > 1000$ \msun ) will at some time form high-mass YSOs with detectable \metho\ masers.  
A population of equal mass clumps that are not capable of producing observable \metho\ masers would decrease the observed fraction and lead us to overestimate the inferred lifetime.
This method also assumes the lifetime of a single \metho\ maser ($\tau_{\rm CH_3OH}$) for all clumps with \metho\ maser associations.
For clumps associated to multiple \metho\ maser sites (i.e., YSOs) an age spread within the clump would increase the observed duration of the clump \metho\ maser phase and lead us to underestimate the phase lifetime when the shorter, single YSO $\tau_{\rm CH_3OH}$ is applied; however, for clumps associated to MMB sources, which have accurate interferometric positions, only $25/272$ ($9.2\%$) are host to more than one \metho\ maser, so this does not likely introduce a strong bias.
To calculate the lifetimes as a function of mass we extend this method to the relative fractions of MC samples in narrow mass bins ($0.0125$ dex logarithmic bins for $10^4$ MC mass samples per clump).
In effect this is the ratio of two mass PDFs each scaled by the sample sizes and then multiplied by $\tau_{\rm CH_3OH}$.

\begin{figure}
\centering
    \includegraphics[width=0.47\textwidth]{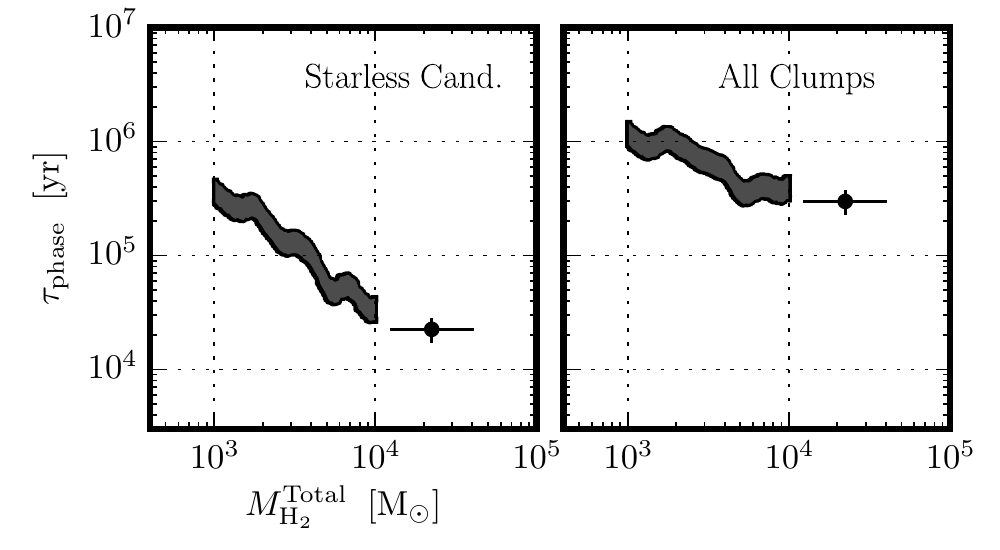}
\caption{
Clump lifetimes as a function of total mass.
Relative lifetimes are computed from the number of MC samples in a narrow mass range and scaled using an absolute lifetime for the Class II \metho\ maser, between $6.1-9.3\times10^4$ yr, corresponding to the upper and lower limits of the filled curves.
The markers show the average values for $M>10^4$ M$_\odot$ where $N \lesssim 30$.
}
\label{fig:Lifetime}
\end{figure}

Figure \ref{fig:Lifetime} shows the lifetime of SCCs as a function of mass.
The starless candidate clump lifetime approximately follows $\tau_{\rm SCC}\sim 0.37 \pm 0.08 \, {\rm Myr} \left(10^3 {\rm M}_{\odot} / M \right)$ for $M > 10^3$ \msun .
The uncertainty corresponds the systematic uncertainty in $\tau_{\rm CH_3OH}$.
It is not advisable to continue our timescale analysis to $< 1000$ M$_{\odot}$ as the assumption that all clumps will form a CH$_3$OH maser is less likely to be true; however, extrapolation of the trend to lower masses indicates that the starless clump lifetime becomes longer than the average clump free-fall time at $< 450$ \msun .  This is approximately twice the median SCC mass.  Thus, most starless clump candidates have lifetimes longer than their average free-fall times of $0.8$ Myr.
In turn, very massive starless clump candidates with $M_{\rm H_2} > 10^4$ M$_\odot$ have very short lifetimes $\tau \lesssim 0.03$ Myr consistent with lack of such massive objects detected in the BGPS survey \citep{ginsburg12}. 
Our lifetimes estimates are $5$ times longer and $3.5$ times longer than the lifetimes estimates of \cite{tackenberg12} for ATLASGAL clumps capable of forming $20$ \msun\ and $40$ \msun\ stars respectively.
The \cite{tackenberg12} estimates are based on large extrapolations from small numbers of clumps ($6$ and $1$ respectively) observed in their $20$ square degree survey area to the entire Galaxy.

\subsection{Origin of Starless vs. Protostellar Clump Mass Difference}\label{ssec:OriginOfMassDifference}
\begin{figure}
\centering
    \includegraphics[width=0.47\textwidth]{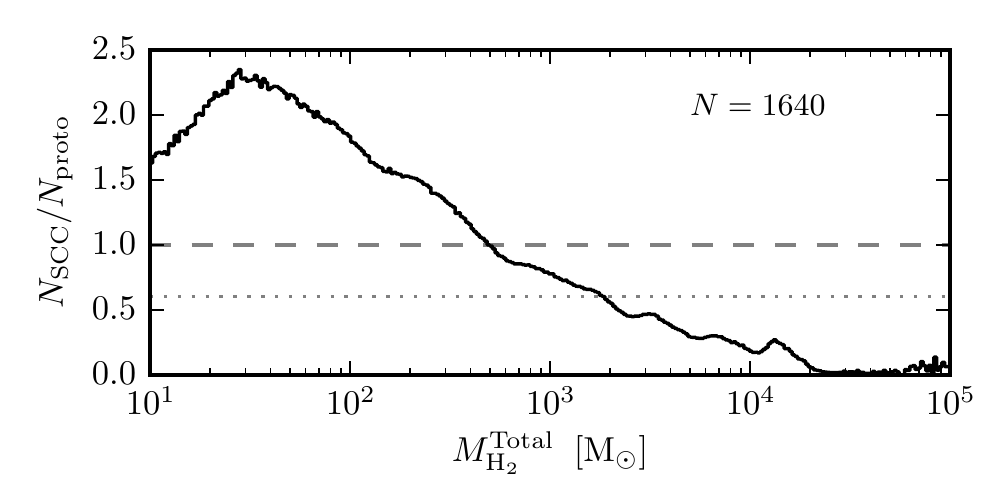}
\caption{
Ratio of starless clump candidates to protostellar clumps by total clump mass drawn from the MC simulations.
The dotted line shows the total sample fraction $613/1027 \approx 0.597$.
The plotted values taken from the Distance Sample are corrected by the factor $1027/613 \approx 1.675$ to be representative of the full survey.
}
\label{fig:MassFraction}
\end{figure}

In \S\ref{ssec:ClumpMass} a systematic difference between the median mass of protostellar clumps and starless clump candidates of $170 - 370$ \msun\ was discussed.
Fundamentally, the median mass difference is due to more starless clump candidates with masses below $M \lesssim 470$ \msun\ and fewer higher mass starless clump candidates than protostellar clumps (see Figure \ref{fig:MassFraction}).
We explore three possible physical explanations for this mass difference: (i) not all starless clump candidates will evolve into detectable protostellar clumps, (ii) clumps grow in mass by accreting surrounding material from their parent GMC, and (iii) clumps evolve at different rates from the starless to protostellar phase in exact proportion to their number statistics. 

\subsubsection{Infertile Starless Clumps}
Perhaps the simplest physical explanation for the systematic mass difference is that there is a large population of clumps among the SCCs that, while starless, will not evolve to form YSOs detectable by our star formation indicators.
In order for this population of objects to remain starless clump candidates, they would either remain ``infertile'' and never form stars or would only be able to form stars with bolometric luminosities that are less than the $30 - 140$ \lsun\ sensitivity (corresponding to $1 - 2$ \msun\ protostars accreting at typical rates) of current far-infrared Galactic plane surveys (\S\ref{sec:DevCatalog}). 
Such a population of low-density and low-mass clumps would bias the SCC distribution to lower masses.  
For a Kroupa IMF, a $230$ \msun\ median mass SCC is capable of forming stars up to $6$ \msun , which is detectable given the sensitivity limits of current far-infrared surveys (see Figure \ref{fig:HigalComplFunc}); however, for SCCs at the $50$\% BGPS mass completeness limit of $70$ \msun , the maximum mass star formed is only $2.5$ \msun\ which is closer to the sensitivity limit of current mid- and far-infrared surveys.

One test of this hypothesis is to check for the presence of dense gas.
The HCO$^+$ $3-2$ transition has an effective excitation density $n_{\rm eff} > 10^4$ cm$^{-3}$ for the typical properties of starless clump candidates \citep{shirley15}.
HCO$^+$ $3-2$ from \cite{shirley13} was detected towards $44\%$ of SCCs and $78\%$ of protostellar clumps in the Distance Sample.
This result indicates that there is either a population of lower density SCCs or that their filling fraction of dense gas is less than toward protostellar clumps.
Since the $3-2$ transition has an upper energy level that is $25.7$ K above ground, the effective excitation density ($n_{\rm eff}$) is a very sensitive function of $T_{\rm K}$ at the median temperature of $12.8$ K measured toward SCCs.
The sensitivity of the $3-2$ transition to $T_{\rm K}$ means that the detection is biased against the SCCs that are colder on average than protostellar clumps.
So this test is biased meaning that since SCCs are colder on average than protostellar clumps, HCO$^+$ $3-2$ requires higher densities in SCCs (by a factor of $\approx 2$) to be excited than toward protostellar clumps.

\begin{figure}
\centering
    \includegraphics[width=0.47\textwidth]{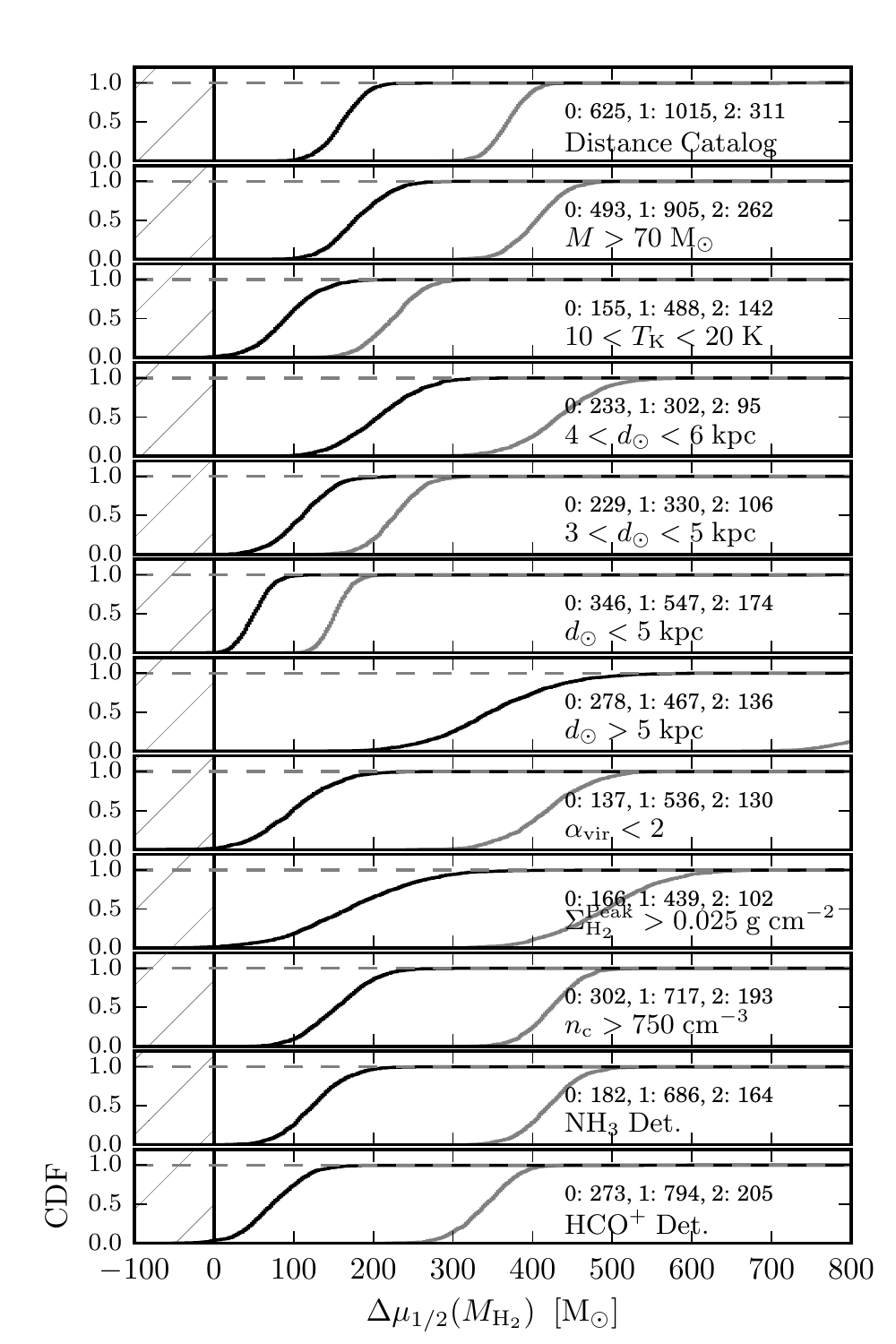}
\caption{
Cumulative distribution functions of the median mass difference $\Delta \mu_{1/2}(M_{\rm H_2})$ drawn from the MC simulations based on different cuts in physical properties.
The black curve shows the difference between the \higal\ Unique and SCCs, and the gray curve shows the difference between the Protostellar and SCCs.
The cut is labeled in the bottom right of each panel and above are the sample numbers for SCC (0), Protostellar (1), and \higal\ $70\um$ Unique (2).
The values $\Delta \mu_{1/2}(M_{\rm H_2})\sim100$ M$_\odot$ agree across a wide range of physical cuts designed to remove low mass, low density, or unbound objects.
}
\label{fig:MassSubsets}
\end{figure}

Another test of this hypothesis is to carefully control systematic effects from sampling clumps with different physical properties by performing a variety of astrophysical cuts on the sample and calculate the distribution of $\Delta\mu_{1/2}(M_{\rm H_2})$ from the MC simulations.
These cuts are designed to remove low-mass, low-density, or unbound objects.
Figure \ref{fig:MassSubsets} shows the cumulative distribution functions (CDFs) for $\Delta\mu_{1/2}(M_{\rm H_2})$ subsamples selected on mass, \amm-derived gas kinetic temperature, heliocentric distance, virial parameter, mass surface density, central density, and molecular detections.
The observed mass differences between categories, $\Delta\mu_{1/2}(M_{\rm H_2})$ are robust to these criteria, with the values calculated with respect to the \higal\ $70\um$ Unique category at $\Delta\mu_{1/2}(M_{\rm H_2})\sim100-200$ M$_\odot$ and the values calculated with respect to the Protostellar category ranging between $\Delta\mu_{1/2}(M_{\rm H_2})\sim200-500$ M$_\odot$.
The robustness of the median mass difference to astrophysical cuts does not support the explanation that a population of low star formation potential SCCs solely drive the increasing trend in clump total mass.

\subsubsection{Possible Clump Mass Growth}\label{ssec:ClumpMassGrowth}
\bgps\ clumps commonly reside in a complex structural hierarchy of clouds and filaments.
When sorted by protostellar indicator, the increasing trend in both $\Sigma$ and $M_{\rm tot}$ but weak increasing trend in $R_{\rm eq}$ are consistent with the physical process of clumps, as focusing points of local gravitational potential minima, accreting mass from the surrounding cloud in a ``conveyor belt'' fashion while maintaining the same approximate volume. 
Clump mass growth via accretion from the surrounding molecular cloud is a possible  mechanism for the difference between SCCs and protostellar clump masses.
An upper-limit to the clump accretion rate can be set by the free-fall time of the highest density sub-component of the clump.
An estimate of \tff\ from the MC simulation shown in Figure \ref{fig:FreeFallMass}, suggest a median $\tff = 0.84$ Myr. 
For a median mass difference between the SCC and $70$ $\mu$m unique clumps or protostellar clumps of $170 - 370$ \msun, this yields an accretion rate of $\approx 200 - 440$ \msun\ Myr$^{-1}$ as an upper limit if this is assumed to be the sole explanation for the mass difference.  
Note that all molecular clouds cannot be collapsing on free-fall timescales and still be consistent with the observed Milky Way SFR \citep{zuckerman74}.
Other mechanisms such as magnetic and turbulent support would increase the clump collapse time and decrease the predicted growth rate.

As a simple estimate, we calculate the total mass available to a clump embedded in a GMC with typical properties.
We assume a GMC with a spherical inflow velocity of $\Delta v \sim 1$ km s$^{-1}$ onto a clump, feeding from a zone of $\Delta R_{\rm feed} \sim 1.0$ pc over $t \sim 1$ Myr.
For median sized clump with $R_{\rm eq} \sim 1$ pc and feeding zone density $n \sim 150$ cm$^{-3}$ (N.B. median $n = 230$ cm$^{-3}$ in GMCs in the Galactic Ring Survey, \citeauthor{romanduval09} \citeyear{romanduval09}),  the homogeneous, spherical-shell reservoir mass would be $M_{\rm res} \sim 250$ \msun.
Higher inflow velocity or density would both increase the available mass.
These values show that reasonable physical conditions within a GMC yield estimates that are consistent with the observed median mass difference between the SCCs and protostellar clumps, suggesting that cloud-to-clump accretion as a plausible physical mechanism.  

There are theoretical models of such large scale clump accretion.
For instance, \cite{vazquezsemadeni09} invoke a model of hierarchical gravitational fragmentation in which multi-scale collapse occurs.
In this picture, the mass accretion rate onto the clumps is determined by the clump tidal radius and initial gravitational well hierarchy (see also \citeauthor{smith09} \citeyear{smith09}).
In another approach, \cite{murray12} have explored a model of Bondi-Hoyle accretion of clumps within GMCs and predict that the clump mass accretion rate is slightly super-linear with mass of the clump ($\dot{M} \sim M_{\rm clump}^{5/4}$).
\cite{maschberger14} instead use a model of stochastic growth with a sub-linear mass growth rate ($\dot{M} \sim M_{\rm sink}^{2/3}$).  
The exact nature of clump accretion and how it depends on the mass of the clump is still under debate; nevertheless, there is theoretical support for cloud-to-clump accretion.

There is also direct observational evidence for large scale flows onto clumps and filaments.
One of the most striking examples is a flow estimated at $2500$ \msun\ Myr$^{-1}$ traced by HCO$^+$ $1-0$ blue asymmetric line profiles along a filamentary complex into the central clump of the source SDC$335.579-0.272$ \citep[SDC335;][]{peretto13}.
The total mass of this region is $5500$ \msun\ which would place this among the most massive clumps in our survey.
Another example of inflow onto clumps is seen in the converging flows toward the DR21 filament complex with inflow rates of $1000$ \msun\ Myr$^{-1}$ onto $4900$ and $3300$ \msun\ clumps from infall profiles of the $1-0$ transitions of HCO$^+$ and CO \citep{schneider10}.
Analysis of velocity centroids can also reveal potential mass inflow onto clumps along filaments \citep[see three examples from the IRDC survey in][]{tackenberg14} although there can be a degeneracy between inflow and outflow in the interpretation \citep{henshaw14}.
These regions are already forming protostars and display hub-filament geometries where filaments of dense gas appear to converge on the central hub, so they are not exact analogues to the massive starless clumps in the BGPS survey; nevertheless, they show that clumps are deep gravitational potentials that draw surrounding material in.   

The examples of large scale inflow also extend to nearby lower mass regions.
For example, inflow signatures are seen in line profiles of HNC $1-0$ toward the Serpens South region indicating mass flow rates of $130$ \msun\ Myr$^{-1}$ onto the filament complex and $28$ \msun\ Myr$^{-1}$ onto the central cluster (\citeauthor{kirk13} \citeyear{kirk13}; for a higher resolution study of the same region, see \citeauthor{fernandezlopez14} \citeyear{fernandezlopez14}).
The central filament (L1495, B213) of the low-mass star-forming Taurus complex also shows direct evidence of a large scale flow in CO $1-0$ of $27-50$ \msun\ pc$^{-1}$ Myr$^{-1}$ perpendicular onto the filament complex \citep{palmeirim13}.
While these last two examples occur in lower mass regions than the typical BGPS starless clumps, they show that flow rates of $10 - 100$ \msun\ Myr$^{-1}$ are possible in low-mass regions.
It therefore seems plausible that higher mass inflow rates of a few $100$ \msun\ Myr$^{-1}$ should be possible in the more massive starless clump regions see in the BGPS.

\subsubsection{Different Starless Clump vs. Protostellar Clump Lifetimes}\label{ssec:DifferentLifetimes}
A third explanation for the observed mass difference is that the lifetime of the SCC phase decreases with increasing total clump mass in exact proportion to the fraction of SCCs at each mass.
The shorter lifetime of massive clumps explains the lack of very massive ($M > 10^4$ \msun ) starless clumps candidates.
We estimate the lifetime of this phase is less than $0.03$ Myr based on the \metho\ maser technique.
Thus protostellar clumps with $> 10^4$ \msun\ require an alternative formation mechanism: the average mass-growth from cloud-to-clump accretion is not sufficient to explain the lack of these massive SCCs.
Using even a high mass infall rate of $\sim10^3$ \msun\ Myr$^{-1}$, the most massive clumps could only accrete a meager few $\sim10^1$ \msun\ over their few $\sim10^{4}$ yr starless-phase lifetimes.

The lifetime ratio of SCCs to Protostellar clumps would have to decrease with increasing mass in proportion to the ratio $N_{\rm SCC}/N_{\rm proto}$ if lifetimes arguments are the sole explanation for the observed median mass difference.  
The ratio of $N_{\rm SCC}/N_{\rm proto}$ scales as $M^{-0.4}$ over the range $100 \leq M \leq 1000$ \msun .
The negative exponent implies that the lifetime scaling for SCCs is a steeper function of mass than the Protostellar clump lifetime.  
The value of the exponent is insensitive to the lower bound of the mass between $100$ and $300$ \msun .
There is no known theoretical prediction of such a lifetime scaling between high-mass SCCs and high-mass Protostellar clumps.

With the current data, it is difficult to disentangle the relative importance of differing lifetimes for SCCs and Protostellar clumps versus the importance of clump mass growth in explaining the median mass difference between SCCs and Protostellar clumps.
It is likely that both processes occur.

\section{Summary}\label{sec:Summary}
We present a sample of \ovsample\ molecular cloud clumps from the \bgps\ sorted by different observational indicators of star formation activity.
We use a variety of Galactic plane surveys in a common overlap region between \lonrange\ that include: $70 \um$ compact sources from \herschel\ \higal, mid-IR color-selected YSOs, \water\ and \metho\ masers, and \uchii\ regions.
We use Monte Carlo random sampling to calculate the clump physical properties using a subsample of \ovbcdpdfs\ well-constrained clump DPDFs.  We also present a catalog of 1663 clump \amm\ \tk\ measurements and 22 GHz \water\ maser observations.

We find the following conclusions:
\begin{enumerate}
  \item We identify a subsample of \ovscc\ dense clumps with no indicators of star formation activity, representing the largest and most robust sample of starless clump candidates from a blind survey to date.
  \item We measure numerous increasing trends in median physical properties from starless candidates to the most extreme indicator of star formation, \uchii\ regions. These include:
    $S^{\rm Total}_{1.1}$ (Jy),
    $\Omega^{\rm Total} / \Omega^{\rm FWHM}$,
    $T_{\rm K}$ (K),
    $\Delta v({\rm NH_3})$ (km s$^{-1}$),
    $R^{\rm Total}_{\rm eq}$ (pc),
    $\Sigma^{\rm FWHM}_{\rm H_2}$ (g cm$^{-2}$),
    $n_{\rm c}$ (cm$^{-3}$), and
    $M^{\rm Total}_{\rm H_2}$ (\msun).
  \item Median mass SCCs ($230$ \msun ) are capable of forming intermediate mass protostars up to $6$ \msun\ (assuming a Kroupa IMF and $\epsilon_{\rm SF}=0.3$) which are detectable by current mid and far-infrared Galactic plane surveys. Mass is well correlated with radius for BGPS clumps, but the observed $M$-$R$ relationship lies above the \cite{kauffmann10} thresholds for massive star formation for only $10.8\%$ of SCCs.
  \item The average SCC free-fall timescale is $\langle t_{\rm ff} \rangle = 0.8$ Myr.  Using CH$_3$OH masers and counting the number of SCCs to CH$_3$OH maser containing clumps, we estimate that the lifetime of massive SCCs decreases as $\sim 0.37 \pm 0.08 \, {\rm Myr} \left(10^3 {\rm M}_{\odot} / M \right)$ for $M > 10^3$ \msun .  There are no SCCs observed with $M > 1.4 \times 10^4$ \msun\ indicating a very short lifetime of $< 0.03$ Myr.  The majority of SCCs (median total mass $230$ $M_\odot$) however have lifetimes that are longer than their average free-fall timescale.
   \item Virial parameters derived from \amm\ show sub-virial clumps with median $\alpha = 0.7$.  More than $75\%$ of \bgps\ clumps are gravitationally bound ($\alpha < 2$). A median-property SCC would be in virial equilibrium if a modest $\sim 50$ $\mu$G magnetic field is present.
  \item We find a median mass difference between the SCC and prostellar categories of $\Delta M^{\rm Total}_{\rm H_2} = 170 - 370$ \msun.  It is unlikely that this mass difference can be solely explained by a population of SCCs which are incapable of forming detectable protostars in current Galactic plane surveys.  
If the observed mass difference is not due to a systematic increase in dust opacity of $1.7-2.5$ from SCCs to protostellar clumps, then there are
two likely explanations: a) SCCs accrete mass from their surroundings and b) that the lifetime of SCCs is proportionally longer than Protostellar clumps at the same mass below $470$ \msun\ and shorter than Protostellar clumps at masses above $470$ \msun .  If mass accretion is the sole explanation then for an average clump free-fall timescale, $\dot{M} \sim 200 - 440 \msun\ \rm{Myr}^{-1}$ is required.  If the variation in SCC and Protostellar clump lifetimes is the sole explanation, then their ratio of lifetimes should scale as $M^{-0.4}$ for $100 \leq M \leq 1000$ \msun .  In reality, both processes likely occur although it is not possible to disentangle them with the current data.
\end{enumerate}

We sincerely thank the observatory staff at the Green Bank Telescope for their help during observing.
We also sincerely thank Shari Breen for sharing coordinates to the MMB catalogue between $20\degree < \ell < 60\degree$ prior to publication.
We would also like to thank Julia Kamenetzky and Kimberly Ward-Duong for many useful comments that benefited this paper.
BES was supported by the National Science Foundation Graduate Research Fellowship under Grant No. DGE-1143953.
YLS and BES were supported in part by the NSF Grants AST-1008577 and AST-1410190.
NJE was supported by NSF grant AST-1109116 to the University of Texas at Austin.
EWR was supported by a Discovery Grant from NSERC of Canada.

%%%%%%%%%%%%%%%%%%%%%%%%%%%%%%%%%%%%%%%%%%%%%%%%%%%%%%%%%%%%%%%%%%%%%%%%%%%%%%%%
%				Bibliography
%%%%%%%%%%%%%%%%%%%%%%%%%%%%%%%%%%%%%%%%%%%%%%%%%%%%%%%%%%%%%%%%%%%%%%%%%%%%%%%%

\bibliographystyle{apj}
\bibliography{mybib}

\newpage

%%%%%%%%%%%%%%%%%%%%%%%%%%%%%%%%%%%%%%%%%%%%%%%%%%%%%%%%%%%%%%%%%%%%%%%%%%%%%%%%
%				  Tables
%%%%%%%%%%%%%%%%%%%%%%%%%%%%%%%%%%%%%%%%%%%%%%%%%%%%%%%%%%%%%%%%%%%%%%%%%%%%%%%%

\clearpage

\begin{turnpage}
\global\pdfpageattr\expandafter{\the\pdfpageattr/Rotate 90}
\begin{deluxetable*}{clccccccrr}
\tabletypesize{\footnotesize}
\tablewidth{0pt}
\tablecaption{\label{tab:mmprops}Observed 1.1 mm Properties}
\tablehead{
% heading names
  \colhead{ID}
& \colhead{}
& \colhead{Peak R.A.}
& \colhead{Peak Decl.}
& \colhead{$\Omega^{\rm FWHM}$}
& \colhead{$\Omega^{\rm Total}$}
& \colhead{$\theta^{\rm FWHM}_{\rm eq}$}
& \colhead{$\theta^{\rm Total}_{\rm eq}$}
& \colhead{$S^{\rm FWHM}_{1.1}$}
& \colhead{$S^{\rm Total}_{1.1}$} \\
% heading units
  \colhead{number}
& \colhead{Source}
& \colhead{(J2000)}
& \colhead{(J2000)}
& \colhead{(sq. \am)}
& \colhead{(sq. \am)}
& \colhead{(\as)}
& \colhead{(\as)}
& \colhead{(mJy)}
& \colhead{(mJy)}
}
\startdata
     2351 &  BGPSv2\_G010.013$+$00.082 &  18 07 29.0 &  -20 14 18.0 &       1.613 &  2.534 &      39.7 &  51.3 &    760 (160) &  1000 (200) \\
     2352 &  BGPSv2\_G010.027$-$00.354 &  18 09 08.3 &  -20 26 15.8 &       1.598 &  3.456 &      39.5 &  60.7 &   1420 (190) &  2110 (290) \\
     2353 &  BGPSv2\_G010.051$-$00.208 &  18 08 38.6 &  -20 20 45.5 &       0.734 &  0.994 &      23.9 &  29.4 &    420 (120) &   490 (140) \\
     2354 &  BGPSv2\_G010.059$-$00.210 &  18 08 40.0 &  -20 20 23.8 &       0.965 &  1.037 &      28.9 &  30.3 &    250 (110) &   290 (120) \\
     2355 &  BGPSv2\_G010.061$-$00.174 &  18 08 32.2 &  -20 19 14.7 &       1.397 &  1.930 &      36.4 &  44.0 &    780 (160) &   930 (180) \\
     2356 &  BGPSv2\_G010.069$-$00.406 &  18 09 25.2 &  -20 25 34.2 &       1.210 &  5.328 &      33.4 &  76.4 &   1050 (160) &  2750 (350) \\
     2357 &  BGPSv2\_G010.079$-$00.414 &  18 09 28.2 &  -20 25 16.6 &       0.778 &  1.080 &      24.9 &  31.1 &    570 (120) &   680 (140) \\
     2358 &  BGPSv2\_G010.079$-$00.194 &  18 08 38.9 &  -20 18 52.9 &       1.670 &  3.096 &      40.5 &  57.2 &   1510 (190) &  2010 (250) \\
\enddata 
\tablecomments{Uncertainties are given in parentheses. \vspace{1mm} \\
(This table is available in its entirety in the online journal.
A portion is shown here for guidance regarding its form and content.)
}
\end{deluxetable*}  % tab:mmprops
\begin{deluxetable*}{rccccccccccc}
\tabletypesize{\footnotesize}
\tablewidth{0pt}
\tablecaption{\label{tab:ammprops}Observed NH$_{3}$ Properties}
\tablehead{
% heading names
  \colhead{ID}
& \colhead{GBT R.A.}
& \colhead{GBT Decl.}
& \colhead{$\theta_{\rm offset}$}
& \colhead{$v_{\rm LSR}$}
& \colhead{$\sigma_{\rm v}$}
& \colhead{$T_{\rm mb}$(1,1)}
& \colhead{$W$(1,1)}
& \colhead{$T_{\rm mb}$(2,2)\tablenotemark{b}}
& \colhead{$W$(2,2)\tablenotemark{b}}
& \colhead{$T_{\rm mb}$(3,3)\tablenotemark{b}}
& \colhead{$W$(3,3)\tablenotemark{b}} \\
% heading units
  \colhead{Number\tablenotemark{a}}
& \colhead{(J2000)}
& \colhead{(J2000)}
& \colhead{(\as)}
& \colhead{(km s$^{-1}$)}
& \colhead{(km s$^{-1}$)}
& \colhead{(K)}
& \colhead{(K km s$^{-1}$)}
& \colhead{(K)}
& \colhead{(K km s$^{-1}$)}
& \colhead{(K)}
& \colhead{(K km s$^{-1}$)}
}
\startdata
     2351 &     18 07 28.3 &     -20 14 10.3 &  12.9 &     30.1300 (0.1600) &        0.6900 (0.1600) &      0.73 (0.20) &     1.57 (0.58) &      0.58 (0.20) &    -0.06 (0.58) &      0.67 (0.22) &     0.83 (0.58) \\
     2356 &     18 09 25.6 &     -20 25 43.6 &  11.0 &     11.7570 (0.0170) &        0.8700 (0.0160) &      3.82 (0.22) &    32.32 (0.64) &      1.37 (0.20) &     3.07 (0.63) &      0.88 (0.22) &     0.21 (0.64) \\
     2358 &     18 08 38.8 &     -20 18 58.9 &  6.1 &     28.2010 (0.0170) &        0.8770 (0.0160) &      3.62 (0.21) &    27.42 (0.62) &      1.29 (0.19) &     3.35 (0.62) &      0.88 (0.23) &     0.20 (0.69) \\
     2360 &     18 09 32.3 &     -20 24 07.7 &  11.5 &     11.7862 (0.0098) &        0.8173 (0.0093) &      6.03 (0.21) &    42.04 (0.67) &      2.39 (0.20) &     4.62 (0.69) &      1.02 (0.21) &     0.05 (0.74) \\
     2361 &     18 08 01.4 &     -20 12 19.1 &  10.8 &     41.8440 (0.0160) &        1.0290 (0.0140) &      4.04 (0.21) &    34.86 (0.64) &      2.04 (0.20) &     5.65 (0.68) &      0.92 (0.22) &    -0.71 (0.74) \\
     2367 &     18 09 27.0 &     -20 19 11.2 &  3.5 &     14.1900 (0.0220) &        2.2420 (0.0210) &      4.92 (0.25) &    68.37 (0.90) &      3.81 (0.22) &    20.81 (0.86) &      4.55 (0.33) &    30.31 (1.28) \\
     2372 &     18 09 36.9 &     -20 18 41.8 &  9.3 &     10.7959 (0.0046) &        0.8328 (0.0041) &     11.89 (0.23) &    90.16 (0.69) &      6.88 (0.21) &    25.18 (0.69) &      3.03 (0.23) &    10.36 (0.76) \\
     2376 &     18 09 05.4 &     -20 13 39.7 &  2.0 &     13.2780 (0.0240) &        0.8210 (0.0240) &      2.59 (0.21) &    16.26 (0.61) &      1.04 (0.20) &     0.18 (0.62) &      0.84 (0.24) &    -0.43 (0.70) \\
\enddata
\tablenotetext{a}{B, C, D, and E denote multiple ammonia pointings that fall within a single 1.1 mm source.}
\tablenotetext{b}{Upper-limits are $T_{mb} < 4\sigma$ and $W < 5\sigma \Delta v  \sqrt{N}$, where $\sigma$ is the rms noise, $\Delta v$ is the width of a single channel in velocity, and $N$ is the number of pixels over which the average RMS was calculated.}
\tablecomments{Uncertainties are given in parentheses. \vspace{1mm} \\
(This table is available in its entirety in the online journal.
A portion is shown here for guidance regarding its form and content.)
}
\end{deluxetable*}  % tab:ammprops
\end{turnpage}

\clearpage

\global\pdfpageattr\expandafter{\the\pdfpageattr/Rotate 0}

\begin{deluxetable*}{rccccccccccc}
\tabletypesize{\footnotesize}
\tablewidth{0pt}
\tablecaption{\label{tab:waterprops}Observed H$_2$O Maser Properties}
\tablehead{
% heading names
  \colhead{ID}
& \colhead{GBT R.A.}
& \colhead{GBT Decl.}
& \colhead{$\theta_{\rm offset}$}
& \colhead{$v_{\rm LSR}$(peak)}
& \colhead{$v_{\rm spread}$}
& \colhead{$T_{\rm mb}$}
& \colhead{$W$}
& \colhead{$N_{\rm lines}$} \\
% heading units
  \colhead{Number}
& \colhead{(J2000)}
& \colhead{(J2000)}
& \colhead{(\as)}
& \colhead{(km s$^{-1}$)}
& \colhead{(km s$^{-1}$)}
& \colhead{(K)}
& \colhead{(K km s$^{-1}$)}
& \colhead{}
}
\startdata
     2380 &     18 07 49.6 &     -20 01 43.2 &   5.9 &        49.97 &         5.59 &         2.94 (0.17) &         6.67 (0.22) &                      2 \\
     2381 &     18 08 41.6 &     -20 07 40.6 &  13.2 &         6.85 &         1.31 &         1.94 (0.17) &         1.98 (0.12) &                      1 \\
     2386 &     18 08 48.1 &     -20 05 55.0 &   9.1 &         6.04 &        12.67 &        25.23 (0.17) &        31.67 (0.26) &                      4 \\
     2392 &     18 09 23.1 &     -20 08 08.7 &  10.9 &        42.93 &        23.69 &        29.36 (0.20) &        34.43 (0.25) &                      3 \\
     2394 &     18 09 02.2 &     -20 05 08.7 &   9.1 &        22.57 &        20.07 &        40.00 (0.17) &        74.63 (0.36) &                      7 \\
     2396 &     18 09 00.6 &     -20 03 34.3 &   2.2 &         9.47 &        32.58 &        15.35 (0.18) &       105.01 (0.43) &                      9 \\
     2406 &     18 09 21.5 &     -20 02 10.0 &  10.9 &        13.30 &         0.98 &         4.40 (0.17) &         2.99 (0.12) &                      1 \\
     2411 &     18 08 45.1 &     -19 54 33.2 &   2.4 &        71.05 &         5.59 &        60.60 (0.20) &        95.22 (0.23) &                      3 \\
\enddata
\tablecomments{Uncertainties are given in parentheses. \vspace{1mm} \\
(This table is available in its entirety in the online journal.
A portion is shown here for guidance regarding its form and content.)
}
\end{deluxetable*}  % tab:waterprops
\begin{deluxetable*}{rrrrrrrrrr}
\tabletypesize{\footnotesize}
\tablewidth{0pt}
\tablecaption{\label{tab:flagstats}Catalog Matching Statistics}
\tablehead{
% heading names
  \colhead{Catalog}
& \colhead{Min. $\ell$}
& \colhead{Max. $\ell$}
& \colhead{$<|b|$}
& \colhead{Total}
& \colhead{Source}
& \colhead{Source}
& \colhead{Source}
& \colhead{Clump}
& \colhead{Clump} \\
% heading units
  \colhead{}
& \colhead{(\degree)}
& \colhead{(\degree)}
& \colhead{(\degree)}
& \colhead{BGPS}
& \colhead{in BGPS}
& \colhead{Match}
& \colhead{Fraction}
& \colhead{Match}
& \colhead{Fraction}
}
\startdata
\multicolumn{1}{l}{\bf Hi-GAL}                &      10 &      65 &  (all BGPS) &          &             &           &          &              &            \\
 HG0                                          &         &         &             &     4683 &     \nodata &   \nodata &  \nodata &         1847 &     0.3944 \\
 HG1                                          &         &         &             &     4683 &     \nodata &   \nodata &  \nodata &          944 &     0.2016 \\
 HG2                                          &         &         &             &     4683 &     \nodata &   \nodata &  \nodata &         1044 &     0.2229 \\
 HG3                                          &         &         &             &     4683 &     \nodata &   \nodata &  \nodata &          665 &     0.1420 \\
 HG4                                          &         &         &             &     4683 &     \nodata &   \nodata &  \nodata &          182 &     0.0389 \\
 CuTeX                                        &         &         &             &     4683 &     \nodata &   \nodata &  \nodata &         1369 &     0.2923 \\
\multicolumn{1}{l}{\bf RMS}                   &      10 &      65 &  (all BGPS) &          &             &           &          &              &            \\
\enddata
\tablecomments{Columns list the (1) catalog matched to the BGPS, (2) minimum Galactic longitude of the survey if applicable, (3) maximum Galactic longitude, (4) range in Galactic latitude, (5) total number of BGPS clumps in the catalog survey region, (6) number of catalog sources in the region with overlapping BGPS coverage, (7) number of catalog sources associated to BGPS clumps, (8) fraction of catalog sources matched to BGPS clumps, (9) number of unique associated BGPS clumps (10) fraction of BGPS clumps with associations.}
\end{deluxetable*}  % tab:flagstats

\clearpage

\begin{turnpage}
\global\pdfpageattr\expandafter{\the\pdfpageattr/Rotate 90}
\begin{deluxetable*}{ccccccccccccccccccccccccccc}
\tabletypesize{\footnotesize}
\tablewidth{0pt}
\tablecaption{\label{tab:SfFlag}Star Formation Indicator Flags}
\tablehead{
% heading names
  \colhead{ID} & $\, $
& \multicolumn{2}{c}{Activity} & $\, $
& \multicolumn{1}{c}{Hi-GAL} & $\, $
& \multicolumn{2}{c}{MIPS} & $\, $
& \multicolumn{4}{c}{mid-IR} & $\, $
& \multicolumn{4}{c}{H$_2$O} & $\, $
& \multicolumn{4}{c}{CH$_3$OH} & $\, $
& \multicolumn{2}{c}{\uchii} \\
  \cmidrule{3-4}
  \cmidrule{8-9}
  \cmidrule{11-14}
  \cmidrule{16-19}
  \cmidrule{21-24}
  \cmidrule{26-27}
  \colhead{Number} &
& \colhead{SF?}
& \colhead{$P_{\rm proto}$} &
& \colhead{Flag} &
& \colhead{Cat.}
& \colhead{Arch.} &
& \colhead{IR?}
& \colhead{R08}
& \colhead{RMS}
& \colhead{EGO} &
& \colhead{H$_2$O?}
& \colhead{GBT}
& \colhead{Arc}
& \colhead{HOPS} &
& \colhead{CH$_3$OH?}
& \colhead{MMB}
& \colhead{Pesta.}
& \colhead{Pand.} &
& \colhead{UCHII?}
& \colhead{Corn.}
}
\startdata
 5096 && 1 & 1.00 && 2 && 5 & 5 && 1 & 1 & 0 & 0 && 1 & 1 & 0 & 0 && 1 & 2 & 0 & 0 && 1 & 1 \\
 5097 && 1 & 1.00 && 4 && 2 & 1 && 1 & 1 & 0 & 0 && 0 & 0 & 0 & 0 && 0 & 0 & 0 & 0 && 0 & 0 \\
 5098 && 1 & 1.00 && 1 && 5 & 7 && 1 & 1 & 0 & 0 && 0 & 0 & 0 & 0 && 0 & 0 & 0 & 0 && 0 & 0 \\
 5099 && 1 & 0.50 && 3 && 2 & 1 && 1 & 1 & 0 & 0 && 0 & 0 & 0 & 0 && 0 & 0 & 0 & 0 && 0 & 0 \\
 5100 && 1 & 1.00 && 1 && 1 & 2 && 1 & 1 & 0 & 0 && 0 & 0 & 0 & 0 && 0 & 0 & 0 & 0 && 0 & 0 \\
 5101 && 1 & 1.00 && 2 && 2 & 8 && 1 & 1 & 0 & 0 && 0 & 0 & 0 & 0 && 1 & 0 & 1 & 0 && 0 & 0 \\
 5102 && 1 & 1.00 && 1 && 3 & 6 && 1 & 1 & 0 & 0 && 0 & 0 & 0 & 0 && 0 & 0 & 0 & 0 && 0 & 0 \\
 5103 && 1 & 1.00 && 2 && 2 & 8 && 2 & 0 & 0 & 0 && 0 & 0 & 0 & 0 && 0 & 0 & 0 & 0 && 0 & 0 \\
\enddata
\tablerefs{see \S\ref{sec:DevCatalog} for catalog descriptions.}
\tablecomments{
Columns list the
(1) BGPS ID number,
(2) protostellar flag,
(3) protostellar probability including R08 AGB contamination,
(4) \higal\ visual inspection flag,
(5-6) number of MIPSGAL Catalog and Archive associations,
(7) mid-IR group, 2 for uniquely R08 AGB,
(8-10) number of mid-IR YSO associations,
(11) H$_2$O group, 2 for uniquely GBT non-detection,
(12-14) number of H$_2$O maser associations,
(15) CH$_3$OH group,
(16-18) number of CH$_3$OH maser associations, 
(19) \uchii\ group, and
(20) number of CORNISH associations. \vspace{1mm} \\
(This table is available in its entirety in the online journal.
A portion is shown here for guidance regarding its form and content.)
}
\end{deluxetable*}  % tab:SfFlag
\end{turnpage}

\clearpage

\global\pdfpageattr\expandafter{\the\pdfpageattr/Rotate 0}

\clearpage

\begin{deluxetable*}{ccccccccccccc}
\tabletypesize{\footnotesize}
\tablewidth{0pt}
\tablecaption{\label{tab:PhysProps}Clump Physical Properties}
\tablehead{
% heading names
  \colhead{ID} & $\, $
& \multicolumn{5}{c}{$d_\odot$ (pc)} & $\, $
& \multicolumn{5}{c}{$M^{\rm Total}_{\rm H_2}$ (M$_\odot$)} \\
  \cmidrule{3-7}
  \cmidrule{9-13}
  \colhead{Number} &
& \colhead{$15.9$}
& \colhead{$50.0$}
& \colhead{$84.1$}
& \colhead{ML}
& \colhead{$\sigma_{\rm 1/2}$} &
& \colhead{$15.9$}
& \colhead{$50.0$}
& \colhead{$84.1$}
& \colhead{ML}
& \colhead{$\sigma_{\rm 1/2}$}
}
\startdata
2409 & & $8073$ & $8630$ & $9264$ & $8711$ & $403.2$ & & $3863$ & $4597$ & $5522$ & $4547$ & $553.2$ \\
2416 & & $4610$ & $5050$ & $5592$ & $4714$ & $324.7$ & & $575.4$ & $719.3$ & $909.8$ & $629.1$ & $108.7$ \\
2418 & & $4602$ & $5036$ & $5581$ & $5187$ & $329.1$ & & $904.2$ & $1107$ & $1369.6$ & $962.5$ & $150.4$ \\
2421 & & $4603$ & $5046$ & $5578$ & $4899$ & $322.0$ & & $1318$ & $1617$ & $2016$ & $1455.5$ & $229.1$ \\
2424 & & $4595$ & $5042$ & $5577$ & $4852$ & $328.1$ & & $3763$ & $4575$ & $5662$ & $4842$ & $625.4$ \\
2425 & & $4594$ & $5037$ & $5576$ & $4800$ & $332.3$ & & $3147$ & $3830$ & $4744$ & $3888$ & $529.0$ \\
2430 & & $1564$ & $2497$ & $3853$ & $2209$ & $637.7$ & & $176.1$ & $449.0$ & $1301$ & $327.2$ & $219.0$ \\
2432 & & $2436$ & $3145$ & $3849$ & $3034$ & $451.4$ & & $165.7$ & $286.1$ & $459.5$ & $207.9$ & $91.93$ \\
\enddata
\tablecomments{
Physical properties calculated from the MC simulations for each clump in the Distance Sample.
The percentiles of the posterior distributions ($15.9-84.1$), maximum likelihood (ML), and median absolute deviation ($\sigma_{\rm 1/2}$) with uncertainties are calculated.
Columns list the
(1) BGPS ID number,
(2-4) heliocentric distance,
(5-8) total mass. \vspace{1mm} \\
(A subset of the derived physical properties ($d_\odot$, $M$) are shown above for guidance regarding the tables form and content.
This table is available in its entirety in the online journal.
}
\end{deluxetable*}  % tab:PhysProps
\LongTables
\begin{deluxetable*}{llrrrrrrr}
\tabletypesize{\footnotesize}
\tablewidth{0pt}
\tablecaption{\label{tab:StagesStats}Statistics of Observed and Physical Properties}
\tablehead{
  \colhead{Property}
& \colhead{Indicator}
& \colhead{$N$}
& \colhead{10}
& \colhead{25}
& \colhead{50 ($\mu_{1/2}$)}
& \colhead{75}
& \colhead{90}
& \colhead{$\sigma_{1/2}$}
\vspace{2mm}
}
\startdata
$\Omega^{\rm Total}$                     & ${\rm Starless \ Cand.}$                 & 2223 & $920.2^{ +3.7 }_{ -3.5 }$ & $1824.2^{ +4.7 }_{ -4.7 }$ & $3620.4^{ +8.2 }_{ -8.3 }$ & $6531^{ +17 }_{ -17 }$ & $10668^{ +38 }_{ -39 }$ & $2110.9^{ +6.2 }_{ -6.2 }$ \\
$({\rm sq.\ arcsec})$                    & ${\rm Protostellar}$                     & 2460 & $2002.9^{ +6.5 }_{ -6.6 }$ & $3678.3^{ +9.8 }_{ -10 }$ & $7114^{ +15 }_{ -16 }$ & $13755^{ +35 }_{ -36 }$ & $23286^{ +65 }_{ -64 }$ & $4191^{ +12 }_{ -12 }$ \\
                                         & ${\rm 70 \ \mu m \ Unique}$              &  995 & $1629.6^{ +7.2 }_{ -7.0 }$ & $2908^{ +11 }_{ -11 }$ & $4985^{ +17 }_{ -17 }$ & $9148^{ +28 }_{ -28 }$ & $14540^{ +63 }_{ -64 }$ & $2626^{ +12 }_{ -12 }$ \\
                                         & ${\rm Mid \ IR}$                         & 1022 & $2360^{ +14 }_{ -14 }$ & $4575^{ +15 }_{ -15 }$ & $8802^{ +42 }_{ -42 }$ & $17114^{ +59 }_{ -63 }$ & $30320^{ +130 }_{ -130 }$ & $5416^{ +22 }_{ -22 }$ \\
                                         & ${\rm H_2O}$                             &  556 & $3129^{ +16 }_{ -17 }$ & $5912^{ +35 }_{ -36 }$ & $12445^{ +49 }_{ -48 }$ & $23076^{ +78 }_{ -74 }$ & $36950^{ +240 }_{ -230 }$ & $7632^{ +36 }_{ -36 }$ \\
                                         & ${\rm CH_3OH}$                           &  298 & $3792^{ +23 }_{ -23 }$ & $7330^{ +49 }_{ -47 }$ & $15526^{ +77 }_{ -75 }$ & $27840^{ +160 }_{ -160 }$ & $40420^{ +270 }_{ -290 }$ & $9135^{ +67 }_{ -66 }$ \\
                                         & ${\rm UCH{\sc II}}$                      &  170 & $4661^{ +42 }_{ -44 }$ & $9635^{ +57 }_{ -54 }$ & $18060^{ +120 }_{ -110 }$ & $31190^{ +170 }_{ -180 }$ & $45480^{ +340 }_{ -330 }$ & $10238^{ +95 }_{ -120 }$ \\
[1ex] \hline \\
\enddata
\tablecomments{Columns list the (1) physical property, unit, and numerical scale factor if applied, (2) star formation indicator category, (3) number of clumps in calculation, (4-8) 10\textsuperscript{th} through 90\textsuperscript{th} percentile, and (9) median absolute deviation. Uncertainties show the $1\sigma$ confidence interval.
}
\end{deluxetable*}
  % tab:StagesStats

\end{document}